\documentclass{article}
\usepackage[english]{babel}
\usepackage[letterpaper,top=2cm,bottom=2cm,left=3cm,right=3cm,marginparwidth=1.75cm]{geometry}
\usepackage{caption}
\usepackage{subcaption}
\usepackage{amsmath}
\usepackage{graphicx}
\usepackage[colorlinks=true, allcolors=blue]{hyperref}
\usepackage{siunitx}
\usepackage{cancel}
\usepackage{graphicx}
\usepackage{pgfplots}
\usepackage{listings}
\usepackage{tikz}
\usepackage{hyperref}
\usepackage{float}
\usepackage{geometry}
\usepackage{lscape}
\usepackage{dirtytalk}
\usepackage{algorithm} 
\usepackage{algpseudocode}
\usepackage{listings}
\usepackage{fancyhdr}
\usepackage{csquotes}
\usepackage{enumitem}

\makeatletter
\def\BState{\State\hskip-\ALG@thistlm}
\makeatother

\pgfplotsset{width=10cm,compat=1.9}
\usepgfplotslibrary{external}
\title{Testing replication for an agent-based model of market fragmentation and latency arbitrage\thanks{C.M.V.O., M.T.T.K., and B.F.T. were supported by MITRE’s Financial Innovation Lab. E.R.C. was supported by the MITRE PhD Fellowship in Computational Finance within the Vermont Complex Systems Institute at the University of Vermont. The authors' affiliation with The MITRE Corporation is provided for identification purposes only, and is not intended to convey or imply MITRE's concurrence with, or support for, the positions, opinions, or viewpoints expressed by the authors.
\newline
©2026 The MITRE Corporation. ALL RIGHTS RESERVED.
Approved for Public Release; Distribution Unlimited. Public Release Case Number 25-1956.
}
}
\author{
    Ethan Ratliff-Crain\\
    \textit{Vermont Complex Systems Institute} \\
    \textit{University of Vermont}\\
    Burlington, VT 05405, USA \\
    \textit{The MITRE Corporation}\\
    McLean, VA 22012, USA \\
    ethan.ratliff-crain@uvm.edu
    \and
    Colin M. Van Oort\\
    \textit{The MITRE Corporation}\\
    McLean, VA 22012, USA
    \and
    Matthew T. K. Koehler\\
    \textit{The MITRE Corporation}\\
    McLean, VA 22012, USA
    \and
    Brian F. Tivnan\\
    \textit{The MITRE Corporation}\\
    McLean, VA 22012, USA \\
    \textit{Vermont Complex Systems Institute} \\
    \textit{University of Vermont}\\
    Burlington, VT 05405, USA
}

\begin{document}

\maketitle
\begin{sloppypar}

\lfoot{\tiny ©2026 The MITRE Corporation. ALL RIGHTS RESERVED.
\newline Approved for Public Release; Distribution Unlimited. Public Release Case Number 25-1956.}

\begin{abstract}
This study strengthens the foundations of multi-venue market modeling by attempting an independent replication of Wah and Wellman’s 2016 model of latency arbitrage in a fragmented market \cite{wah_latency_2016}.
We find that faithful replication is hindered by missing implementation details in the original paper and limited quantitative reporting. 
We demonstrate that increasing the number of simulation runs beyond the original design allows for the creation of bootstrap confidence intervals to support rigorous tests of quantitative alignment, compensating for lacking distributional information (e.g. variance).
We also demonstrate that increased complexity across the modeled scenarios corresponds with increased difficulty aligning to the original results.
We draw on a codebase released by the original authors in connection with a later paper to recover additional implementation details; however, we reject quantitative alignment between that codebase and the published results. 
Combining information from the paper and the released code, we achieve relational equivalence for most metrics but reject quantitative alignment for model settings where latency is non-zero.
We show that many of the qualitative takeaways from the original paper on the effects of market fragmentation and latency arbitrage are sensitive to the specifics of a `greedy strategy' extension given to the zero-intelligence (ZI) trader agents. Under an alternative interpretation of this strategy, we find that market fragmentation decreases execution times in all experiments and increases trader welfare in most experiments.
Finally, to facilitate future replication, critique, and extension, we provide an ODD (Overview, Design concepts, Details) protocol for our implementations of the model.
\end{abstract}
\newpage

\section{Introduction}
\label{intro}

The main regulator of U.S. financial markets, the Securities and Exchange Commission (SEC), has recently stated that existing market-simulation frameworks are not sufficiently mature to guide its policymaking \cite{us_securities_and_exchange_commission_regulation_2024}.
They point to a lack of existing frameworks for studying the structural considerations they are trying to address, and they warn:
\begin{displayquote}
Challenges associated with simulations include model accuracy and complexity, data quality, computational limitations, uncertainty and sensitivity, validation and verification, scalability, and the challenges associated with applying simulations to human behavior which is inherently unpredictable, and the challenges associated with modeling dynamic and evolving systems like financial markets \cite{us_securities_and_exchange_commission_regulation_2024} (p.328).
\end{displayquote}
Essential steps towards addressing these concerns are the development of computational models that better capture the complexity of real-world markets and independent replication to establish the rigor of insights from such models.

Towards these ends, we present here the results from our attempt to replicate Wah and Wellman's influential agent-based model (ABM) of latency arbitrage in a fragmented market \cite{wah_latency_2013,wah_latency_2016}.
To our knowledge, their simulation framework was the first to include multiple exchanges and communication latency---defining features of the modern U.S. stock market, which comprises 17 exchanges as of 2025\footnote{The 18th stock exchange is planned to begin operating in 2026 \cite{texas_stock_exchange_txse_2025}.}.
Under this framework, Wah and Wellman report that market fragmentation and latency arbitrage negatively impact market liquidity and efficient allocation of value to traders.
The model is an important building block for empirically-relevant market models \cite{duffin_agent-based_2018,tivnan_towards_2017,van_oort_adaptive_2023}, and it has been cited in a number of studies and policy briefs discussing alternative market designs 
(e.g. \cite{baldauf_high-frequency_2020,brolley_order-flow_2020,budish_high-frequency_2015,chicago_stock_exchange_inc_comment_2016,securities_and_exchange_commission_self-regulatory_2016,securities_and_exchange_commission_self-regulatory_2017}).

Our motivation was to use a successful replication of the model as a rigorous baseline for planned extensions targeting specific aspects of fragmentation in the modern U.S. stock market.
We found, however, that full replication is hampered by undocumented implementation choices and sparse statistical reporting.
Our main results are the following:
\begin{enumerate}
    \item Increasing the number of simulation runs an order of magnitude above the original experimental design helps rigorously test quantitative alignment, compensating for lacking distributional information (e.g. variance).
    \item Increased complexity across the modeled scenarios corresponds with increased difficulty aligning to the original results
    \item Qualitative alignment (i.e. relational equivalence) across liquidity and efficiency metrics can be achieved only after consulting with a codebase released by the original authors connected to a later study.
    \item The top-level takeaways from the model are sensitive to the specifics of a `greedy strategy' extension given to the zero-intelligence (ZI) traders; under an alternative implementation, we find that market fragmentation overall decreases execution times and increases trader welfare.
\end{enumerate}
Taken together, these findings highlight the value of independent replication in the modeling process while underscoring the need for more computational study of market fragmentation and improved reporting of market model assumptions and results.

In the next section, we give background on the role of replication in the agent-based modeling and agent-based market modeling literature (Section \ref{background}).
We then provide an overview of Wah and Wellman’s model (Section \ref{model}). 
We introduce our quantitative alignment methodology in Section \ref{quantitative_alignment} and use this methodology to test three implementations of the model against the canonical results. 
We compare the qualitative behavior of these implementations to the original results in Section \ref{qualitative} before concluding in (Section \ref{conclusion}).
We also provide an ODD (Overview, Design concepts, Details) protocol \cite{grimm_standard_2006,grimm2010odd} for the model in our appendix, to facilitate future replication, critique, and extension.

\section{Background}
\label{background}

Independent replication is a cornerstone of scientific practice.
It verifies whether the methods described by the original study are sufficient to produce the results claimed and allows for the extension of these methods for further study \cite{axelrod_advancing_1997,vermeer_leveraging_2020}.
While still relatively rare in the agent-based modeling literature \cite{axelrod_advancing_1997,heath_survey_2009}, its value to the practice has been demonstrated numerous times.
Unsuccessful replication highlights where the documented description is not sufficient to reproduce the original findings \cite{will_replication_2008}, ideally leading to clarification and further analysis \cite{macy_surprising_2010,will_resolving_2009}.
Successful replication efforts often identify sensitivity of the original findings to certain implementation details, thereby refining our understanding of the model's assumptions and scope \cite{edmonds_replication_2003,merlone_horizontal_2008,thiele_replicating_2015}.
The replicated model can be checked for robustness to additional experiments, analysis, and extensions \cite{burman_call_2010,hauke_theory_2020,thiele_replicating_2015}.
More broadly, the practices of replication, extension, and alignment allow for the cumulative development of more complex models built on robust foundations \cite{axelrod_advancing_1997,axtell_aligning_1996}.

In their seminal study on aligning computational models, Axtell et al. \cite{axtell_aligning_1996} introduce a comparison framework with three different levels of `equivalence' between models.
The strongest level of equivalence is numerical identity, where both models produce identical outputs given the same settings. This will generally not be feasible for models with stochastic components unless the replication shares roughly the same code (including pseudorandom number generators and seeds).
The next level is distributional equivalence, where the distributions of the two sets of results cannot be distinguished statistically from one another.
The weakest level is relational equivalence, under which the models produce the same qualitative relationships between inputs and outputs.
Axtell et al. highlight that distributional detail of the reported results needs to be reported by the original modeler in order for equivalence to be statistically tested in a replication effort.

Axtell et al. \cite{axtell_aligning_1996} also note that a precise, detailed description of the model is vital to replication.
Major effort has been undertaken to encourage rigorous and standardized model descriptions to aid in this process \cite{collins_call_2015,edmonds_replication_2003,grimm_standard_2006,vincenot_how_2018}.
A prevailing format is the Overview, Design concepts and Details (ODD) protocol introduced and refined by Grimm et al. \cite{grimm_standard_2006,grimm2010odd}.
Szangolies et al. \cite{szangolies_visual_2024} propose a standardized visual ODD (vODD) format to provide a quick overview of the model to supplement the more detailed written description.
Grimm et al. \cite{grimm_using_2025} recently demonstrated how the ODD protocol can be used to aid in the replication process, ambitiously aiming to replicate 18 models and succeeding to at least some degree for 15 of them.

\subsection{Replication of agent-based market models}
\label{abmm_replication}
Independent replication is particularly rare for agent-based market models, and published claims have generally not exceeded relational equivalence.
Preis et al. \cite{preis_multi-agent-based_2006} assert that the Farmer et al. `zero-intelligence' model (ZIM) \cite{daniels_quantitative_2003,smith_statistical_2003} is a special case of their more complex market model, claiming reproduction of the ZIM results without providing evidence in the paper.
Tivnan et al. \cite{tivnan_adding_2011} integrate the ZIM market environment with a richer trader ecology from Ghoulmie et al. \cite{ghoulmie_heterogeneity_2005}. Tivnan et al. note that they lacked the code and quantitative results from the original models and therefore could not test distributional equivalence. They claim success in establishing relational equivalence but do not display the results of that exercise in the paper due to space constraints.
Bookstaber et al. \cite{bookstaber_toward_2016} assert replication and extension of the Preis et al. model, showing their base replication produces similar stylized facts to the Preis et al. model but not providing evidence of equivalence in the paper.
Duffin and Cartlidge \cite{duffin_agent-based_2018} replicate and extend Wah and Wellman’s 2013 two-market model \cite{wah_latency_2013}, demonstrating relational equivalence for a single metric. Similar to Tivnan et al., they could not test distributional equivalence due to lack of quantitative results from the original paper.

Another route in developing cumulative progress in the agent-based market modeling space has been through sharing model code as open-source.
An early success in this vein was the Artificial Stock Market (ASM) \cite{lebaron_building_2002,palmer_artificial_1999}, which was built in the 1990's by researchers at the Sante Fe Institute, released online by the original team, and reimplemented in different coding languages by others \cite{ehrentreich_agent-based_2008,johnson_agent-based_2002,polhill_lessons_2005}. Multiple versions allowed for comparison between implementations, including analysis by Polhill et al. \cite{polhill_lessons_2005,polhill_ghost_2005} on how differences in floating point arithmetic can produce divergent behavior.
A number of open-source financial ABM platforms have been released in recent years \cite{belcak_fast_2021,byrd_abides_2019,cliff_open-source_2018,de_luca_open_2012,mascioli_financial_2024}.
This allows for the development of computational experiments upon common frameworks, but it does not mean that subsequent works are necessarily replicating and verifying the results of studies that came before.
Also, while multiple of these open-source platforms tout the ability to implement multiple exchanges in a market \cite{byrd_abides_2019,mascioli_financial_2024}, they have not generally been used to model such a scenario.

\section{The model}
\label{model}

\begin{figure}
    \centering
    \includegraphics[width=0.6\textwidth]{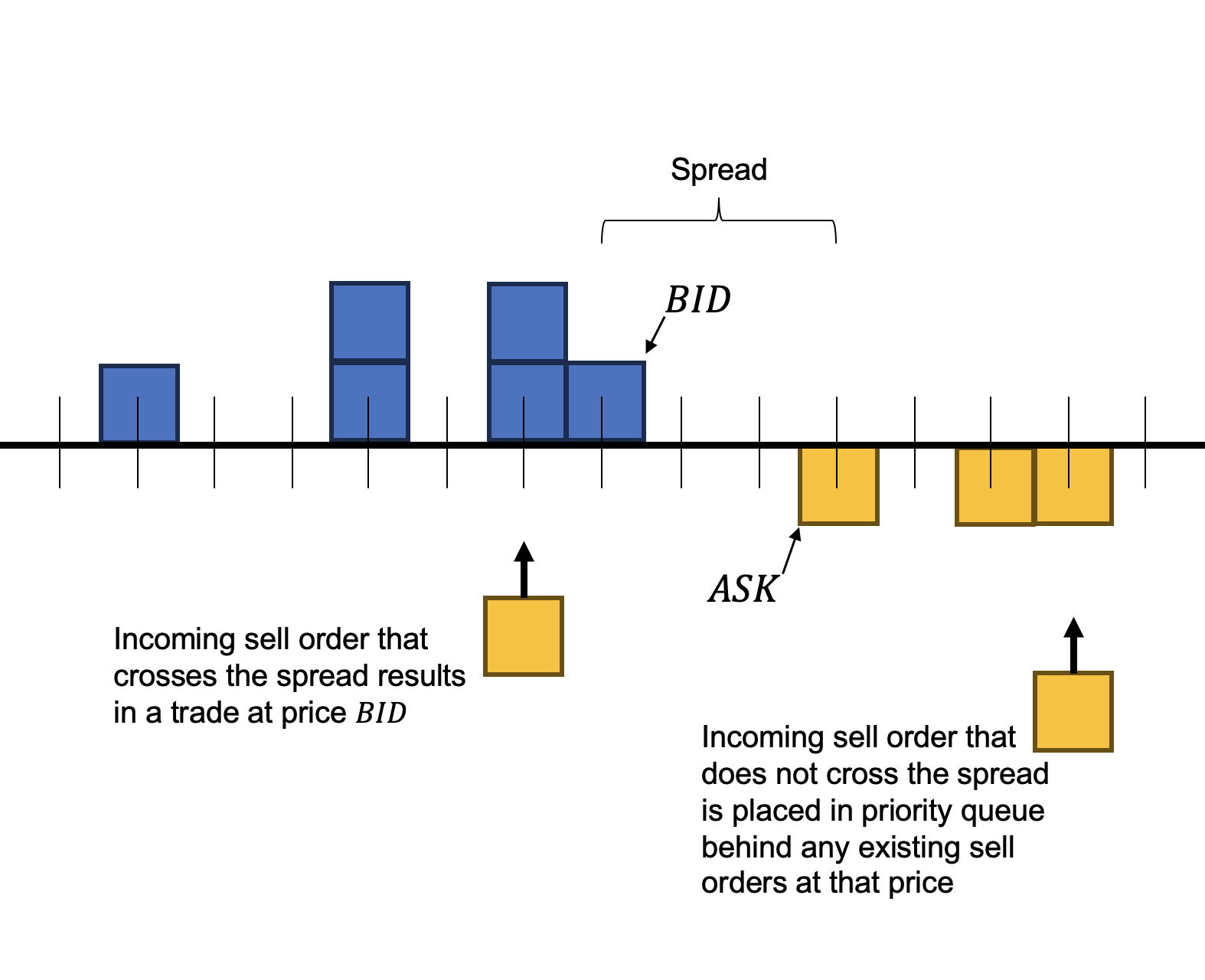}
    \caption{High-level logic for a simple continuous double auction (CDA) market. Buy and sell limit orders are shown as blue and yellow boxes, respective. Example logic for incoming sell limit orders is shown, with one order resulting in a trade and the other resulting in a new resting order, queued behind an existing sell order at the same price. Note that in this example the spread ($ASK-BID$) is of width three and will increase to four after the trade. Logic for incoming buy limit orders is analogous. Each of the 17 national exchanges in the U.S. stock market currently operates a CDA. The Wah and Wellman \cite{wah_latency_2013,wah_computational_2016} model is one of only a few agent-based market models to feature a fragmented market with multiple exchanges operating CDAs.}
    \label{fig:cda}
\end{figure}

The Wah and Wellman model \cite{wah_latency_2013,wah_latency_2016} examines the effects of fragmentation, latency, and latency arbitrage in a financial market.
The model features a single asset traded by investor (`background') agents in a market with one or more exchange.
A consolidated market with a single exchange is simulated first, after which an additional exchange, communication latency, and an arbitrage agent are incrementally added to examine their respective effects.
The specific type of arbitrage examined is an idealized form of latency arbitrage, in which a trader agent with faster access to information in the market takes advantage of price discrepancies to earn a risk-free profit.
The slower background traders get cross-market information at a lag from the Security Information Processor (SIP), a data consolidator for the market.
The hypothesis is that fragmentation and unequal access to cross-market information impact the efficient allocation of value to investors and the liquidity of the market overall.

Wah and Wellman's original 2013 model \cite{wah_latency_2013} features zero-intelligence (ZI) traders that each act exactly once per simulation according to a fixed, uniform trading strategy based on Gode and Sunder \cite{gode_allocative_1993}. 
The following features were varied: the number of exchanges, the amount of SIP latency, and the absence or presence of a latency arbitrage (LA) agent. All other parameters were fixed across the experiments.
Wah and Wellman \cite{wah_latency_2013} ran 200 simulations for each experiment and reported that latency arbitrage in fragmented markets reduces total surplus and has mixed effects on liquidity.
Duffin and Cartlidge \cite{duffin_agent-based_2018} replicated and extended the 2013 model in 2018, demonstrating relational equivalence for one metric (trader surplus) but not numerical agreement. They show sensitivity of the model results to the ZI shading parameter and obtain qualitatively different results when the background traders are updated to use `zero-intelligence plus' (ZIP) \cite{cliff_zero_1997} strategies.

Wah and Wellman substantially expanded the model in 2016 \cite{wah_latency_2016}. In the updated version, ZI traders participate multiple times per simulation, shade orders using more sophisticated surplus-seeking logic, and have heterogeneous strategy parameters.
For each experiment, Wah and Wellman \cite{wah_latency_2016} compute an equilibrium distribution of ZI trader strategies, where each agent aims to maximize its expected payoff (surplus).
They ran the model for three sets of environment parameters, varying market configurations for each environment and running 50,000 simulations for each experiment. They found latency arbitrage reduces surplus and liquidity in the updated model, while fragmentation has mixed effects.

We focus on replicating the 2016 version \cite{wah_latency_2016} of the model here. 
This decision is based on the 2016 study's more advanced ZI trader behavior and richer experimental design relative to the original 2013 study \cite{wah_latency_2013}.
The 2013 study features only a single set of environment parameters and presents results only graphically, without quantitative summary measures.
Duffin and Cartlidge \cite{duffin_agent-based_2018} similarly only report results visually rather than quantitatively.
Wah and Wellman \cite{wah_latency_2016} ran the model for three environments and reported mean surplus values for each experiment, providing more information to benchmark our replication against.
The 2016 study's results are also based on 50,000 simulations for each experiment, orders of magnitude more than the 200 simulations run per experiment in the 2013 study.
Finally, the extended ZI agent strategy and mixture of strategy parameters used in the 2016 model differ from the basic ZI agents critiqued by Duffin and Cartlidge \cite{duffin_agent-based_2018}, making it unclear how their critique applies to the updated model.

We refer to the 2016 study \cite{wah_latency_2016} simply as WW going forward.
We detail the high-level logic of the WW model in the following sections.
We follow an abbreviated version of the ODD (Overview, Design concepts, Details) protocol \cite{grimm_standard_2006,grimm2010odd} here, highlighting ambiguities from the original text and corresponding design choices we made.
We also provide a high-level visualization of the model following the vODD format \cite{szangolies_visual_2024} in Fig.~\ref{fig:visual_ODD}.
We give a more complete accounting of the model following the ODD protocol in Appendix \ref{ww_ODD}.

\begin{figure}
    \centering
    \includegraphics[width=1.0\textwidth]{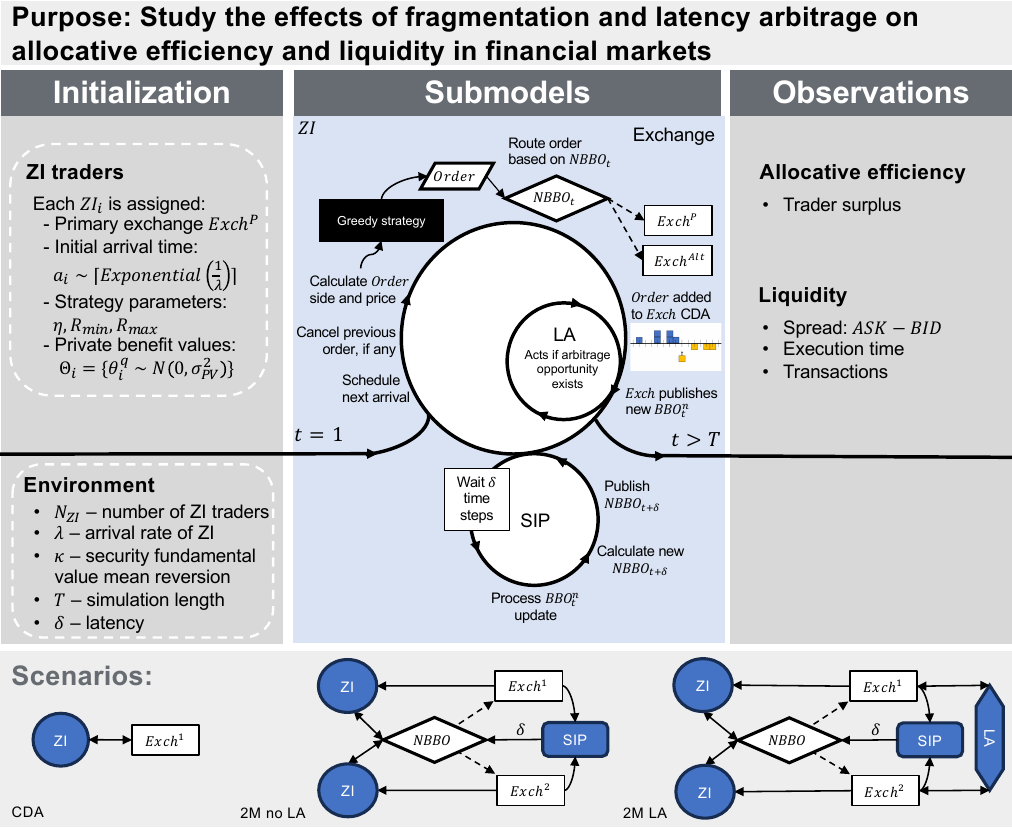}
    \caption{Visualization of WW following the vODD format \cite{szangolies_visual_2024}. The model examines the effects of fragmentation and latency arbitrage in financial markets by varying the number of exchanges, SIP latency, and presence of a latency arbitrage (LA) trader in the market. At initialization, a market is set up according to the experimental settings. Then, $N_{ZI}$ zero-intelligence (ZI) traders are created. Each is assigned a primary exchange, private strategy features, and an initial arrival time. ZI arrivals drive the events of the model. Upon arrival, a ZI trader will submit an order to an exchange, resulting in BBO feed updates, possible arbitrage action by the LA agent (if it is present), and updates to the SIP NBBO feed at some latency $\delta\geq 0$. Metrics of market efficiency and liquidity are averaged over many runs of the model for each experiment and then compared across the different scenarios.}
    \label{fig:visual_ODD}
\end{figure}

\subsection{Entities, state variables, and scales}
\label{entities}

\begin{table}[H]
\centering
\renewcommand{\arraystretch}{1.05}
\begin{tabular}{p{0.26\textwidth}p{0.69\textwidth}}
\hline
\textbf{Entities} & \textbf{Scheduler}: schedules and kicks off events in the simulation \\
& \textbf{Exchange}: market operating a continuous double auction (CDA) \\
& \textbf{Security}: fungible asset traded on each exchange \\
& \textbf{Security Information Processor (SIP)}: data consolidator for the market; maintains and publishes the National Best Bid and Offer (NBBO) \\
& \textbf{Traders:} \\
& Zero Intelligence (ZI) background traders \\
& Latency Arbitrage (LA) trader \\
\hline
\textbf{State variables} & \textbf{Scheduler} \\
& Current time $t$ \\
& List of scheduled events (ZI arrivals and SIP updates) \\[0.3em]
& \textbf{Exchange} \\
& Limit order book (LOB): all resting orders with prices, quantities, and timestamps \\
& Best bid and offer (BBO): highest buy price and lowest sell price on the exchange \\
& BBO subscribers \\[0.3em]
& \textbf{SIP} \\
& Exchange BBO(s): Current BBO feeds from the exchange(s) \\
& National Best Bid and Offer (NBBO) \\
& NBBO subscribers \\[0.3em]
& \textbf{Security} \\
& Fundamental value $r_t$ \\[0.3em]
& \textbf{ZI trader} \\
& Valuation vector $\Theta_i$: marginal private benefits for buying and selling the asset \\
& Strategy parameters $\{R_{\min}, R_{\max}, \eta\}$ \\
& Primary exchange \\
& Primary exchange BBO \\
& Current holdings: quantity $q_i$ the trader is long or short \\
& Outstanding orders \\
& Profits from trading \\[0.3em]
& \textbf{LA trader (if present)} \\
& Exchange BBOs \\
& Profits from trading \\[0.3em]
\hline
\textbf{Scales} & $T$: Simulation length \\
& $N_{ZI}$: Number of ZI traders \\
& $N_{exch}\in[1,2]$: Number of exchanges \\
& $\delta\geq 0$: SIP latency \\
& $N_{LA}\in [0,1]$: Presence or absence of the LA trader \\
& $\lambda$: ZI arrival rate \\
& $q_{\max}$: maximum shares ZI trader can be long or short the asset \\
\hline
\end{tabular}
\caption{ODD protocol entities, state variables, and scales for WW.}
\label{tab:entities}
\end{table}

\subsection{Process overview and scheduling}
\label{process_overview}

Time is discrete, with steps $t\in\{1,\ldots,T\}$. A scheduler maintains a queue of events, ordered by time and then by insertion order (first-in, first-out). Only two types of events go through the scheduler queue: ZI trader arrivals and SIP NBBO update events (scheduled with latency $\delta$ after a quote change).
All other actions in the model are triggered immediately (with zero modeled latency) as consequences of scheduled events. Non-scheduled actions include order submissions and cancellations, exchange matching and trade execution, exchange BBO updates, LA trader responses, and notifications of trades.

\subsection{Submodels}

\subsubsection{Exchanges}

The market can be configured with one or two exchanges.
Each exchange in the model operates a continuous double auction (CDA)\footnote{
Wah and Wellman \cite{wah_latency_2013,wah_latency_2016} also examined an alternative clearing mechanism, periodic call auctions, as a possible remedy to latency arbitrage. 
All national exchanges in the U.S. stock market currently use CDAs, however, and we leave the periodic call auction models out of scope for now.
} for the asset.
A CDA is a two-sided market where traders submit orders to buy and sell the asset. 
The market is `continuous' in the sense that orders and trades occur as soon as they are received and processed by the exchange's matching engine\footnote{There is a time resolution below which the time of two different messages cannot be differentiated, but for modern exchanges this is on the magnitude of microseconds or even nanoseconds \cite{angel_when_2014,van_oort_adaptive_2023}.}.
A limit order\footnote{
A market order is an alternative type of order which executes immediately at the best available price. All agents in the model exclusively submit limit orders, however, and so market orders are out of scope.
} to buy (sell) the stock specifies a maximum (minimum) price the agent is willing to trade at and a quantity to buy (sell). 
In the model, all orders are for a single unit of the stock. 
Each exchange maintains a limit order book (LOB), constructed with the following elements:
\begin{enumerate}
    \item Orders: map from order IDs to their respective orders
    \item Bids: map from price to the queue of buy order IDs at that price
    \item Asks: map from price to the queue of sell order IDs at that price
\end{enumerate}
The exchange publishes a BBO feed to its subscribers, containing the best bid and offer prices in the LOB at that moment. 
LOBs are initialized with no bids or asks at the start. The BBO has null prices for its best available bid and ask whenever the respective side of the LOB is empty (including at initialization).

When an exchange receives an add message for order $O_i$ with limit price $p$, it does the following. First, it checks whether the added price can be matched against existing orders in the book. More explicitly, if order $O_i$ is a buy order, the exchange checks whether there is a resting sell order $O_j$ with price $q\leq p$. Similarly, if $O_i$ is a sell order, the exchange checks whether there is a resting buy order $O_j$ with price $q\geq p$. In either of these cases, the orders are matched at price $q$, resulting in a transaction. If multiple resting orders exist on the LOB that could match with the incoming order, the oldest order is given priority (i.e. price-time priority). In this model, all orders are for a single quantity of the asset, and so each trade is always for a single unit. Receipts are sent to the traders associated with $O_i$ and $O_j$, and $O_j$ is removed from the LOB.
If $O_i$ instead does \textit{not} immediately match, it is added to the LOB. 
A trader can request to withdraw a resting order from the LOB. Upon receiving a withdraw request, the exchange removes the corresponding order from the LOB.

After any change to the exchange's LOB due to a withdraw, add, or trade, the exchange publishes its updated best bid and ask prices to its subscribers. This is published regardless of whether the best bid or ask prices were changed by the LOB update.
The exchange loops over its subscribers in the order they subscribed to the exchange's BBO feed, directly calling the subscriber's \texttt{update\_bbo()} function.


\subsubsection{Security Information Processor (SIP)}

The SIP consolidates information for the market. It subscribes to the BBO feed from each exchange, publishing the following information in its National Best Bid and Offer (NBBO) feed: the overall highest bid, the exchange holding the highest bid, the overall lowest ask, and the exchange holding the lowest ask. 
Any time an exchange publishes a BBO update, the SIP schedules an NBBO update event with the new information to occur at $t+\delta$, where $t$ is the time the BBO was received and $\delta$ is the SIP latency parameter.
When the scheduled NBBO update occurs, the updated feed is instantaneously communicated to all SIP subscribers.
We assume that when latency $\delta=0$ the SIP should update and publish its NBBO without going through the scheduler.

\subsubsection{Security}

The security being traded in the market is mostly abstract, except it possesses an evolving `common fundamental value' known by the traders. The fundamental time series follows a mean-reverting random walk:
$$r_t = \max \{0, \kappa \overline{r} + (1 - \kappa)r_{t-1} + u_t\},$$
where parameter $\kappa\in[0,1]$ specifies the strength of reversion to the mean $\overline{r}$, and $u_t \sim N(0, \sigma_{\text{shock}}^2)$ is a Gaussian random shock at each time step.
WW do not specify what the initial conditions are for the process, so we assume $r_0=\overline{r}$, and we then calculate $r_i$ for each $i\in[1,T]$.

The ZI traders use the estimated terminal fundamental value of the security in their trading strategy. We model the fundamental value logic as a function of the security, which the ZI agents call.
The estimate is based on the fundamental value $r_t$ at the time $t$ the trader looks it up. The estimated terminal value $\hat{r}_t$ is then the expected fundamental value at time $T$, accounting for the mean-reversion process.
WW do not specify how $\hat{r}_t$ is made to be an integer, nor whether each component to the price should be converted to int or just the final price. We assume the estimated terminal value should be rounded after the calculation of $\hat{r}_t$. In other words,
$$\hat{r}_t = \texttt{round}\left(\left(1 - (1 - \kappa)^{T-t}\right)\overline{r} + (1 - \kappa)^{T-t}r_t\right).$$

\subsubsection{Latency arbitrageur (LA) trader}
\label{la_main}
The latency arbitrageur (LA) agent enacts a cross-market arbitrage strategy, taking advantage of disparities in price arising from fragmentation of orders across multiple exchanges.
At initialization, the LA agent subscribes to the BBO feeds from both exchanges. Over time, it tracks the overall best prices available in the market with zero latency.
Each time the LA agent receives a new BBO update from an exchange, it evaluates whether to act. 
Let $BID^i$ and $ASK^i$ denote the best bid and ask at exchange $i$ at a given point in time. Then let $BID^*=\text{max}\{BID^1, BID^2\}$ and $ASK^*=\text{min}\{ASK^1,ASK^2\}$. The LA agent has a threshold parameter $\alpha$ that determines whether an arbitrage opportunity is worth acting on.
If $BID^*>(1+\alpha)ASK^*$, the LA agent submits an order to buy at the midpoint price $\left\lfloor \frac{BID^*+ASK^*}{2}\right\rfloor$ at the exchange holding the best ask and an order to sell at midpoint price $\left\lceil \frac{BID^*_t+ASK^*_t}{2}\right\rceil$ at the exchange holding the best bid.

\subsubsection{Zero-intelligence (ZI) traders}
\label{zi_main}

The zero-intelligence (ZI) traders in the model enact a strategy in the family of trader strategies introduced by Gode and Sunder \cite{gode_allocative_1993}.
A ZI trader in the model acts according to its private valuation of the asset at time $t$ based on the common fundamental value $r_t$ and the private marginal benefit for trading with its current net position.
WW cite Goettler et al. \cite{goettler_informed_2009} as introducing a similar trading strategy based on marginal private benefits.
The trader `shades' its order price by some amount, aiming to achieve surplus value from buying (selling) at a lower (higher) price than its internal valuation of the asset.

\paragraph{Initialization}
Each ZI trader agent is assigned a `primary' exchange. We assume the ZI traders are evenly split across the exchanges for a given market configuration.
Each ZI trader subscribes to the SIP NBBO and to the BBO feed from its primary exchange.
Its holdings, profits, and number of transactions are all initialized to zero.
Next, for a given trader $ZI_i$, we construct $\Theta_i$, the vector of marginal private benefits for changes to the trader's net position:
$$\Theta_i=\left(\theta_i^{-q_{\max}+1}, \ldots,\theta_i^0,\theta_i^{+1},\ldots,\theta_i^{q_{\max}}\right),$$where $q_{\max}$ is the maximum number of shares the trader can be long or short the asset.
The vector $\Theta_i$ is generated from $2q_{\max}$ independent draws from $N(0,\sigma_{PV}^2)$, sorted such that $\theta^q\geq \theta^{q+1}$ $\forall q$. 
If $ZI_i$ has position $q$, buying an additional share yields a marginal private benefit $\theta_i^{q+1}$, while selling a share yields a marginal private benefit $\theta_i^q$.

The initial ZI arrival times $\{a_i\}$ are each drawn independently from an exponential distribution, $a_i\sim\lceil Exponential(1/\lambda)\rceil$. WW do not specify how to make the Poisson process discrete. We choose to take the ceiling in order for individual trader interarrival times to be greater than zero.

\paragraph{Arrival and base trading strategy}

Upon arrival at time $t$, the first thing trader $ZI_i$ does is schedules its next arrival for $t+b_i$, where again $b_i\sim\lceil Exponential(1/\lambda)\rceil$.
Next, the trader cancels any unfilled orders it submitted previously.
Then, it randomly decides whether to buy or sell (with equal probability).
If the trading decision would result in $ZI_i$'s net position exceeding $q_{\max}$ shares long or short, it exits and does not submit an order on this turn.
Otherwise, the trader calculates its current private valuation $v_i(t)$ of the asset:
    \[ v_i(t)=\hat{r}_t + \begin{cases} 
          \theta_i^{q_t+1} & \text{if buying one unit} \\
          \theta_i^{q_t} & \text{if selling one unit} \\
       \end{cases}.
    \]
WW do not specify how the price should be made to be discrete. We choose to round $v_i(t)$ to the nearest integer.
The trader then determines its order price $p$ as:
    \[ p\sim \begin{cases} 
        \max\left(0, U[v_i(t)-R_{max},v_i(t)-R_{min}]\right) & \text{if buying} \\
        \max\left(0, U[v_i(t)+R_{min},v_i(t)+R_{max}]\right) & \text{if selling} \\
       \end{cases},
    \]
where $U$ here denotes an integer uniformly chosen at random within the specified range.
The requested surplus for $ZI_i$ is then $s_i(t)=|v_i(t)-p_i(t)|$.

\paragraph{Greedy strategy extension}
WW introduce an additional strategy extension, which we refer to as the `greedy strategy'\footnote{The `greedy' terminology is taken from the \textit{MarketSim} codebase and not from the original paper: \url{https://github.com/egtaonline/market-sim/blob/marketsim1/hft-sim/src/entity/agent/BackgroundAgent.java\#L274}.} going forward. Below is the entirety of the detail given in the paper about this strategy:
\begin{displayquote}
	We extend ZI by including a threshold parameter $\eta \in [0, 1]$, whereby if the agent could achieve a fraction $\eta$ of its requested surplus at the current price quote, it would simply take that quote rather than posting a limit order to the book. Setting $\eta = 1$ is equivalent to the strategy without employing the threshold  \cite{wah_latency_2016} (p.75).
\end{displayquote}
Note also that WW separately say,
\begin{displayquote}
    Background traders have direct access to the quotes on their primary market and the NBBO, but not to those on the alternate market \cite{wah_latency_2016} (p.76).
\end{displayquote}
Thus, in our `best guess' (\textit{BestGuess}) implementation of the model, we implement this strategy as the following:
based on both the BBO feed from their primary exchange and the NBBO feed from the SIP, let \texttt{max\_bid} and \texttt{min\_ask} be the best market prices observed by $ZI_i$.
The trader then evaluates whether there is a greedy opportunity, updating its price if so:
    \begin{lstlisting}
    requested_surplus = |v_i(t) - p|
    if side == `BUY':
        if requested_surplus * eta <= v_i(t) - min_ask:
            p = min_ask
    else: // `SELL'
        if requested_surplus * eta <= max_bid - v_i(t):
            p = max_bid
    \end{lstlisting}

\paragraph{Routing strategy}

Trader $ZI_i$ will submit its order to its primary exchange unless the NBBO indicates the order would immediately execute at a better price at the alternate exchange.
Specifically, let \texttt{BBO} refer to the primary exchange BBO, and assume \texttt{order.exchange} is originally set to the $ZI_i$'s primary exchange. Then the order routing logic is as follows:
    \begin{lstlisting}
    route_order(order):
        NBBO_price_better = False
        will_transact = False
        if order.type == `BUY':
            if (NBBO.ask.price != Null and (BBO.ask == Null or 
                    NBBO.ask.price < BBO.ask)):
                NBBO_price_better = True
            if order.price >= NBBO.ask.price:
                will_transact = True
            alt_ex = NBBO.ask.exchange
        else: // `SELL'
            if (NBBO.bid.price != Null and (BBO.bid == Null or 
                    NBBO.bid.price > BBO.bid)):
                NBBO_price_better = True
            if order.price <= NBBO.bid.price:
                will_transact = True
            alt_ex = NBBO.bid.exchange
    
        if NBBO_price_better and will_transact:
            order.exchange = alt_ex
            
        order.exchange.submit_order(order)
    \end{lstlisting}

\subsection{Model experiments}
\label{experiments_main}

\begin{table}[H]
\centering
\begin{tabular}{|r|r|r|r|r|r|}
\hline
Environment & $N_{ZI}$ & $\lambda$ & $\kappa$ & $T$ \\
\hline
    1 & 24 & 0.05 & 0.05 & 15000 \\
    2 & 238 & 0.005 & 0.02 & 10000 \\
    3 & 58 & 0.005 & 0.02 & 5000 \\
\hline
\end{tabular}
\caption{Parameter settings for the three market environments from WW. }
\label{table:ww16_env_main}
\end{table}

\begin{table}[H]
\centering
\begin{tabular}{|p{0.15\textwidth}|p{0.15\textwidth}|p{0.6\textwidth}|}
\hline
Parameter &  Default &  Description \\
\hline
    $\overline{r}$ & 100,000 & Mean fundamental value of security. \\
    $\sigma_{shock}^2$ & 5,000,000 & Variance of the random shock to the security's fundamental value. \\
    $\sigma_{PV}^2$ & 5,000,000 & Variance of ZI private valuations; used to draw private marginal utility of the agent's net position. \\
    $\alpha$ & 0.001 & Threshold from which the LA agent determines whether an opportunity is worth pursuing. \\
    $q_{\max}$ & 10 & Maximum position (long or short) for ZI traders \\
\hline
\end{tabular}
\caption{Default parameters used across model configurations. These were specified by WW except for $\alpha$, which is based on the 2013 paper \cite{wah_latency_2013} and \textit{MarketSim}.}
\label{table:default_params_main}
\end{table}

WW conduct experiments across three environments. The parameters varied across these environments are shown in Table \ref{table:ww16_env_main}. The parameter $N_{ZI}$ is the number of ZI traders, $\lambda$ is the arrival rate of ZI traders, $\kappa$ is the strength of mean reversion of the security's fundamental value, and $T$ is the simulation length. The fixed parameters across environments are shown in Table \ref{table:default_params_main}.
Within an environment, there are three possible configurations of the market: single exchange (\textit{CDA}); two exchanges with no LA agent (\textit{2M no LA}); and two exchanges with LA agent (\textit{2M LA}). In the two-market configurations, latency $\delta$ is also varied, over a different set of values for each environment.
Specifically:
\begin{enumerate}[label=\textbf{Environment \arabic*:}, leftmargin=*, labelwidth=4cm, labelsep=0.5em, align=left]
  \item $\delta\in\{0,100,200,300,400,600,700,900\}$
  \item $\delta\in\{0,50,100\}$
  \item $\delta\in\{0,25,50,75,100\}$
\end{enumerate}
An experiment in the model is therefore defined by the environment, market configuration, and latency setting.

\begin{table}[H]
\centering
\begin{tabular}{|l|r|r|r|}
\hline
ID & $R_{min}$ & $R_{max}$ & $\eta$ \\
\hline
$ZI_1$ & 0 & 125 & 1 \\
$ZI_2$ & 0 & 250 & 1 \\
$ZI_3$ & 0 & 500 & 1 \\
$ZI_4$ & 250 & 500 & 1 \\
$ZI_5$ & 0 & 1000 & 1 \\
$ZI_6$ & 500 & 1000 & 0.4 \\
$ZI_7$ & 500 & 1000 & 1 \\
$ZI_8$ & 0 & 1500 & 0.6 \\
$ZI_9$ & 1000 & 2000 & 0.4 \\
$ZI_{10}$ & 0 & 2500 & 0.4 \\
$ZI_{11}$ & 0 & 2500 & 1 \\
\hline
\end{tabular}
\caption{ZI strategy combinations, taken from WW \cite{wah_latency_2016} (p. 80).}
\label{Table:zi_strategies_main}
\end{table}

The allowed ZI strategies considered by WW are given in Table \ref{Table:zi_strategies_main}. In a run of the model, each ZI agent has a fixed strategy from this set of options. To narrow down the space of possible ZI strategy assignments, WW use an approach called `empirical game-theoretic analysis' (EGTA) \cite{cassell_egtaonline_2012,wellman_empirical_2025}. 
They assume a given experimental setting is a game in which the ZI agents aim to maximize their expected payoff (surplus) based on the strategies of other players.
A `strategy profile' specifies a probability vector over the ZI strategies.
WW approximate the full game with fewer players to find the Nash equilibria strategy profiles for the game.
Then, they simulate the full game by independently sampling a `mixture' of ZI strategies from the equilibrium strategy profile that maximizes total ZI surplus.
A mixture is a draw of strategy assignments for all $N_{ZI}$ agents from the equilibrium profile; for each mixture, 100 independent simulation runs are conducted.
For a given experiment, they sample 500 mixtures, thus running the model 50,000 times for each experiment.

WW note that limiting their main analysis to these selected profiles allows them to \say{focus on trader behavior in equilibrium, when all market participants are best responding to each others' strategies in order to optimize their own gains from trade} \cite{wah_latency_2016} (p.77).
We focus on these same `equilibrium' scenarios in order to assess the fidelity of our replication to the results presented by WW.
We therefore do not repeat WW's EGTA here and instead take their selected strategy profiles as specified. We summarize these strategy profiles for each experiment in our supplemental material in Table~\ref{table:env_mixtures}. We sample mixtures from these profiles and run simulations similar to the main results presented by WW.
We sample 5,000 mixtures and simulate 100 runs per mixture for each experiment, as discussed further in Section \ref{quantitative_alignment_methodology}.

To summarize, the model experiments are run as follows, with $\{M=500, R=100\}$ for WW and $\{M=5000, R=100\}$ for our replication attempts:
\begin{lstlisting}
    For each experiment (environment+configuration+latency):
        Sample M strategy mixtures from equilibrium strategy profile
        For each mixture m:
            Run R independent simulations
\end{lstlisting}

\subsection{Output metrics}

WW calculate the mean value for each metric taken across 50,000 simulations (500 mixtures, 100 runs per mixture) for each experiment. The metrics are calculated and saved at the end of each simulation. The means are calculated on the saved values across simulations (i.e. averaged over all 50,000 runs). WW report the quantitative value for the mean ZI trader surplus and LA surplus (if any) for each experiment. All other metrics are reported visually.

\paragraph{Surplus}

At the end of the simulation, a ZI agent $ZI_i$ will have a net terminal position $q_i^T\in[-q_{\max}, q_{\max}]$.
The terminal surplus $S_i$ for $ZI_i$ is the sum of the public and private value of its net position.
Let $r_T$ be the fundamental value of the asset at time $T$, and let $\Theta_i=\{\theta_j\}$ be the marginal private benefits vector for $ZI_i$ constructed in Section \ref{zi_main}.
Also, let $c_i$ be the net cash flow of $ZI_i$ from its trades over the course of the simulation.
Terminal surplus $S_i$ is then calculated at the end of the simulation as:
\[ S_i = c_i^T + q_i^Tr_T +
    \begin{cases} 
        \sum_{j=1}^{q_i^T}\theta_i^j & \text{if } q_i^T>0 \\
        0 & \text{if } q_i^T=0 \\
        \sum_{j=q_i^T+1}^0 -\theta_i^j & \text{if } q_i^T<0 \\
    \end{cases}.
\]
The surplus for the LA agent is simply its net cash flow from trading. This is because the LA agent never has a non-zero net position after it is done executing its strategy; it buys and sells one unit of the asset each time it acts.

\paragraph{Spread}

The spread is the amount by which the best ask price ($ASK$) exceeds the best bid price ($BID$); in other words, $spread=ASK - BID$.
WW take the median spread over the course of the simulation, then take the mean over all simulations.
The median NBBO spread is the median amount the NBBO ask exceeds the NBBO bid over the simulation.
The BBO spread is the median spread from the exchange BBO. When there are two exchanges, this is the mean of the median spreads from the individual exchange BBOs.

WW do not address how to handle a locked or crossed NBBO (i.e. where $ASK\leq BID$) when calculating spread, nor what to do when there are no quotes on one side of the book.
WW also do not clarify whether the median is calculated on the spread for each quote update or if the spread should be calculated once per time step. We assume it should be calculated for each update (i.e. tick time, rather than clock time). 
The NBBO median spread is thus taken on the spreads calculated for each NBBO update where the $BID$ and $ASK$ are both non-empty and $BID\leq ASK$.
Similarly, we calculate the NBBO median spread as the median of all spreads on BBO updates where the $BID$ and $ASK$ are both non-empty ($BID<ASK$ at the exchange-level by definition of the CDA), and the medians are averaged across all exchanges in the market.

\paragraph{Execution time}

WW describe execution time as `the interval between order submission and transaction for orders that eventually trade' \cite{wah_latency_2016} (p.79).
We assume the execution time should be calculated for both legs of the trade, therefore meaning at least one leg of each trade will have an execution time of zero.
The mean execution time is then the average of the execution times minus submission times for all orders that resulted in trades.

\paragraph{Transactions}

Finally, WW present the number of transactions from the model but do not elaborate on how they calculate this. Based on Fig. 11 in the original study \cite{wah_latency_2016} (p.90) and the behavior of the model, we infer the transactions reported by WW are the total number of \textit{orders} that resulted in trades throughout the simulation, separated by the type of trader that submitted the order. A trade between two ZI traders thus results in two ZI transactions, while a trade between a ZI trader and the LA agent is logged as one ZI transaction and one LA transaction.

\section{Quantitative alignment}
\label{quantitative_alignment}

\subsection{Methodology}
\label{quantitative_alignment_methodology}

A critical piece of the replication process is having a clear and robust replication standard---the criteria used to determine whether replication has been achieved to the desired level \cite{axtell_aligning_1996,wilensky_making_2007}.
This requires a reliable measure to assess alignment between two sets of results.
As discussed in Section \ref{background}, Axtell et al. \cite{axtell_aligning_1996} provide a framework to evaluate the extent two model implementations produce equivalent behavior.
The amount and type of information provided by the original study determines what tests of equivalence can be employed by the replicator \cite{radax_prospects_2010}.
We do not have the random seeds used by WW for their simulation runs and thus inherently rule out achieving numerical identity of results.
The next level of equivalence in the Axtell et al. framework is distributional equivalence. This requires distributional information about the reported measurements \cite{axtell_aligning_1996}.
In our case, the only quantitative information reported by WW are the mean ZI surplus values and the mean LA surplus (if any) from the 50,000 simulations (500 mixtures, 100 runs per mixture) for each experiment.
Other results in the paper are reported graphically, allowing us to test relational equivalence but not quantitative alignment.

The mean surplus values provide quantitative information we can test alignment against beyond just relational equivalence, however. 
For example, two sets of results could be relationally equivalent while being an order of magnitude different quantitatively from one another.
The amount of information provided by WW falls somewhere between the level of detail needed for relational and distributional equivalence.
In theory, two models could produce sample means indistinguishable from one another while having statistically distinguishable distributional details (e.g. variance) if that information was known.
We will refer to such checks simply as quantitative alignment going forward, with the understanding that what we are able to test here falls short of distributional equivalence.

In order to make as complete a use of the source information provided as possible, we seek to determine a meaningful test of quantitative alignment against the mean surplus values before moving on to assessing relational equivalence.
We have the sample mean values from WW, while we have the full set of results from running our implementations of the model.
A naive approach to test quantitative alignment in this circumstance is to repeat their analysis and run our code for 500 agent mixtures (100 runs per mixture) and then test our distribution of values against the reported mean from WW.
One possibility is to apply a one-sample t-test to test the hypothesis that our mean surplus value differs from the WW mean.
This approach treats the sample mean reported by WW as if it is approximately the population mean for surplus values generated by the model for the specified experimental setting.
Our criteria could then be to reject quantitative alignment for a given experimental setting if the t-test p-value is below a selected threshold $\rho$.

We validate our alignment methodology by testing our implementation of the model against itself before comparing it to WW.
As we do not have the random seeds used by WW, our measure of alignment should ideally not fail due to simply using a different random seed if all other logic is the same.
In order to have confidence in our measure of quantitative alignment, our expectation is the following: if we apply the test to two sets of results produced by the exact same code but with different random seeds, we should reject quantitative alignment approximately $100\rho$\% of the time.
To test this expectation, we ran our \textit{BestGuess} model for 5,000 mixtures, 100 runs per mixture (500,000 runs in total) for each experiment (environment+configuration+latency).
For a test, we randomly select results from 500 mixtures (100 runs per mixture) as our target results $x_{\text{target}}$. Let $\mu_{\text{target}}$ be the mean ZI surplus value from $x_{\text{target}}$.
We then randomly select, with replacement, 500 mixtures from the remaining 4,500 mixtures and let that be our comparison sample $x^*$.
We then check whether we would reject quantitative alignment between $x^*$ and the target mean $\mu_{\text{target}}$ using the t-test criteria given above for some threshold.
We repeat this reliability test 1,000 times, each time selecting a new target mean and comparison sample.

The results are shown in Table \ref{tab:self_test_dist_eq_bestguess} for thresholds of $\rho=0.05$ and $\rho=0.01$.
Note that the target mean and sample results for this test were produced by the same code with the same parameters except for the simulation random seeds.
We nevertheless see that under this one-sample t-test methodology we reject the results from \textit{BestGuess} being quantitatively aligned to itself 17.0--48.8\% of the time for $\rho=0.05$ and 6.3--35.9\% of the time for $\rho=0.01$. This rejection rate is quite variable and much higher than expected for the specified significance levels. 
Based on these results, we may falsely reject two sets of results from the same code being quantitatively aligned one-quarter to almost one-half of the time.
We therefore determine we need a better measure of quantitative alignment before we can reliably test against the original model.

We next consider an alternative quantitative alignment criteria, this time using a bootstrap approach.
Such an approach was suggested by Axtell et al. when there is \say{no solid reason to assume a convenient (e.g., Gaussian) form of the underlying distributions} \cite{axtell_aligning_1996} (p.140).
We can utilize the full set of results from running our \textit{BestGuess} implementation for 5,000 agent mixtures, 100 runs per mixture.
To construct a bootstrap sample, we randomly select, with replacement, the results from 500 of these mixtures.
Constructing 1,000 bootstrap samples, we can calculate percentile confidence interval of the bootstrap sample means.
We can then check whether the target mean value falls within this bootstrap confidence interval.
If the target mean falls outside this interval, we reject the hypothesis that the surplus values produced by the two versions of the model are quantitatively aligned.

We again sanity check this approach for our \textit{BestGuess} implementation against itself for the mean ZI surplus values.
For a single test, we randomly select 500 mixtures (100 runs per mixture) to calculate our target mean $\mu_{\text{target}}$. 
We then apply the updated bootstrap methodology, testing whether the target mean falls within the 95\% confidence interval (CI) calculated from 1,000 bootstrap samples from the remaining 4,500 mixtures.
We repeat this test 1,000 times, randomly selecting a new target and sample population each time.
The results of this test are again shown in Table \ref{tab:self_test_dist_eq_bestguess}. 
Under the bootstrap methodology the false rejection rate (4.7--7.8\%) for the 95\%-CI threshold is much lower and less variable than under the one-sample t-test methodology, albeit still slightly higher on average than the expected 5\%.
This gives us greater confidence in the bootstrap confidence interval methodology than the t-test, as we want to minimize the risk we falsely reject quantitative alignment while still meaningfully comparing our simulated results to the canonical values.

\begin{table}[H]
    \centering
\begin{tabular}{rlr|rr|rr}
\hline
    env & model & latency & \multicolumn{2}{|c|}{t-test rej. \%} & \multicolumn{2}{c}{Bootstrap rej. \%} \\
     &  &  & $\rho=0.05$ & $\rho=0.01$ & 95\% CI & 99\% CI \\
\hline
  1 &      CDA &        0 &               24.70 &               13.00 &                   5.60 &                   1.80 \\
  1 & 2M no LA &        0 &               19.80 &                7.10 &                   6.40 &                   2.50 \\
  1 & 2M no LA &      100 &               36.10 &               22.60 &                   6.70 &                   2.10 \\
  1 & 2M no LA &      200 &               38.50 &               25.00 &                   7.70 &                   2.10 \\
  1 & 2M no LA &      300 &               27.10 &               13.40 &                   7.80 &                   2.40 \\
  1 & 2M no LA &      400 &               39.30 &               27.70 &                   5.90 &                   1.70 \\
  1 & 2M no LA &      600 &               24.00 &               12.90 &                   6.70 &                   1.70 \\
  1 & 2M no LA &      700 &               22.20 &               12.20 &                   6.70 &                   1.60 \\
  1 & 2M no LA &      900 &               19.20 &                8.10 &                   7.10 &                   1.80 \\
  1 &    2M LA &      100 &               38.00 &               23.20 &                   6.30 &                   1.30 \\
  1 &    2M LA &      200 &               41.40 &               26.10 &                   7.80 &                   1.80 \\
  1 &    2M LA &      300 &               35.90 &               22.50 &                   5.40 &                   1.10 \\
  1 &    2M LA &      400 &               31.40 &               18.30 &                   5.90 &                   1.70 \\
  1 &    2M LA &      600 &               17.80 &                8.40 &                   5.70 &                   0.90 \\
  1 &    2M LA &      700 &               48.80 &               35.90 &                   7.00 &                   1.60 \\
  1 &    2M LA &      900 &               17.70 &                6.90 &                   6.90 &                   1.80 \\
  2 &      CDA &        0 &               17.80 &                7.70 &                   5.40 &                   1.60 \\
  2 & 2M no LA &        0 &               18.00 &                9.30 &                   6.80 &                   1.70 \\
  2 & 2M no LA &       50 &               17.50 &                7.60 &                   5.80 &                   1.40 \\
  2 & 2M no LA &      100 &               19.00 &                7.80 &                   6.70 &                   1.80 \\
  2 &    2M LA &       50 &               18.40 &                8.00 &                   7.50 &                   1.30 \\
  2 &    2M LA &      100 &               30.00 &               15.40 &                   5.80 &                   1.80 \\
  3 &      CDA &        0 &               31.10 &               18.00 &                   5.50 &                   1.20 \\
  3 & 2M no LA &        0 &               18.30 &                8.50 &                   6.50 &                   1.30 \\
  3 & 2M no LA &       25 &               17.00 &                7.60 &                   6.80 &                   1.60 \\
  3 & 2M no LA &       50 &               18.00 &                6.30 &                   6.60 &                   0.90 \\
  3 & 2M no LA &       75 &               20.80 &                9.00 &                   6.60 &                   1.70 \\
  3 & 2M no LA &      100 &               21.10 &                8.10 &                   6.10 &                   1.80 \\
  3 &    2M LA &       25 &               32.40 &               18.90 &                   4.70 &                   1.30 \\
  3 &    2M LA &       50 &               21.00 &               10.40 &                   6.20 &                   1.60 \\
  3 &    2M LA &       75 &               29.70 &               15.70 &                   7.70 &                   1.70 \\
  3 &    2M LA &      100 &               33.20 &               18.70 &                   5.30 &                   1.70 \\
\hline
\end{tabular}
    \caption{
    False rejection rates for quantitative self-alignment of \textit{BestGuess} ZI surplus values under two testing methodologies. For each of 1,000 trials for a given experiment, results from 500 mixtures (100 runs each) are held out to calculate the target mean, and alignment is tested against a bootstrap sample of 500 mixtures drawn from the remaining 4,500 mixtures. 
    The one-sample t-test methodology rejects quantitative alignment if the t-test p-value is below the selected threshold $\rho$.
    The bootstrap confidence interval (CI) methodology rejects quantitative alignment if the target mean falls outside the selected CI (95\% or 99\%).
    Experiments are defined by environment, market configuration, and latency, as specified in Section~\ref{experiments_main}.
    }
    \label{tab:self_test_dist_eq_bestguess}
\end{table}

\subsection{\textit{BestGuess} versus WW}
\label{bestguess_vs_ww}

Having validated our bootstrap quantitative alignment methodology, we now apply it to test our results against the canonical WW values.
As mentioned in the previous section, we ran our \textit{BestGuess} replication model for 5,000 mixtures, 100 runs per mixture, for a total of 500,000 runs per experiment (environment+configuration+latency).
We apply our bootstrap quantitative alignment test for these results against the mean ZI and LA surplus values reported by WW.
Using the full set of results, we construct 1,000 bootstrap samples of 500 mixtures, 100 runs per mixture, each.
To construct a bootstrap sample of our results, we randomly select, with replacement, a sample of 500 mixtures from the 5,000 total mixtures. We calculate the mean surplus value (ZI or LA depending on which we are testing) for the selected bootstrap sample.
Repeating this 1,000 times, we get a range of means produced by our implementation of the model for a given experiment.

\begin{table}[H]
    \centering
    \begin{tabular}{rlr|r|rrrr}
    \hline
    env & model & latency & WW ZI & \multicolumn{4}{c}{Our Bootstrap ZI Surplus} \\
    &  &  & mean & mean & SE & 95\% CI diff. & 99\% CI diff. \\
    \hline
  1 &      CDA &        0 &              10383 &        10424.98 &          25.16 &       \textbf{(-8.56, 91.70)} &     \textbf{(-19.40, 105.87)} \\
  1 & 2M no LA &        0 &              11807 &        11646.95 &          20.46 &   (-201.47, -119.01) &   (-213.62, -108.64) \\
  1 & 2M no LA &      100 &              10373 &        12352.73 &          30.11 &   (1917.68, 2039.47) &   (1898.85, 2055.08) \\
  1 & 2M no LA &      200 &              10621 &        12432.74 &          28.11 &   (1759.90, 1865.58) &   (1745.67, 1883.28) \\
  1 & 2M no LA &      300 &              11244 &        13334.28 &          22.59 &   (2048.73, 2133.00) &   (2034.82, 2144.56) \\
  1 & 2M no LA &      400 &              10438 &        12011.45 &          31.21 &   (1514.59, 1631.01) &   (1494.74, 1646.98) \\
  1 & 2M no LA &      600 &              11128 &        13219.34 &          23.00 &   (2044.96, 2137.10) &   (2032.95, 2147.65) \\
  1 & 2M no LA &      700 &              11302 &        13416.63 &          21.97 &   (2073.08, 2158.08) &   (2067.02, 2174.30) \\
  1 & 2M no LA &      900 &              12358 &        14370.30 &          19.96 &   (1973.33, 2052.03) &   (1960.74, 2066.15) \\
  1 &    2M LA &      100 &               5919 &        11087.45 &          31.05 &   (5108.51, 5227.72) &   (5092.44, 5246.67) \\
  1 &    2M LA &      200 &               6358 &        10914.22 &          28.98 &   (4499.37, 4613.63) &   (4483.80, 4636.16) \\
  1 &    2M LA &      300 &               6398 &        11007.58 &          28.25 &   (4552.43, 4663.06) &   (4539.36, 4678.94) \\
  1 &    2M LA &      400 &               6130 &        12006.76 &          25.31 &   (5826.29, 5923.12) &   (5810.17, 5941.75) \\
  1 &    2M LA &      600 &               7459 &        13077.26 &          19.19 &   (5581.28, 5655.00) &   (5573.17, 5671.59) \\
  1 &    2M LA &      700 &               5256 &         9894.58 &          34.86 &   (4568.50, 4706.73) &   (4550.74, 4722.00) \\
  1 &    2M LA &      900 &               6819 &        13113.03 &          20.10 &   (6256.27, 6334.16) &   (6241.98, 6345.99) \\
  2 &      CDA &        0 &             136140 &       136131.59 &          65.51 &    (\textbf{-132.27, 118.40)} &    \textbf{(-182.89, 152.51)} \\
  2 & 2M no LA &        0 &             134339 &       134497.56 &          68.15 &      (30.47, 288.08) &       (1.55, 330.69) \\
  2 & 2M no LA &       50 &             135789 &       152272.08 &          60.61 & (16359.26, 16598.50) & (16321.55, 16624.47) \\
  2 & 2M no LA &      100 &             136542 &       150308.53 &          63.92 & (13628.88, 13885.56) & (13593.96, 13918.30) \\
  2 &    2M LA &       50 &             133177 &       151116.85 &          63.00 & (17815.33, 18057.21) & (17785.51, 18092.99) \\
  2 &    2M LA &      100 &             124012 &       139444.74 &          79.90 & (15269.38, 15587.25) & (15223.51, 15648.18) \\
  3 &      CDA &        0 &              27482 &        27472.00 &          40.48 &      \textbf{(-92.02, 66.62)} &     \textbf{(-107.86, 90.36)} \\
  3 & 2M no LA &        0 &              29424 &        29929.43 &          31.43 &     (442.42, 569.11) &     (425.19, 588.16) \\
  3 & 2M no LA &       25 &              29347 &        32160.12 &          29.81 &   (2758.37, 2869.43) &   (2738.92, 2893.16) \\
  3 & 2M no LA &       50 &              29479 &        32518.84 &          27.80 &   (2987.94, 3094.59) &   (2972.80, 3115.38) \\
  3 & 2M no LA &       75 &              29271 &        32020.52 &          30.45 &   (2691.58, 2810.43) &   (2678.22, 2821.88) \\
  3 & 2M no LA &      100 &              29277 &        31964.32 &          30.76 &   (2624.09, 2747.89) &   (2609.72, 2765.10) \\
  3 &    2M LA &       25 &              26612 &        29915.89 &          40.24 &   (3218.13, 3380.89) &   (3192.64, 3400.87) \\
  3 &    2M LA &       50 &              27953 &        31863.76 &          31.61 &   (3849.29, 3971.74) &   (3829.06, 3988.38) \\
  3 &    2M LA &       75 &              26388 &        30566.72 &          36.17 &   (4104.39, 4243.83) &   (4085.11, 4273.02) \\
  3 &    2M LA &      100 &              25070 &        29576.74 &          41.90 &   (4428.30, 4592.02) &   (4408.80, 4619.89) \\
    \hline
\end{tabular}
    \caption{
    Test of quantitative alignment of ZI surplus results from our \textit{BestGuess} replication versus WW.
    Each WW ZI value is the mean ZI surplus reported by WW \cite{wah_latency_2016} (pp.~83,~85) for 500 mixtures (100 simulation runs each) sampled according to the welfare-maximizing strategy profile.
    Our results are calculated from 5,000 mixtures (100 runs each) based on the same strategy profile. From these, we construct 1,000 bootstrap samples of 500 mixtures each to produce confidence intervals for the mean difference. 
    We report the mean and standard error of the bootstrap sample means, along with the 95\% and 99\% confidence intervals for the difference between our bootstrap sample means and the reported WW mean.
    Confidence intervals that contain zero are shown in \textbf{bold}, indicating where quantitative alignment is not rejected.
    }
    \label{tab:ww_quant_equivalence_bestguess}
\end{table}

Table \ref{tab:ww_quant_equivalence_bestguess} reports the 500-mixture mean ZI surplus values reported by WW, the mean of our bootstrap sample means, the bootstrap standard error (SE), and the 95\% and 99\% confidence intervals for the difference between our bootstrap sample means and the reported WW mean.
We highlight in bold the experiments where the bootstrap confidence intervals contain the original mean, indicating where quantitative alignment is not rejected. We do not reject quantitative alignment at the 95\% level for each of the consolidated \textit{CDA} experiments.
The canonical means fall outside the confidence intervals for each of the other fragmented market configurations.
The canonical means are also further away from our range of results when latency is nonzero in the fragmented market configurations.

Table \ref{tab:ww_la_quant_equivalence_bestguess} similarly reports the 500-mixture mean LA surplus values reported by WW, the mean and SE of our bootstrap sample means, and the 95\% and 99\% confidence intervals for the difference between our bootstrap sample means and the reported WW mean.
Note that these values are limited to the experiments where there is an LA agent.
We see further difference between our implementation and the original results here. In general, our mean LA surplus values are much lower than the values reported by WW. Note that our ZI surplus values exceed those reported by WW in Table~\ref{tab:ww_quant_equivalence_bestguess} for the experiments with LA, and by a larger margin than in the experiments with non-zero latency without LA.
Thus, we reject quantitative alignment for all but the simplest configuration of the model (\textit{CDA}), and we appear to get further away with increased environmental complexity.

\begin{table}[H]
    \centering
    \begin{tabular}{rlr|r|rrrr}
    \hline
    env & model & latency & WW LA & \multicolumn{4}{c}{Our Bootstrap LA Surplus} \\
    &  &  & mean & mean & SE & 95\% CI diff. & 99\% CI diff. \\
    \hline
  1 & 2M LA &      100 &               3487 &          477.46 &           4.11 & (-3017.45, -3001.26) & (-3020.43, -2999.39) \\
  1 & 2M LA &      200 &               3164 &          526.96 &           5.02 & (-2646.79, -2627.22) & (-2650.33, -2624.88) \\
  1 & 2M LA &      300 &               3224 &          561.61 &           5.48 & (-2673.32, -2651.84) & (-2675.87, -2648.27) \\
  1 & 2M LA &      400 &               4018 &          552.45 &           4.44 & (-3474.20, -3456.73) & (-3476.67, -3453.40) \\
  1 & 2M LA &      600 &               4349 &         1194.00 &           3.72 & (-3162.28, -3147.79) & (-3164.54, -3146.08) \\
  1 & 2M LA &      700 &               2958 &          395.65 &           4.66 & (-2571.55, -2552.57) & (-2574.25, -2550.72) \\
  1 & 2M LA &      900 &               4825 &         1120.33 &           4.72 & (-3713.31, -3695.89) & (-3716.12, -3692.78) \\
  2 & 2M LA &       50 &               2417 &         1107.80 &           2.95 & (-1314.93, -1303.54) & (-1317.40, -1302.08) \\
  2 & 2M LA &      100 &               2888 &         1662.21 &           3.87 & (-1233.64, -1218.41) & (-1235.39, -1216.72) \\
  3 & 2M LA &       25 &                538 &          229.84 &           1.46 &   (-310.99, -305.24) &   (-311.72, -304.05) \\
  3 & 2M LA &       50 &               1154 &          448.30 &           2.04 &   (-709.58, -701.78) &   (-710.43, -700.34) \\
  3 & 2M LA &       75 &               1470 &          556.20 &           2.38 &   (-918.63, -909.38) &   (-920.13, -907.60) \\
  3 & 2M LA &      100 &               1763 &          632.59 &           2.62 & (-1135.45, -1125.67) & (-1137.51, -1123.92) \\
    \hline
\end{tabular}
    \caption{Test of quantitative alignment of LA surplus results from our \textit{BestGuess} replication versus WW.
    Each WW LA value is the mean LA surplus reported by WW \cite{wah_latency_2016} (pp.~83,~85) for 500 mixtures (100 simulation runs each) sampled according to the welfare-maximizing strategy profile.
    Our results are calculated from 5,000 mixtures (100 runs each) based on the same strategy profile. From these, we construct 1,000 bootstrap samples of 500 mixtures each to produce confidence intervals for the mean difference. 
    We report the mean and standard error of the bootstrap sample means, along with the 95\% and 99\% confidence intervals for the difference between our bootstrap sample means and the reported WW mean. No confidence intervals contain zero, and thus quantitative alignment is rejected for each experiment.
    }
    \label{tab:ww_la_quant_equivalence_bestguess}
\end{table}

\subsection{Using an existing codebase: \textit{MarketSim}}
\label{marketsim}

WW note that the market simulation system used in the original study \say{has been extended by other members of the Strategic Reasoning Group at the University of Michigan and employed in several other studies} \cite{wah_latency_2016} (p.78).
The Strategic Reasoning Group made the codebase \textit{MarketSim}\footnote{\url{https://github.com/egtaonline/market-sim/tree/marketsim1}} available in 2016, and it was cited in a 2017 study by Wah, Wright, and Wellman (WWW) \cite{wah_welfare_2017} on the effects of market making in a CDA model.
Wah's dissertation \cite{wah_computational_2016} describes common modeling assumptions for WW and WWW and does not appear to differentiate these models aside from the following: the first examines latency arbitrage in a market with multiple exchanges, while WWW looks at market making scenarios in a market with a single exchange.
The ZI agent strategies in particular are described with common assumptions and logic between the two models.
\textit{MarketSim} also features a SIP and is configurable to run with multiple exchanges and SIP latency.
We therefore downloaded the \textit{MarketSim} codebase to try to reproduce results from WW as well as WWW.

\textit{MarketSim} is mainly coded in Java, with Python wrapper scripts to create the JSON model configuration files and bash scripts to kick off the Java processes.
The \textit{MarketSim} GitHub repo comes with a build tutorial explaining how to download, build, and run \textit{MarketSim}.
The tutorial and repo include example model configurations related to the WWW model experiments, but it does not include all configurations needed to replicate WWW and includes none of the experiment configurations to run the experiments from WW \cite{wah_latency_2016}.
We forked\footnote{Our fork is available here: \url{https://github.com/eratlif1/market-sim-WW-replication/tree/marketsim1-upstream}} \textit{MarketSim} in order to update the Python scripts, add the WW experiment configurations, and create a processing script to format the JSON files output from runs of the model.
We made no changes to the Java code containing the \textit{MarketSim} model logic.

To ensure proper implementation, we first ran \textit{MarketSim} with the experimental configurations specified in WWW for the experiments with only ZI agents (i.e. no market maker).
This exercise and evidence are detailed more fully in Appendix~\ref{WWW}.
Despite some uncertainty about WWW's experimental design, we confirm quantitative alignment to the original results for seven out of eight experiments using the bootstrap quantitative alignment methodology from Section \ref{quantitative_alignment_methodology}
We also achieve relational equivalence on two other metrics for almost all experiments.
This gives us confidence \textit{MarketSim} is as specified by WWW and that we are correctly interpreting how to run it for a given market configuration and mixture of ZI strategy profiles.

We next ran \textit{MarketSim} for the experiments specified in WW, running the model for 5,000 mixtures, 100 runs per mixture, similar to our independent replication (Section \ref{bestguess_vs_ww}).
For each experiment we sample 1,000 bootstrap samples of 500 mixtures, 100 runs per mixture, in order to create 95\% and 99\% bootstrap confidence intervals.
We then compare these confidence intervals to the mean ZI surplus values reported by WW, which were calculated from a sample of 500 mixtures, 100 runs per mixture.
Shown in Table~\ref{tab:ww_quant_equivalence_marketsim}, we see large differences between the original ZI surplus results and our results from \textit{MarketSim}. Even for the simplest configuration of the model---the consolidated \textit{CDA} settings---the \textit{MarketSim} results are $\sim 10$\% greater than the results reported by WW.

\begin{table}[H]
    \centering
    \begin{tabular}{rlr|r|rrrr}
    \hline
    env & model & latency & WW & \multicolumn{4}{c}{Our Bootstrap ZI Surplus} \\
     &  &  & mean & mean & SE & lower CI diff. & upper CI diff. \\
    \hline
  1 &      CDA &        0 &              10383 &        12289.73 &          26.74 &   (1855.59, 1959.12) &   (1846.53, 1973.57) \\
  1 & 2M no LA &        0 &              11807 &        14019.45 &          19.41 &   (2173.89, 2250.87) &   (2163.41, 2263.96) \\
  1 & 2M no LA &      100 &              10373 &        11493.66 &          30.53 &   (1064.36, 1184.35) &   (1040.53, 1201.47) \\
  1 & 2M no LA &      200 &              10621 &        11772.13 &          31.48 &   (1090.31, 1214.70) &   (1070.75, 1236.44) \\
  1 & 2M no LA &      300 &              11244 &        12527.27 &          24.46 &   (1237.25, 1330.07) &   (1223.00, 1339.55) \\
  1 & 2M no LA &      400 &              10438 &        11612.09 &          32.19 &   (1110.35, 1234.00) &   (1091.75, 1259.64) \\
  1 & 2M no LA &      600 &              11128 &        12293.56 &          25.01 &   (1117.65, 1215.88) &   (1105.85, 1227.51) \\
  1 & 2M no LA &      700 &              11302 &        12573.54 &          24.60 &   (1222.85, 1319.70) &   (1213.19, 1336.33) \\
  1 & 2M no LA &      900 &              12358 &        13942.99 &          19.74 &   (1545.23, 1623.89) &   (1534.78, 1637.40) \\
  1 &    2M LA &      100 &               5919 &        10320.99 &          33.90 &   (4335.02, 4466.33) &   (4309.45, 4496.33) \\
  1 &    2M LA &      200 &               6358 &        10335.76 &          31.19 &   (3913.24, 4038.55) &   (3899.93, 4059.57) \\
  1 &    2M LA &      300 &               6398 &        10484.96 &          30.72 &   (4024.74, 4146.65) &   (4010.35, 4161.10) \\
  1 &    2M LA &      400 &               6130 &        11163.95 &          26.74 &   (4985.30, 5083.69) &   (4965.27, 5104.31) \\
  1 &    2M LA &      600 &               7459 &        12350.82 &          20.70 &   (4852.06, 4932.84) &   (4840.08, 4944.95) \\
  1 &    2M LA &      700 &               5256 &         9106.39 &          43.77 &   (3765.96, 3934.92) &   (3733.69, 3961.31) \\
  1 &    2M LA &      900 &               6819 &        12370.84 &          20.19 &   (5512.02, 5591.30) &   (5497.10, 5602.61) \\
  2 &      CDA &        0 &             136140 &       152444.71 &          60.11 & (16188.98, 16419.37) & (16149.99, 16452.03) \\
  2 & 2M no LA &        0 &             134339 &       153540.75 &          61.67 & (19079.45, 19322.28) & (19030.06, 19360.26) \\
  2 & 2M no LA &       50 &             135789 &       153328.55 &          60.02 & (17420.62, 17660.11) & (17377.15, 17679.95) \\
  2 & 2M no LA &      100 &             136542 &       152249.01 &          62.77 & (15587.16, 15837.89) & (15549.02, 15884.94) \\
  2 &    2M LA &       50 &             133177 &       151447.36 &          59.22 & (18154.99, 18387.69) & (18112.46, 18420.29) \\
  2 &    2M LA &      100 &             124012 &       138244.54 &          90.23 & (14047.07, 14410.17) & (13991.34, 14460.46) \\
  3 &      CDA &        0 &              27482 &        30428.98 &          43.63 &   (2859.82, 3035.40) &   (2838.94, 3057.55) \\
  3 & 2M no LA &        0 &              29424 &        32620.14 &          29.97 &   (3140.55, 3254.04) &   (3119.73, 3274.39) \\
  3 & 2M no LA &       25 &              29347 &        32249.51 &          30.11 &   (2843.98, 2959.56) &   (2820.94, 2976.45) \\
  3 & 2M no LA &       50 &              29479 &        32432.73 &          30.05 &   (2896.54, 3014.19) &   (2875.82, 3032.04) \\
  3 & 2M no LA &       75 &              29271 &        32122.41 &          29.37 &   (2793.07, 2906.29) &   (2780.78, 2919.21) \\
  3 & 2M no LA &      100 &              29277 &        32162.18 &          28.95 &   (2829.27, 2943.19) &   (2812.02, 2961.46) \\
  3 &    2M LA &       25 &              26612 &        29014.75 &          46.33 &   (2310.24, 2489.45) &   (2287.27, 2512.28) \\
  3 &    2M LA &       50 &              27953 &        31366.24 &          32.38 &   (3353.59, 3477.03) &   (3332.15, 3502.22) \\
  3 &    2M LA &       75 &              26388 &        29840.00 &          41.25 &   (3370.17, 3531.57) &   (3341.22, 3551.45) \\
  3 &    2M LA &      100 &              25070 &        28743.86 &          45.32 &   (3587.38, 3757.65) &   (3563.13, 3785.74) \\
    \hline
\end{tabular}
    \caption{Test of quantitative alignment of ZI surplus from \textit{MarketSim}'s results versus  WW. 
    Each WW ZI value is the mean ZI surplus reported by WW \cite{wah_latency_2016} (pp.~83,~85) for 500 mixtures (100 simulation runs each) sampled according to the welfare-maximizing strategy profile.
    Our results are calculated from 5,000 mixtures (100 runs each) based on the same strategy profile. From these, we construct 1,000 bootstrap samples of 500 mixtures each to produce confidence intervals for the mean difference. 
    We report the mean and standard error of the 1,000 bootstrap sample means, along with the 95\% and 99\% confidence intervals for the difference between our bootstrap sample means and the reported WW mean. No confidence intervals contain zero, and thus quantitative alignment is rejected for each experiment.}
    \label{tab:ww_quant_equivalence_marketsim}
\end{table}

The quantitative alignment tests for mean LA surplus from \textit{MarketSim} versus WW similarly show differences, shown in Table~\ref{tab:ww_la_quant_equivalence_marketsim}.
Interestingly, the range of LA surplus values from \textit{MarketSim} are similar to those we saw from our \textit{BestGuess} implementation, with differences versus WW of similar magnitude. The \textit{MarketSim} mean LA surplus values are significantly lower than those reported by WW for all but one experiment (\textit{2M LA} with $\delta=25$ in Environment 3). This is despite the \textit{MarketSim} ZI surplus values being much larger than those reported by WW.
As we were successful at achieving quantitative alignment with \textit{MarketSim} in most experiments from WWW but not WW, these results suggests some difference in model logic between the current codebase and what was used for the WW study.

\begin{table}[H]
    \centering
    \begin{tabular}{rlr|r|rrrr}
    \hline
    env & model & latency & WW LA & \multicolumn{4}{c}{Our Bootstrap LA Surplus} \\
    &  &  & mean & mean & SE & 95\% CI diff. & 99\% CI diff. \\
    \hline
  1 & 2M LA &      100 &               3487 &          582.32 &           6.65 & (-2918.07, -2891.66) & (-2922.02, -2887.51) \\
  1 & 2M LA &      200 &               3164 &          642.50 &           7.71 & (-2536.68, -2506.74) & (-2542.10, -2502.44) \\
  1 & 2M LA &      300 &               3224 &          679.99 &           7.46 & (-2558.63, -2528.98) & (-2564.73, -2525.40) \\
  1 & 2M LA &      400 &               4018 &          726.28 &           6.94 & (-3305.61, -3278.20) & (-3309.42, -3274.14) \\
  1 & 2M LA &      600 &               4349 &         1488.15 &           5.86 & (-2871.72, -2848.81) & (-2875.02, -2845.20) \\
  1 & 2M LA &      700 &               2958 &          437.99 &           6.62 & (-2533.06, -2506.82) & (-2536.59, -2504.53) \\
  1 & 2M LA &      900 &               4825 &         1382.05 &           6.44 & (-3455.26, -3430.16) & (-3459.43, -3425.37) \\
  2 & 2M LA &       50 &               2417 &         1842.37 &           3.99 &   (-582.45, -566.88) &   (-585.29, -564.66) \\
  2 & 2M LA &      100 &               2888 &         1552.83 &           3.51 & (-1342.13, -1328.32) & (-1343.29, -1326.44) \\
  3 & 2M LA &       25 &                538 &          658.55 &           2.94 &     (114.82, 126.18) &     (113.50, 128.41) \\
  3 & 2M LA &       50 &               1154 &          859.93 &           2.80 &   (-299.76, -288.58) &   (-300.62, -286.63) \\
  3 & 2M LA &       75 &               1470 &          841.49 &           3.10 &   (-634.26, -622.17) &   (-635.90, -620.59) \\
  3 & 2M LA &      100 &               1763 &          821.54 &           3.53 &   (-948.24, -934.72) &   (-950.18, -931.89) \\
    \hline
\end{tabular}
    \caption{Test of quantitative alignment of LA surplus from \textit{MarketSim}'s results versus WW.
    Each WW LA value is the mean LA surplus reported by WW \cite{wah_latency_2016} (pp.~83,~85) for 500 mixtures (100 simulation runs each) sampled according to the welfare-maximizing strategy profile.
    Our results are calculated from 5,000 mixtures (100 runs each) based on the same strategy profile. From these, we construct 1,000 bootstrap samples of 500 mixtures each to produce confidence intervals for the mean difference. 
    We report the mean and standard error of the bootstrap sample means, along with the 95\% and 99\% confidence intervals for the difference between our bootstrap sample means and the reported WW mean. No confidence intervals contain zero, and thus quantitative alignment is rejected for each experiment.
    }
    \label{tab:ww_la_quant_equivalence_marketsim}
\end{table}

\subsubsection{Differences between \textit{MarketSim} and our \textit{BestGuess} implementation}

Given the overlap in timing and authorship for \textit{MarketSim} and the WW model, we felt it may hold insights for the replication despite also having apparent differences.
Not knowing what differs between the two versions of the model makes it difficult to determine what design choices from \textit{MarketSim} are (1) necessary to recreate the original results, (2) responsible for the differences in results, or (3) immaterial.
We identified two key differences between the \textit{MarketSim} logic and our \textit{BestGuess} implementation.
The first is a difference in the information used by the ZI traders for their `greedy strategy'. The second is an apparent bug in the order routing logic.

\paragraph{Greedy strategy}
The ZI agent strategy used in the WW and WWW model is referred to as `ZIRP' (zero-intelligence with re-submission plus) in \textit{MarketSim}.
Its base strategy\footnote{The core ZIRP strategy, including its greedy strategy logic, can be found here: \url{https://github.com/egtaonline/market-sim/blob/marketsim1/hft-sim/src/entity/agent/BackgroundAgent.java\#L274}.} 
appears to be implemented similar to our logic detailed in Section \ref{zi_main}.
A difference is found when we examine the greedy strategy logic, however.
Below is the greedy logic from \textit{MarketSim}. Let \texttt{BBO} again refer to the BBO for the trader $ZI_i$'s primary exchange, $v_i(t)$ denote private valuation of the asset, $p$ the order price from the base strategy, and $\eta$ the trader's greedy threshold parameter:
    \begin{lstlisting}
    max_bid = BBO.BID
    min_ask = BBO.ASK
    requested_surplus = |v_i(t) - p|
    if side == `BUY':
        if requested_surplus * eta <= v_i(t) - min_ask:
            p = v_i(t)
    else: // `SELL'
        if requested_surplus * eta <= max_bid - v_i(t):
            p = v_i(t)
    \end{lstlisting}
There are two key differences to note between the above logic and our \textit{BestGuess} interpretation (Section \ref{zi_main}). First, in \textit{MarketSim} the trader only uses the primary exchange BBO to determine if a greedy opportunity exists; it does not look at the NBBO.
Second, if a greedy opportunity exists, the trader sets its price to be its \textit{private valuation} $v_i(t)$, rather than equal to the BBO $ASK$ or $BID$.
In a single-exchange market (inherently with no latency), this logic will be equivalent to our interpretation, as the $BBO$ and $NBBO$ are equivalent, and the greedy buy (sell) order will execute at the $ASK$ ($BID$) price.
However, if there are two exchanges, greedy opportunities at the alternative exchange will not be acted on in the \textit{MarketSim} logic.
If there are two exchanges and latency is non-zero, the order may be re-routed to the alternative exchange based on stale information, potentially resulting in the worst case scenario where the order executes (or becomes a limit order) with price $v_i(t)$, resulting in zero achieved surplus for the trader.
Setting the order price to $v_i(t)$ is overly aggressive if the goal is to achieve at least fraction $\eta$ of its requested surplus $s_i(t)$ (i.e. $\eta s_i(t)$).

WW say the ZI agents have access to quotes on their primary market and the NBBO. It would seem strange for the ZI agents to then only use the primary market quotes to inform their greedy order strategy, particularly if a greedy order can still be rerouted to the alternative exchange if the NBBO specifies a better price there.
WW also say, \say{Setting $\eta=1$ is equivalent to the strategy without employing the threshold} \cite{wah_latency_2016} (p.75). However, in the \textit{MarketSim} implementation, a trader with $\eta=1$ could still end up submitting a greedy order at price $v_i(t)$ that then gets misrouted to the other exchange due to a stale NBBO. Thus, setting $\eta=1$ is not equivalent to not employing the greedy strategy in this implementation.

\paragraph{Order routing}

The other main difference we found between \textit{MarketSim} and our interpretation of the model is the logic used to route ZI orders. 
\textit{MarketSim} has the ZI agent's primary exchange reroute the order based on the NBBO, rather than having the trader make the routing decision.
This makes some sense, as exchanges in the NMS are required to reroute orders based on best execution rules \cite{us_securities_and_exchange_commission_regulation_2005}.
At the same time, WW's description of the model implies the traders do the routing. Also, if the trader does not use the SIP NBBO in its pricing strategy, and it does not use it to route its own order, why does the trader need access to the NBBO?
However, there is no latency between a trader submitting an order and the exchange receiving it in the model, and as such the exchange versus the trader doing the routing should not have any effect.

Within the exchange order routing logic in \textit{MarketSim}, however, we find a possible bug.
Below is the \texttt{submitNMSOrder} function (in its original Java) called by the ZIRP traders in \textit{MarketSim} when they submit an order to their primary exchange:
\begin{lstlisting}
public void submitNMSOrder(Agent agent, OrderType type, Price price,
    int quantity, TimeStamp currentTime, TimeStamp duration) {
    checkNotNull(type, "Order type");
    checkArgument(quantity > 0, "Quantity must be positive");
    
    BestBidAsk nbbo = sip.getNBBO();
    Market bestMarket = this;

    if (quantity > 0) { // buy
        boolean nbboBetter = nbbo.getBestAsk() != null
            && nbbo.getBestAsk().lessThan(quote.getAskPrice());
        boolean willTransact = price.greaterThan(nbbo.getBestAsk());
        if (nbboBetter && willTransact)
            bestMarket = nbbo.getBestAskMarket();
    } else { // sell
        boolean nbboBetter = nbbo.getBestBid() != null
                && nbbo.getBestBid().greaterThan(quote.getBidPrice());
        boolean willTransact = price.lessThan(nbbo.getBestBid());
        if (nbboBetter && willTransact)
            bestMarket = nbbo.getBestBidMarket();
    }

    if (!bestMarket.equals(this))
        log.log(INFO, "Routing %s %s %d @ %s from %s %s to NBBO %s %s",
            agent, type, quantity, price, this, quote, bestMarket, nbbo);

    bestMarket.submitOrder(agent, type, price, quantity, currentTime,
        duration);
}
\end{lstlisting}
The exchange checks the NBBO and determines whether the order would execute immediately at a better price at the alternative exchange, rerouting the order if so.
The function contains a check asserting that the order's \texttt{quantity} `must be positive'. It then uses the test \texttt{quantity > 0} to determine if the order is a buy order as opposed to a sell order. Elsewhere in \textit{MarketSim}, orders have a \texttt{type} that equals \texttt{OrderType.BUY} or \texttt{OrderType.SELL} (see the code referenced in the ZIRP strategy logic above for an example). It appears the \texttt{submitNMSOrder} function is using the wrong criteria to determine order type, and in fact it will treat every incoming order as a buy order when determining the routing decision.
This means an incoming sell order may be misrouted to the alternative exchange based on the ask prices at that exchange, or it may fail to be rerouted to the alternative exchange when the order may have executed there at a better price than at the primary exchange.

\subsection{\textit{BestGuess}+\textit{MarketSim}}

We next took the greedy strategy logic from \textit{MarketSim} highlighted above and integrated it into our \textit{BestGuess} model. 
We will refer to this implementation of the model as \textit{BestGuess+MS} going forward.
All logic in this implementation is as specified in our ODD except for the greedy strategy logic, which now follows the above detail from \textit{MarketSim}.
We again ran this model for 5,000 mixtures, 100 runs per mixture, and employ the quantitative alignment testing outlined in Section \ref{quantitative_alignment_methodology}.

Our quantitative alignment tests for mean ZI surplus under the \textit{BestGuess+MS} implementation are shown in Table~\ref{tab:ww_quant_equivalence_bestguess_ms_alt}.
We find smaller differences compared to the original results across more experiments in these results than we saw for our \textit{BestGuess} results (Table~\ref{tab:ww_quant_equivalence_bestguess}).
The 95\% confidence intervals from \textit{BestGuess+MS} ZI surplus means contain the WW ZI surplus means for the consolidated \textit{CDA} configurations and the fragmented market configurations without latency from Environments 2 and 3. The canonical ZI surplus mean for the fragmented market configuration without latency in Environment 1 is within the 99\% confidence intervals for this implementation but not the 95\% confidence intervals.
We still reject quantitative alignment for all configurations of the model with non-zero latency, however.

\begin{table}[H]
    \centering
    \begin{tabular}{rlr|r|rrrr}
    \hline
    env & model & latency & WW & \multicolumn{4}{c}{Our Bootstrap ZI Surplus} \\
     &  &  & mean & mean & SE & 95\% CI diff. & 99\% CI diff. \\
    \hline
  1 &      CDA &        0 &              10383 &        10425.08 &          25.05 &       \textbf{(-6.59, 91.70)} &     \textbf{(-20.88, 107.22)} \\
  1 & 2M no LA &        0 &              11807 &        11848.60 &          20.29 &        (3.85, 79.76) &       \textbf{(-7.32, 91.41)} \\
  1 & 2M no LA &      100 &              10373 &        10165.29 &          29.04 &   (-261.75, -151.72) &   (-281.56, -136.20) \\
  1 & 2M no LA &      200 &              10621 &        10374.49 &          29.28 &   (-304.84, -190.37) &   (-319.17, -174.62) \\
  1 & 2M no LA &      300 &              11244 &        11066.88 &          23.35 &   (-220.67, -129.55) &   (-229.91, -113.65) \\
  1 & 2M no LA &      400 &              10438 &        10215.79 &          31.54 &   (-283.98, -161.31) &   (-305.11, -145.09) \\
  1 & 2M no LA &      600 &              11128 &        10905.40 &          23.52 &   (-269.76, -178.31) &   (-286.20, -165.97) \\
  1 & 2M no LA &      700 &              11302 &        11132.93 &          23.40 &   (-212.51, -124.40) &   (-229.52, -110.91) \\
  1 & 2M no LA &      900 &              12358 &        12188.92 &          19.28 &   (-206.59, -132.95) &   (-217.71, -119.99) \\
  1 &    2M LA &      100 &               5919 &         4885.89 &          43.02 &  (-1117.64, -953.04) &  (-1139.36, -920.91) \\
  1 &    2M LA &      200 &               6358 &         5311.49 &          41.96 &  (-1124.50, -965.41) &  (-1161.70, -946.49) \\
  1 &    2M LA &      300 &               6398 &         5412.35 &          40.00 &  (-1061.06, -906.79) &  (-1080.34, -876.78) \\
  1 &    2M LA &      400 &               6130 &         5189.03 &          41.57 &  (-1022.13, -860.66) &  (-1045.77, -831.78) \\
  1 &    2M LA &      600 &               7459 &         6510.35 &          34.76 &  (-1019.09, -885.71) &  (-1037.37, -863.14) \\
  1 &    2M LA &      700 &               5256 &         4311.61 &          48.33 &  (-1033.21, -844.46) &  (-1061.59, -818.67) \\
  1 &    2M LA &      900 &               6819 &         6435.05 &          37.25 &   (-452.60, -311.27) &   (-478.14, -283.28) \\
  2 &      CDA &        0 &             136140 &       136130.24 &          65.87 &    \textbf{(-136.49, 118.04)} &    \textbf{(-165.45, 156.12)} \\
  2 & 2M no LA &        0 &             134339 &       134304.59 &          66.93 &     \textbf{(-159.13, 96.80)} &    \textbf{(-202.71, 140.73)} \\
  2 & 2M no LA &       50 &             135789 &       134176.50 &          67.52 & (-1749.85, -1483.62) & (-1784.80, -1445.79) \\
  2 & 2M no LA &      100 &             136542 &       134830.23 &          68.94 & (-1847.38, -1579.04) & (-1884.78, -1543.89) \\
  2 &    2M LA &       50 &             133177 &       131877.57 &          66.75 & (-1429.96, -1168.87) & (-1454.77, -1117.89) \\
  2 &    2M LA &      100 &             124012 &       121634.05 &          86.93 & (-2550.99, -2210.60) & (-2596.45, -2155.82) \\
  3 &      CDA &        0 &              27482 &        27473.78 &          41.34 &      \textbf{(-91.20, 70.80)} &     \textbf{(-114.83, 99.48)} \\
  3 & 2M no LA &        0 &              29424 &        29449.31 &          31.17 &      \textbf{(-38.14, 80.64)} &     \textbf{(-55.39, 100.15)} \\
  3 & 2M no LA &       25 &              29347 &        29173.53 &          30.08 &   (-232.63, -115.76) &    (-245.77, -96.16) \\
  3 & 2M no LA &       50 &              29479 &        29256.27 &          29.72 &   (-279.13, -164.12) &   (-299.10, -148.92) \\
  3 & 2M no LA &       75 &              29271 &        29052.93 &          30.68 &   (-278.61, -157.04) &   (-303.42, -141.71) \\
  3 & 2M no LA &      100 &              29277 &        29038.14 &          29.87 &   (-295.51, -174.62) &   (-315.24, -164.33) \\
  3 &    2M LA &       25 &              26612 &        26357.77 &          40.65 &   (-335.19, -173.46) &   (-365.85, -149.73) \\
  3 &    2M LA &       50 &              27953 &        27538.00 &          36.79 &   (-485.88, -342.43) &   (-508.52, -312.64) \\
  3 &    2M LA &       75 &              26388 &        25830.41 &          41.05 &   (-640.52, -474.18) &   (-659.90, -449.40) \\
  3 &    2M LA &      100 &              25070 &        24449.34 &          46.87 &   (-715.92, -530.06) &   (-738.23, -506.31) \\
    \hline
\end{tabular}
    \caption{Test of quantitative alignment of ZI surplus from our \textit{BestGuess+MS} implementation results versus WW. 
    Each WW ZI value is the mean ZI surplus reported by WW \cite{wah_latency_2016} (pp.~83,~85) for 500 mixtures (100 simulation runs each) sampled according to the welfare-maximizing strategy profile.
    Our results are calculated from 5,000 mixtures (100 runs each) based on the same strategy profile. From these, we construct 1,000 bootstrap samples of 500 mixtures each to produce confidence intervals for the mean difference. 
    We report the mean and standard error of the 1,000 bootstrap sample means, along with the 95\% and 99\% confidence intervals for the difference between our bootstrap sample means and the reported WW mean.
    Confidence intervals that contain zero are shown in \textbf{bold}, indicating where quantitative alignment is not rejected.}
    \label{tab:ww_quant_equivalence_bestguess_ms_alt}
\end{table}

Our quantitative alignment results for mean LA surplus under the \textit{BestGuess+MS} implementation are similarly shown in Table~\ref{tab:ww_la_quant_equivalence_bestguess_ms_alt}. The differences between the bootstrap sample LA surplus means and the canonical means is much smaller than we saw for either of the previous sets of results from \textit{BestGuess} and \textit{MarketSim}. However, we still reject quantitative alignment for each experiment with LA. Note that the LA achieves much higher surplus on average in the \textit{BestGuess+MS} implementation than in our original \textit{BestGuess} implementation. This shows the latency arbitrage opportunities are more prevalent, larger, or both under the implementation of the ZI greedy strategy taken from \textit{MarketSim} versus our \textit{BestGuess} greedy strategy. This fits with our expectation, as the ZI agents use more of their available information in the \textit{BestGuess} implementation and act more strategically.

\begin{table}[H]
    \centering
    \begin{tabular}{rlr|r|rrrr}
    \hline
    env & model & latency & WW LA & \multicolumn{4}{c}{Our Bootstrap LA Surplus} \\
    &  &  & mean & mean & SE & 95\% CI diff. & 99\% CI diff. \\
    \hline
  1 & 2M LA &      100 &               3487 &         4279.07 &          27.57 & (738.41, 846.33) & (721.62, 865.13) \\
  1 & 2M LA &      200 &               3164 &         3971.93 &          26.61 & (756.94, 855.87) & (738.61, 880.99) \\
  1 & 2M LA &      300 &               3224 &         4010.53 &          27.37 & (731.13, 839.26) & (716.75, 853.23) \\
  1 & 2M LA &      400 &               4018 &         4771.45 &          29.85 & (694.47, 810.59) & (673.47, 828.50) \\
  1 & 2M LA &      600 &               4349 &         5115.70 &          25.89 & (715.43, 816.48) & (701.50, 834.79) \\
  1 & 2M LA &      700 &               2958 &         3721.58 &          26.71 & (710.13, 816.54) & (698.10, 830.06) \\
  1 & 2M LA &      900 &               4825 &         5113.45 &          26.97 & (235.29, 341.54) & (225.39, 363.58) \\
  2 & 2M LA &       50 &               2417 &         2355.86 &           8.93 & (-78.15, -43.91) & (-82.22, -38.47) \\
  2 & 2M LA &      100 &               2888 &         3246.52 &          23.37 & (312.28, 404.25) & (301.52, 419.73) \\
  3 & 2M LA &       25 &                538 &          616.55 &           8.17 &   (63.29, 94.03) &   (58.82, 98.98) \\
  3 & 2M LA &       50 &               1154 &         1337.14 &          14.71 & (155.44, 212.36) & (147.12, 225.31) \\
  3 & 2M LA &       75 &               1470 &         1787.98 &          20.78 & (277.83, 357.39) & (268.44, 372.37) \\
  3 & 2M LA &      100 &               1763 &         2173.46 &          26.17 & (359.47, 462.56) & (340.67, 474.26) \\
    \hline
\end{tabular}
    \caption{Test of quantitative alignment of LA surplus results from our \textit{BestGuess+MS} replication versus WW.
    Each WW LA value is the mean LA surplus reported by WW \cite{wah_latency_2016} (pp.~83,~85) for 500 mixtures (100 simulation runs each) sampled according to the welfare-maximizing strategy profile.
    Our results are calculated from 5,000 mixtures (100 runs each) based on the same strategy profile. From these, we construct 1,000 bootstrap samples of 500 mixtures each to produce confidence intervals for the mean difference. 
    We report the mean and standard error of the bootstrap sample means, along with the 95\% and 99\% confidence intervals for the difference between our bootstrap sample means and the reported WW mean. No confidence intervals contain zero, and thus quantitative alignment is rejected for each experiment.
    }
    \label{tab:ww_la_quant_equivalence_bestguess_ms_alt}
\end{table}

\subsubsection{A note on the order routing bug}

We also considered including the order routing bug from \textit{MarketSim}'s \texttt{submitNMSOrder} function, in case this was important to drive the behavior reported by WW.
We ran this alternative version of the model (call it \textit{BestGuess+MS+bug})  again for 5,000 mixtures, 100 runs per mixture, and compared against the original results.
We overall found that our hybrid version without the bug (\textit{BestGuess+MS}) aligned more closely to the original results than the version with the bug in most experimental settings. We present these results in full in Appendix~\ref{alt_vs_buggy}. For space considerations, and because it does not appear the bug helps us better replicate the canonical results, we focus on the \textit{BestGuess}, \textit{MarketSim}, and \textit{BestGuess+MS} implementations of the model going forward.

\section{Qualitative behavior of the model}
\label{qualitative}

We next look at the qualitative behavior of our implementations of the model (\textit{BestGuess}, \textit{MarketSim}, and \textit{BestGuess+MS}), comparing against the results plotted by WW to assess relational equivalence.
WW plot the results across experimental settings for surplus, execution time, spread (BBO and NBBO), and the number of transactions.
In each figure caption, we reference the corresponding page and figure numbers from WW to which our results should be compared.
Each metric from our simulations are averaged over all 5,000 mixtures, 100 runs per mixture, for a given experiment, versus 500 mixtures for the WW results.

\begin{figure}[H]
\centering
\begin{subfigure}[t]{0.3\textwidth}
    \centering
    \includegraphics[width=1\textwidth]{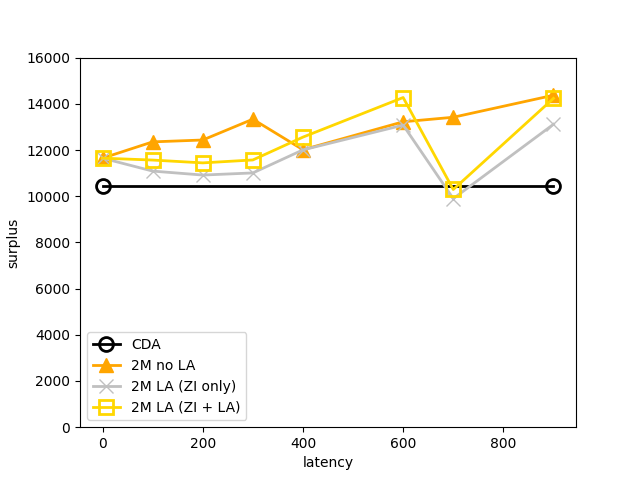}
    \caption{Env. 1 -- \textit{BestGuess}}
    \label{fig:env1_surplus}
\end{subfigure}%
\begin{subfigure}[t]{.3\textwidth}
    \centering
    \includegraphics[width=1\textwidth]{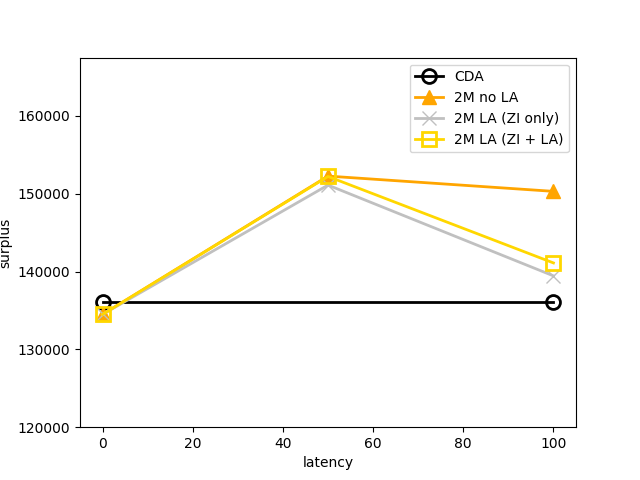}
    \caption{Env. 2 -- \textit{BestGuess}}
    \label{fig:env2_surplus}
\end{subfigure}%
\begin{subfigure}[t]{0.3\textwidth}
    \centering
    \includegraphics[width=1\textwidth]{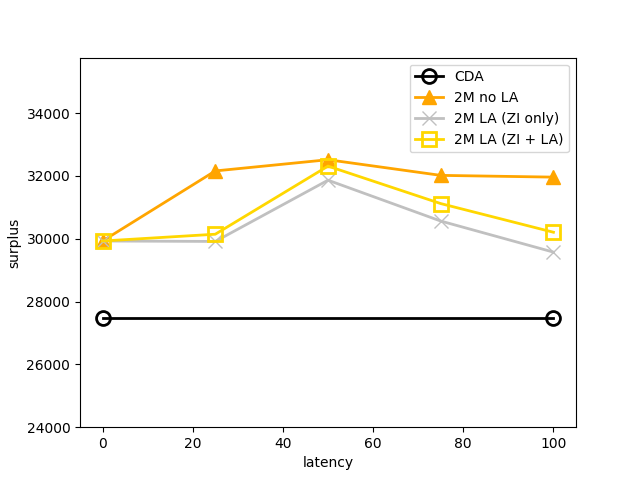}
    \caption{Env. 3 -- \textit{BestGuess}}
    \label{fig:env3_surplus}
\end{subfigure}%
\hspace{1em}
\begin{subfigure}[t]{0.3\textwidth}
    \centering
    \includegraphics[width=1\textwidth]{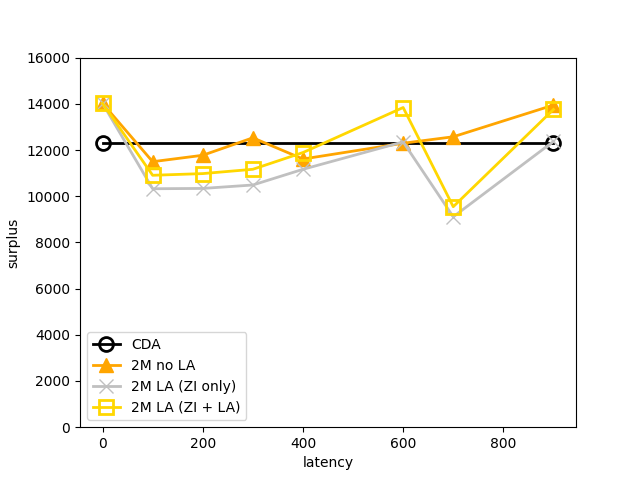}
    \caption{Env. 1 -- \textit{MarketSim}}
    \label{fig:env1_surplus_ms}
\end{subfigure}%
\begin{subfigure}[t]{.3\textwidth}
    \centering
    \includegraphics[width=1\textwidth]{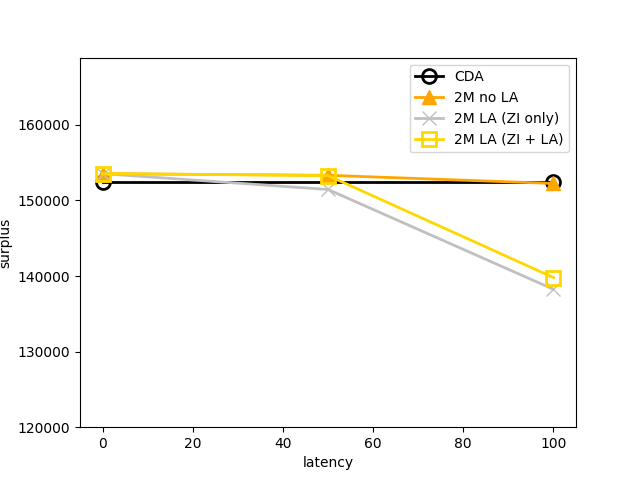}
    \caption{Env. 2 -- \textit{MarketSim}}
    \label{fig:env2_surplus_ms}
\end{subfigure}%
\begin{subfigure}[t]{0.3\textwidth}
    \centering
    \includegraphics[width=1\textwidth]{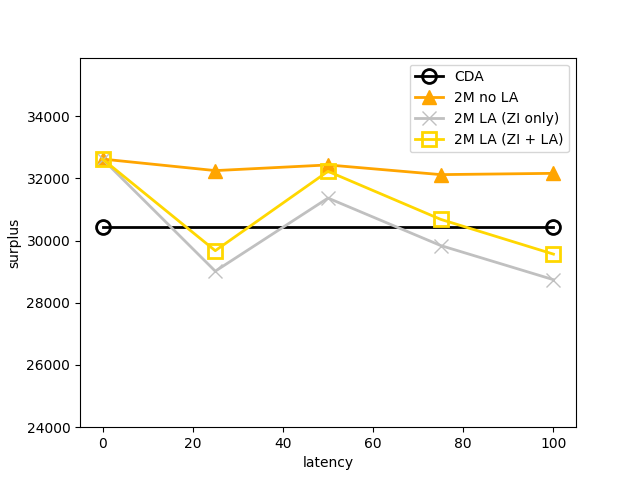}
    \caption{Env. 3 -- \textit{MarketSim}}
    \label{fig:env3_surplus_ms}
\end{subfigure}%
\hspace{1em}
\begin{subfigure}[t]{0.3\textwidth}
    \centering
    \includegraphics[width=1\textwidth]{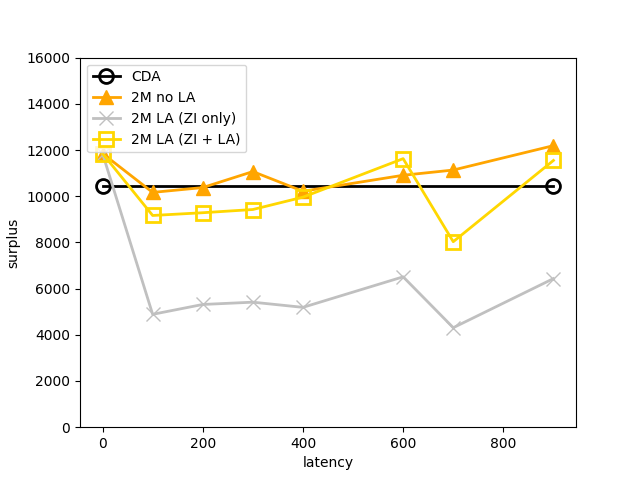}
    \caption{Env. 1 -- \textit{BestGuess+MS}}
    \label{fig:env1_surplus_alt}
\end{subfigure}%
\begin{subfigure}[t]{.3\textwidth}
    \centering
    \includegraphics[width=1\textwidth]{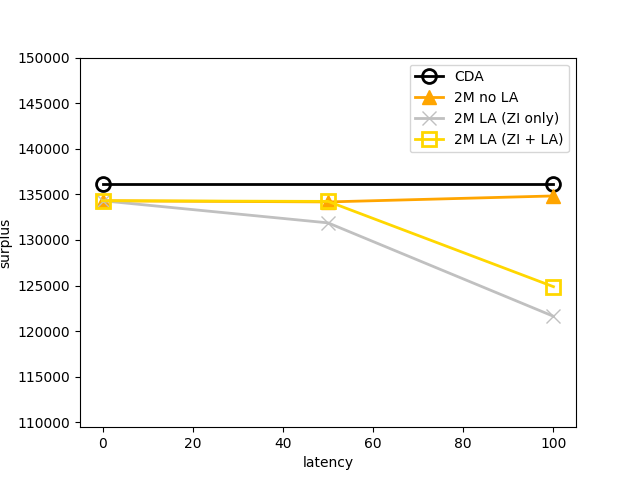}
    \caption{Env. 2 -- \textit{BestGuess+MS}}
    \label{fig:env2_surplus_alt}
\end{subfigure}%
\begin{subfigure}[t]{0.3\textwidth}
    \centering
    \includegraphics[width=1\textwidth]{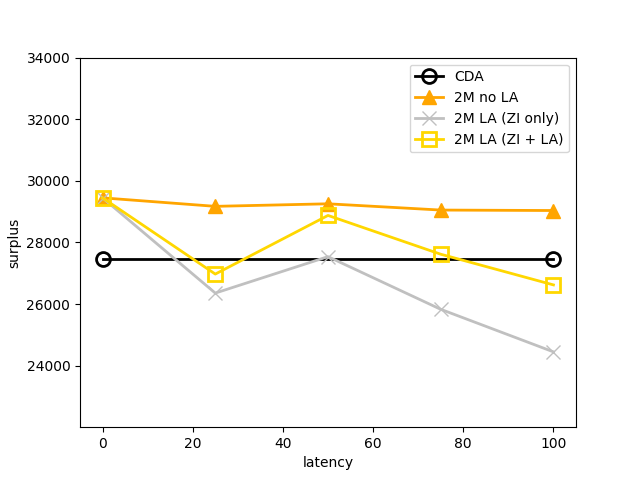}
    \caption{Env. 3 -- \textit{BestGuess+MS}}
    \label{fig:env3_surplus_alt}
\end{subfigure}%
\caption{Mean surplus results from our implementations of the model and \textit{MarketSim}.
These results should be compared to the corresponding original figures reported by WW \cite{wah_latency_2016} (Fig. 4, p.86). 
Results from our simulations are averaged over 5,000 mixtures, 100 runs per mixture, for each experiment.}
\label{fig:ww_surplus}
\end{figure}

\paragraph{Surplus}
We start with the average total surplus values across the different experiments (environment+configuration+latency). 
As shown in Fig.~\ref{fig:ww_surplus}, our \textit{BestGuess} results exhibit clear differences from the canonical results. 
The WW results show mixed effects on surplus from fragmentation, and the introduction of the LA agent harms surplus in most settings.
Our \textit{BestGuess} results show almost every fragmented market configuration, with or without LA, outperform the conslidated \textit{CDA} configurations.
The only exceptions are the \textit{2M LA} results for $\delta=900$ in Env. 1 and the \textit{2M no LA} results with $\delta=0$ in Env. 2. 
In all other cases, surplus is higher with fragmentation, and non-zero latency results in higher surplus relative to no latency.
In most of the non-zero-latency settings, the presence of the LA agent reduces surplus, but surplus is still higher in those settings than the single-CDA market.

The \textit{MarketSim} results achieve relational equivalence in Environments 1 and 3 but not in Environment 2. In the Environment 2 \textit{MarketSim} results, the \textit{2M no LA} with $\delta=0$ surplus is greater than the \textit{CDA} surplus and decreases as latency is increased. In WW, the \textit{2M no LA} surplus with $\delta=0$ is lower than surplus in the consolidated \textit{CDA} configuration and increase as latency increases.
Our \textit{BestGuess+MS} implementation meanwhile achieves relational equivalence for all three environments; the only relational difference is the \textit{2M no LA} surplus for $\delta=100$ in Environment 2 is slightly lower than the \textit{CDA} surplus rather than slightly above.
Overall, these results add support to our previous evidence that the hybrid \textit{BestGuess+MS} model implementation is the closest to the original model.

\begin{figure}[H]
\centering
\begin{subfigure}[t]{0.3\textwidth}
    \centering
    \includegraphics[width=1\textwidth]{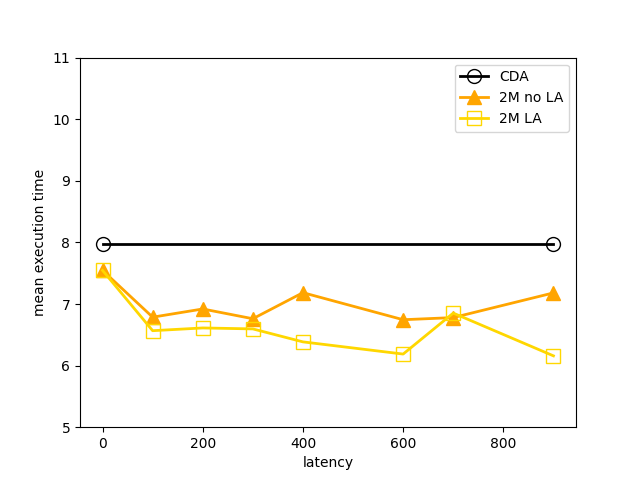}
    \caption{Env. 1 -- \textit{BestGuess}}
    \label{fig:env1_ex_time}
\end{subfigure}%
\begin{subfigure}[t]{.3\textwidth}
    \centering
    \includegraphics[width=1\textwidth]{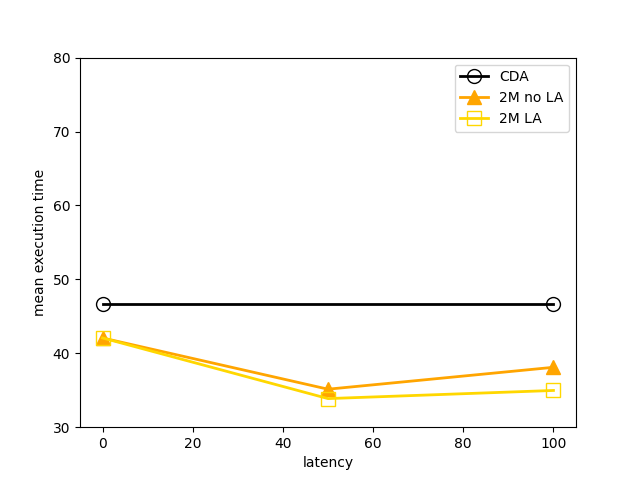}
    \caption{Env. 2 -- \textit{BestGuess}}
    \label{fig:env2_ex_time}
\end{subfigure}%
\begin{subfigure}[t]{0.3\textwidth}
    \centering
    \includegraphics[width=1\textwidth]{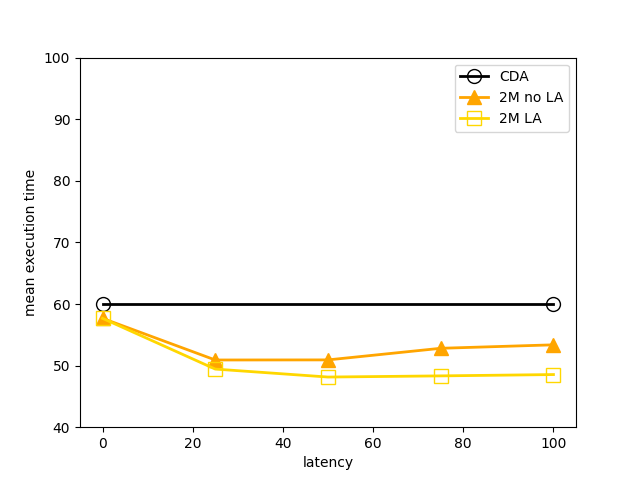}
    \caption{Env. 3 -- \textit{BestGuess}}
    \label{fig:env3_ex_time}
\end{subfigure}%
\hspace{1em}
\begin{subfigure}[t]{0.3\textwidth}
    \centering
    \includegraphics[width=1\textwidth]{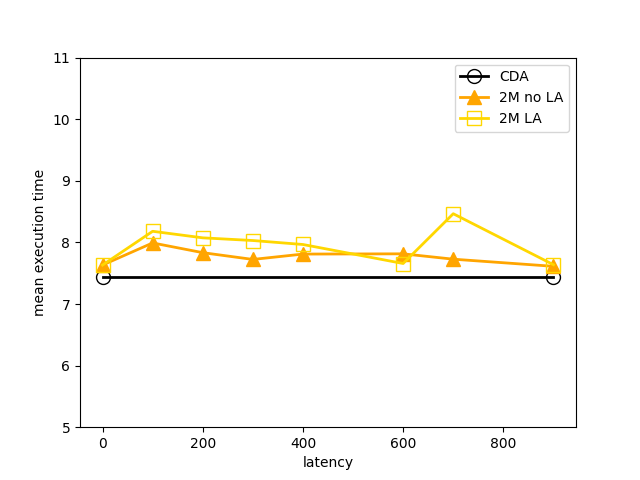}
    \caption{Env. 1 -- \textit{MarketSim}}
    \label{fig:env1_ex_time_ms}
\end{subfigure}%
\begin{subfigure}[t]{.3\textwidth}
    \centering
    \includegraphics[width=1\textwidth]{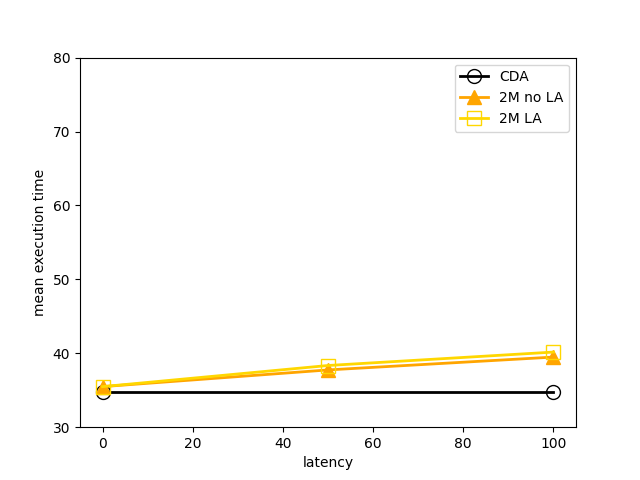}
    \caption{Env. 2 -- \textit{MarketSim}}
    \label{fig:env2_ex_time_ms}
\end{subfigure}%
\begin{subfigure}[t]{0.3\textwidth}
    \centering
    \includegraphics[width=1\textwidth]{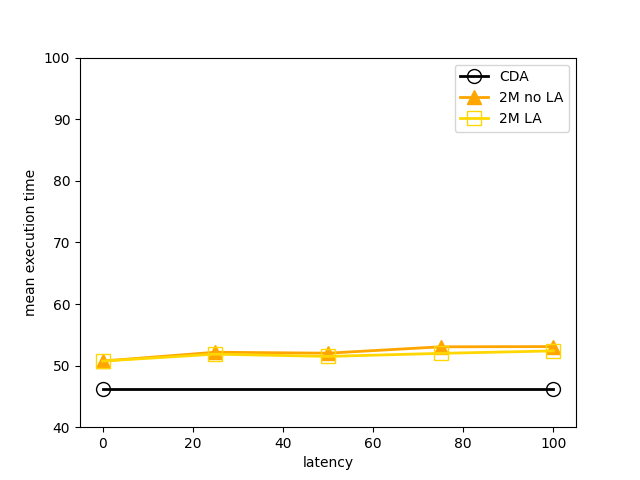}
    \caption{Env. 3 -- \textit{MarketSim}}
    \label{fig:env3_ex_time_ms}
\end{subfigure}%
\hspace{1em}
\begin{subfigure}[t]{0.3\textwidth}
    \centering
    \includegraphics[width=1\textwidth]{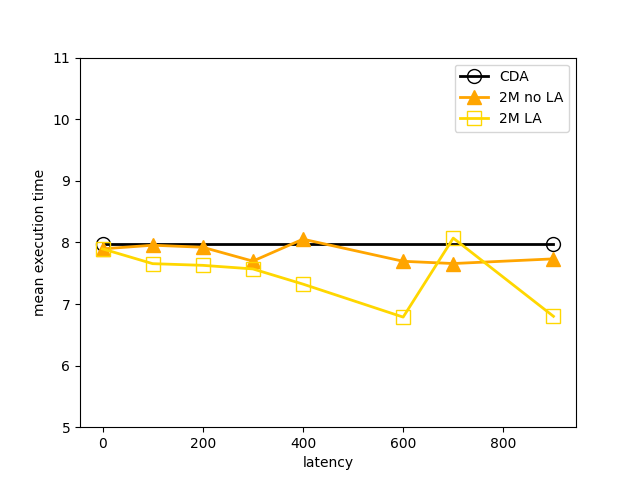}
    \caption{Env. 1 -- \textit{BestGuess+MS}}
    \label{fig:env1_ex_time_alt}
\end{subfigure}%
\begin{subfigure}[t]{.3\textwidth}
    \centering
    \includegraphics[width=1\textwidth]{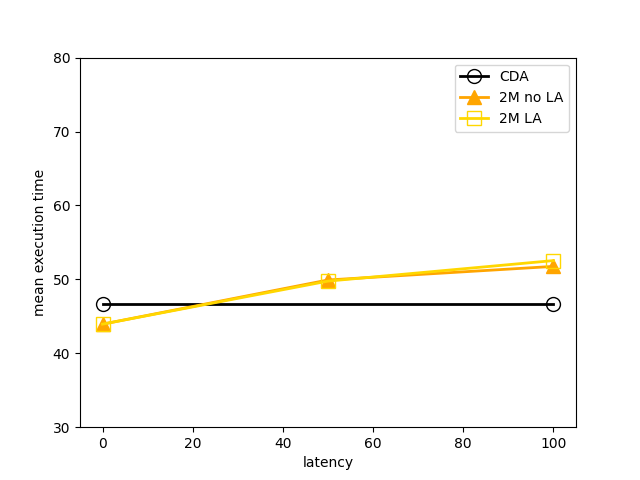}
    \caption{Env. 2 -- \textit{BestGuess+MS}}
    \label{fig:env2_ex_time_alt}
\end{subfigure}%
\begin{subfigure}[t]{0.3\textwidth}
    \centering
    \includegraphics[width=1\textwidth]{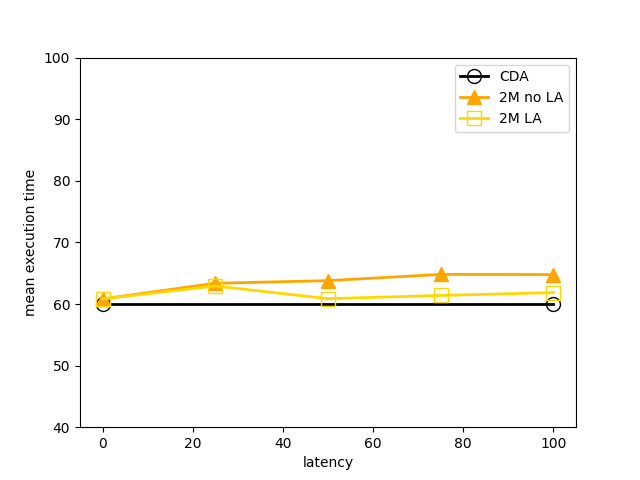}
    \caption{Env. 3 -- \textit{BestGuess+MS}}
    \label{fig:env3_ex_time_alt}
\end{subfigure}%
\caption{Mean execution time results from our implementations of the model and \textit{MarketSim}.
These results should be compared to the corresponding original figures reported by WW \cite{wah_latency_2016} (Fig. 6, p.86)
Results from our simulations are averaged over 5,000 mixtures, 100 runs per mixture, for each experiment.}
\label{fig:ww_extime}
\end{figure}

\paragraph{Execution time}
In Fig. \ref{fig:ww_extime}, we show our mean execution time results. 
WW found that execution time in their model was roughly the same or higher with LA than without LA in each experiment except $\delta=50$ in Environment 3. They also found that fragmentation did not generally improve execution time.
In our \textit{BestGuess} implementation results, fragmentation reduces execution time in every configuration of the model, non-zero latency generally improves execution time, and the presence of the LA agent also improves execution time in almost every non-zero latency setting.
We therefore again have material differences in qualitative behavior between our \textit{BestGuess} implementation and the original results.

Execution time from \textit{MarketSim} has notable differences as well, but in a different direction than our \textit{BestGuess} results. In \textit{MarketSim}, execution time is increased by fragmentation relative to the consolidated \textit{CDA} configuration. The presence of the LA agent increases execution time in most cases in Environments 1 and 2, but it has minimal effect in Environment 3.
We also see that the \textit{CDA} execution times differ quantitatively from the original results, while the \textit{CDA} results from our other implementations cannot be visually distinguished from WW.

Our \textit{BestGuess+MS} results resemble the WW results in most but not all configurations. In particular, we achieve relational equivalence in Env. 2; we achieve relational equivalence for the \textit{2M no LA} results but not the \textit{2M LA} results in Env. 1; and we achieve relational equivalence for the \textit{2M LA} results but not the \textit{2M no LA} results in Env. 3.
Thus, we do not have as consistent of alignment for execution time as we did for surplus between \textit{BestGuess+MS} and WW, but it remains our most faithful replication.

\paragraph{Spread}

\begin{figure}[H]
\centering
\begin{subfigure}[t]{0.3\textwidth}
    \centering
    \includegraphics[width=1\textwidth]{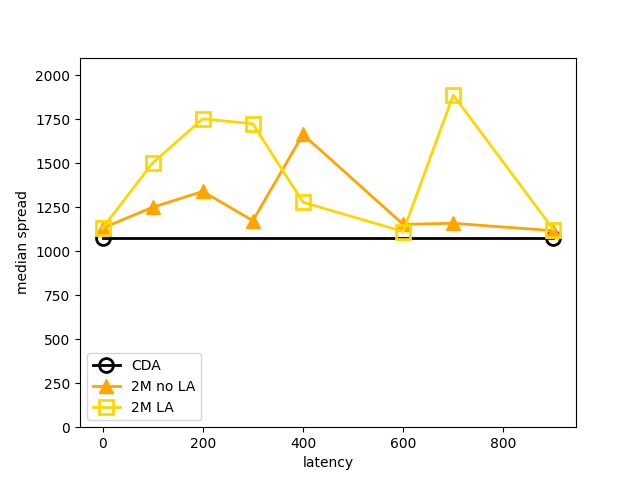}
    \caption{Env. 1 -- \textit{BestGuess}}
    \label{fig:env1_bbo_bg}
\end{subfigure}%
\begin{subfigure}[t]{.3\textwidth}
    \centering
    \includegraphics[width=1\textwidth]{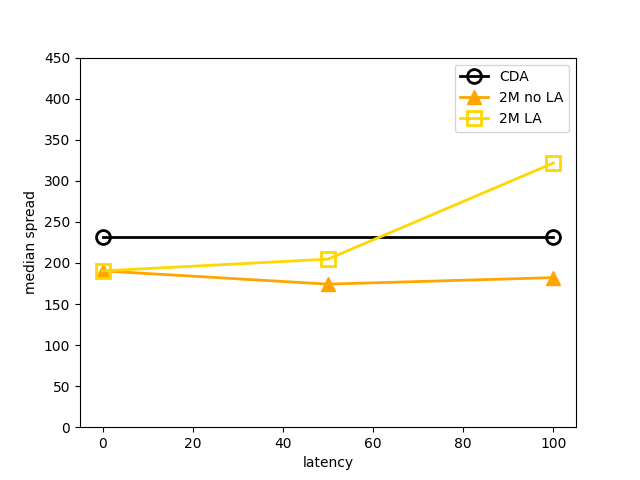}
    \caption{Env. 2 -- \textit{BestGuess}}
    \label{fig:env2_bbo_bg}
\end{subfigure}%
\hspace{1em}
\begin{subfigure}[t]{.3\textwidth}
    \centering
    \includegraphics[width=1\textwidth]{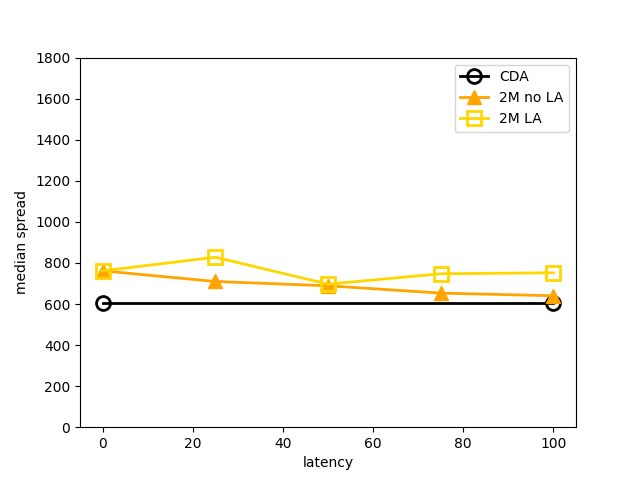}
    \caption{Env. 3 -- \textit{BestGuess}}
    \label{fig:env3_bbo_bg}
\end{subfigure}%
\hspace{1em}
\begin{subfigure}[t]{0.3\textwidth}
    \centering
    \includegraphics[width=1\textwidth]{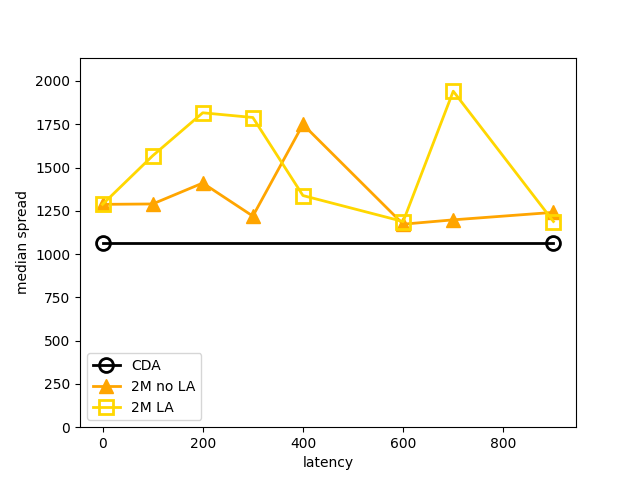}
    \caption{Env. 1 -- \textit{MarketSim}}
    \label{fig:env1_bbo_ms}
\end{subfigure}%
\begin{subfigure}[t]{.3\textwidth}
    \centering
    \includegraphics[width=1\textwidth]{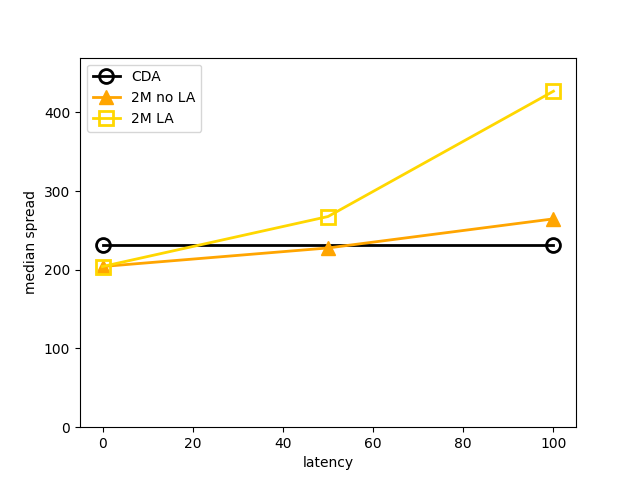}
    \caption{Env. 2 -- \textit{MarketSim}}
    \label{fig:env2_bbo_ms}
\end{subfigure}%
\begin{subfigure}[t]{.3\textwidth}
    \centering
    \includegraphics[width=1\textwidth]{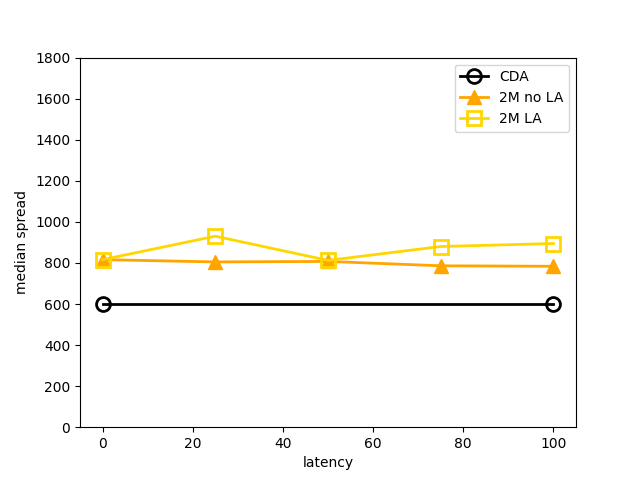}
    \caption{Env. 3 -- \textit{MarketSim}}
    \label{fig:env3_bbo_ms}
\end{subfigure}%
\hspace{1em}
\begin{subfigure}[t]{0.3\textwidth}
    \centering
    \includegraphics[width=1\textwidth]{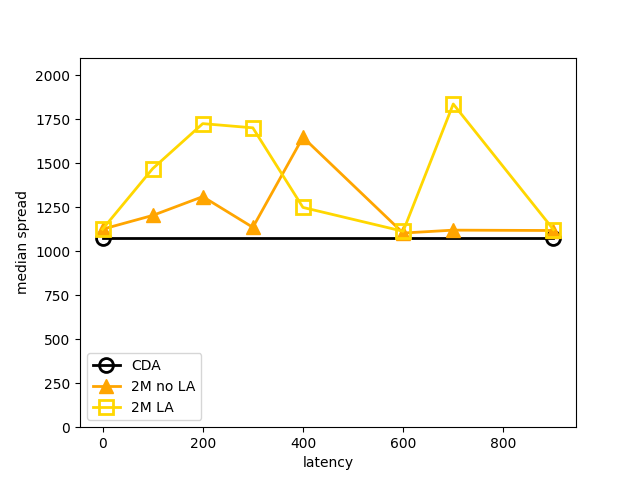}
    \caption{Env. 1 -- \textit{BestGuess+MS}}
    \label{fig:env1_bbo_bg_ms_alt}
\end{subfigure}%
\begin{subfigure}[t]{.3\textwidth}
    \centering
    \includegraphics[width=1\textwidth]{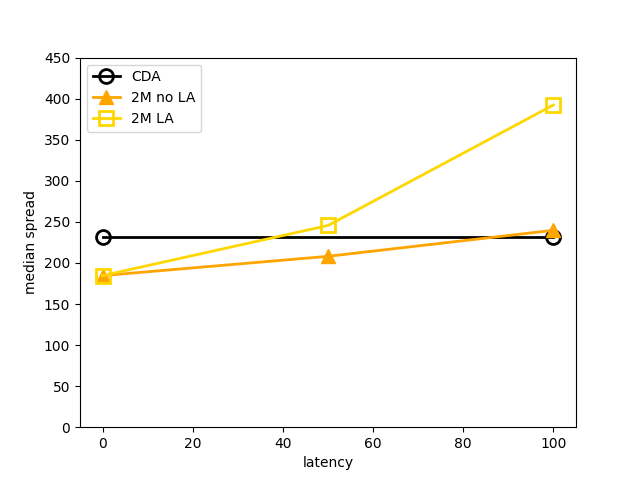}
    \caption{Env. 2 -- \textit{BestGuess+MS}}
    \label{fig:env2_bbo_bg_ms_alt}
\end{subfigure}%
\begin{subfigure}[t]{.3\textwidth}
    \centering
    \includegraphics[width=1\textwidth]{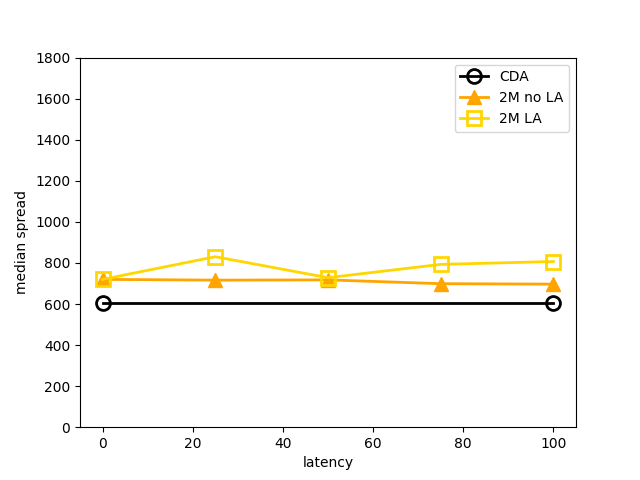}
    \caption{Env. 3 -- \textit{BestGuess+MS}}
    \label{fig:env3_bbo_bg_ms_alt}
\end{subfigure}%
\caption{Mean median spreads from the exchange BBOs for our implementations of the model and \textit{MarketSim}.
These results should be compared to the corresponding original figures reported by WW \cite{wah_latency_2016} (Fig. 7, p.87). 
BBO spreads are averaged across the different exchanges in a given simulation run. Results from our simulations are averaged over 5,000 mixtures, 100 runs per mixture, for each experiment.}
\label{fig:ww_bbo_spreads}
\end{figure}

\begin{figure}[H]
\centering
\begin{subfigure}[t]{0.3\textwidth}
    \centering
    \includegraphics[width=1\textwidth]{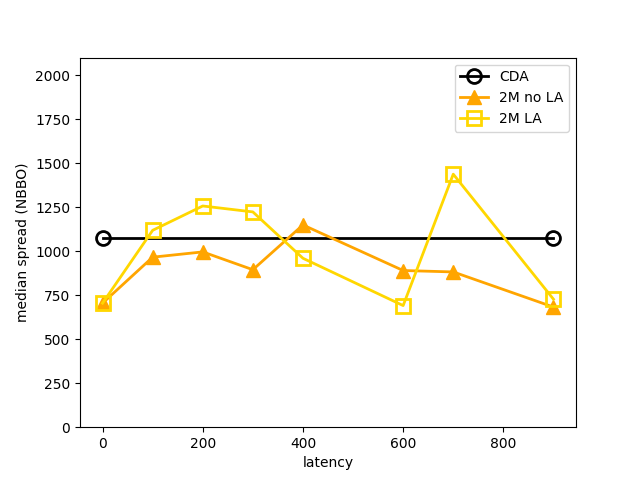}
    \caption{Env. 1 -- \textit{BestGuess}}
    \label{fig:env1_nbbo_bg}
\end{subfigure}%
\begin{subfigure}[t]{.3\textwidth}
    \centering
    \includegraphics[width=1\textwidth]{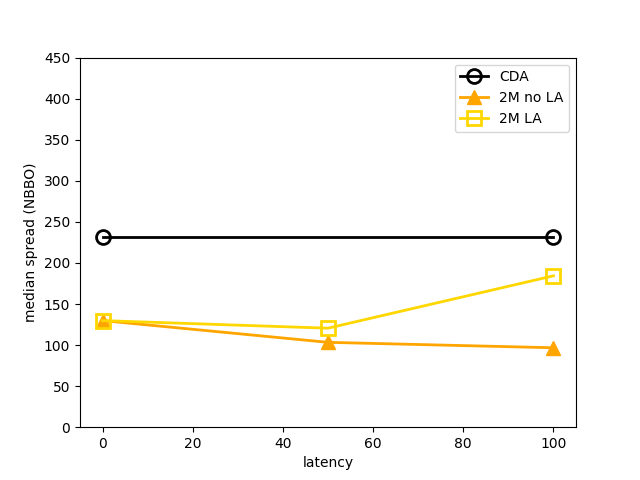}
    \caption{Env. 2 -- \textit{BestGuess}}
    \label{fig:env2_nbbo_bg}
\end{subfigure}%
\begin{subfigure}[t]{.3\textwidth}
    \centering
    \includegraphics[width=1\textwidth]{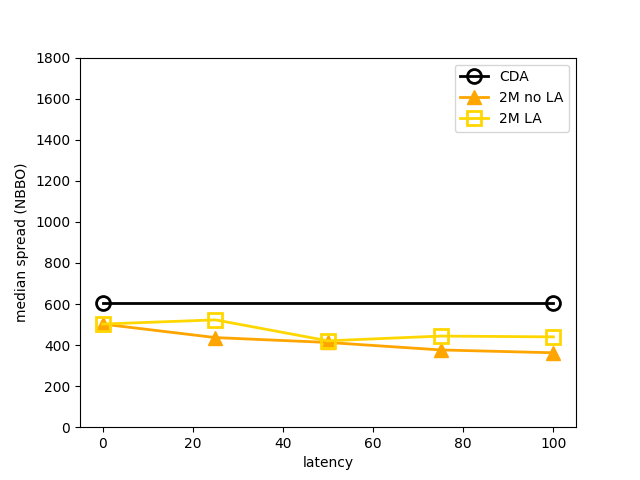}
    \caption{Env. 3 -- \textit{BestGuess}}
    \label{fig:env3_nbbo_bg}
\end{subfigure}%
\hspace{1em}
\begin{subfigure}[t]{0.3\textwidth}
    \centering
    \includegraphics[width=1\textwidth]{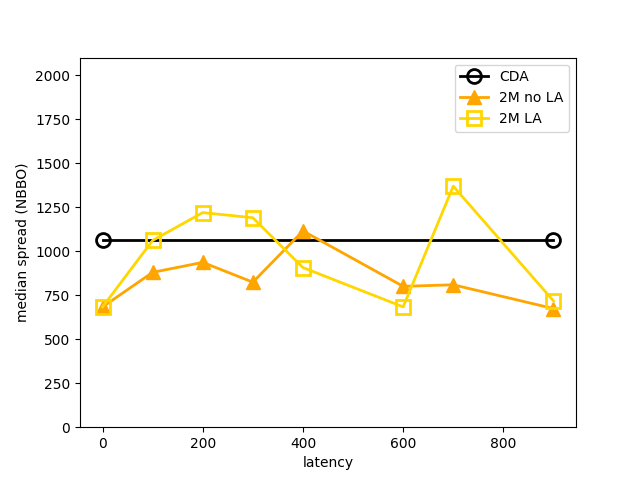}
    \caption{Env. 1 -- \textit{MarketSim}}
    \label{fig:env1_nbbo_ms}
\end{subfigure}%
\begin{subfigure}[t]{.3\textwidth}
    \centering
    \includegraphics[width=1\textwidth]{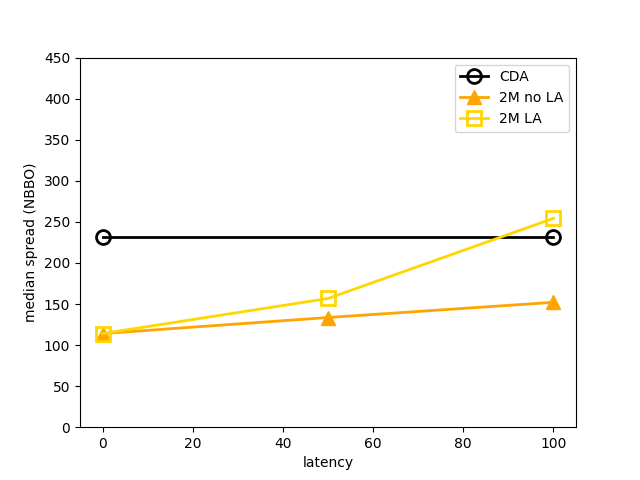}
    \caption{Env. 2 -- \textit{MarketSim}}
    \label{fig:env2_nbbo_ms}
\end{subfigure}%
\begin{subfigure}[t]{.3\textwidth}
    \centering
    \includegraphics[width=1\textwidth]{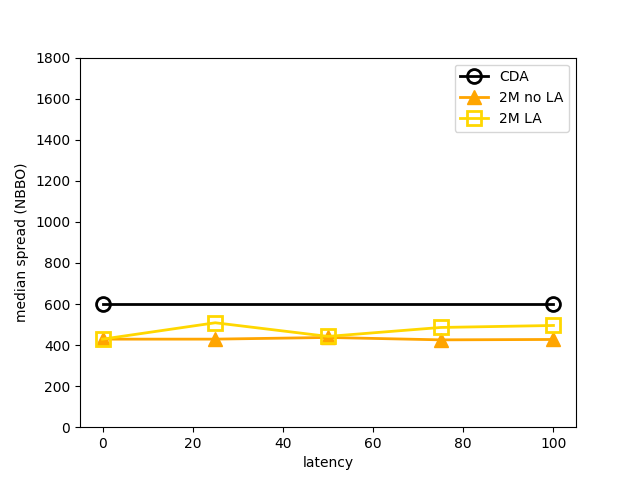}
    \caption{Env. 3 -- \textit{MarketSim}}
    \label{fig:env3_nbbo_ms}
\end{subfigure}%
\hspace{1em}
\begin{subfigure}[t]{0.3\textwidth}
    \centering
    \includegraphics[width=1\textwidth]{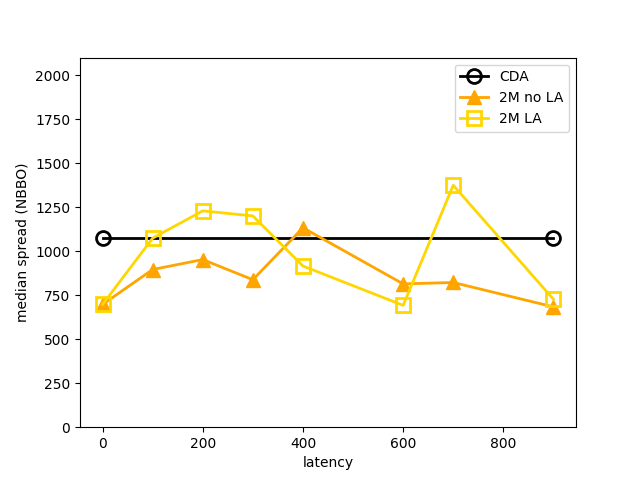}
    \caption{Env. 1 -- \textit{BestGuess+MS}}
    \label{fig:env1_nbbo_alt}
\end{subfigure}%
\begin{subfigure}[t]{.3\textwidth}
    \centering
    \includegraphics[width=1\textwidth]{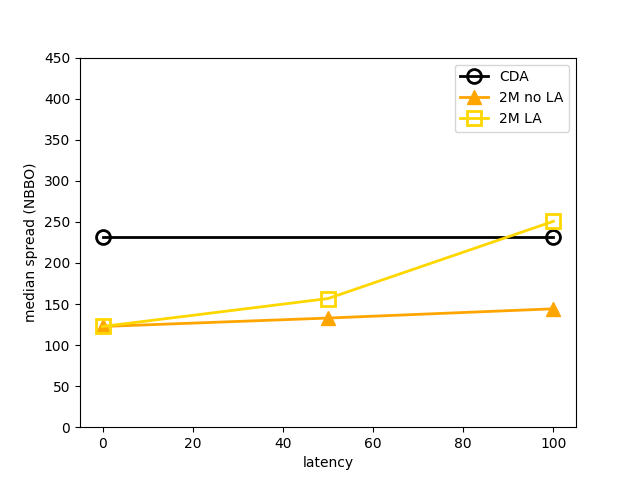}
    \caption{Env. 2 -- \textit{BestGuess+MS}}
    \label{fig:env2_nbbo_alt}
\end{subfigure}%
\begin{subfigure}[t]{.3\textwidth}
    \centering
    \includegraphics[width=1\textwidth]{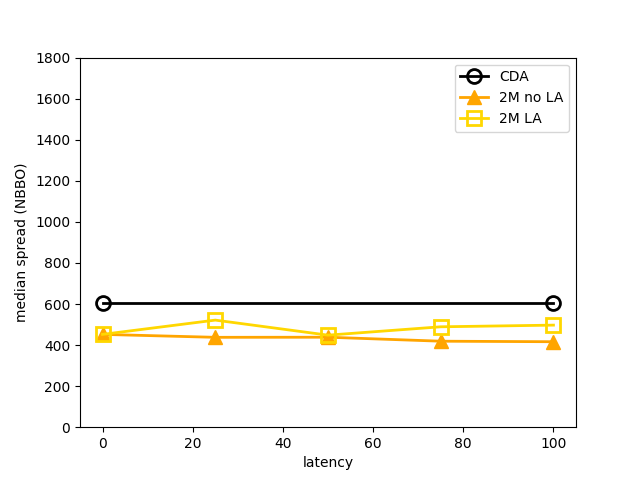}
    \caption{Env. 3 -- \textit{BestGuess+MS}}
    \label{fig:env3_nbbo_alt}
\end{subfigure}%
\caption{Mean median SIP NBBO spread for our implementations of the model and \textit{MarketSim}.
These results should be compared to the corresponding original figures reported by WW \cite{wah_latency_2016} (Fig. 7, p.87). 
Results from our simulations are averaged over 5,000 mixtures, 100 runs per mixture, for each experiment.}
\label{fig:ww_nbbo_spreads}
\end{figure}

In Figs. \ref{fig:ww_bbo_spreads} and \ref{fig:ww_nbbo_spreads}, respectively, we present the BBO and NBBO spreads generated by our simulation. 
In WW, latency arbitrage increases spreads in most settings, while fragmentation and latency in general have mixed effects relative to the consolidated \textit{CDA} market configurations.
We are not able to visually distinguish between the \textit{CDA} spreads from WW and our replication results, including for the \textit{MarketSim} results.
The presence of the LA agent generally exacerbates BBO spreads in our results, similar to WW.
The BBO spreads are otherwise relatively consistent across the three replications, and all differ from the WW results in similar ways (e.g. $\delta=0$ in Environment 1 and $\delta=50$ in Environment 3). 

The NBBO spreads for Environment 1 roughly match between each of our replications and WW.
Our \textit{BestGuess} results differ somewhat from WW in Environments 2 and 3, with a consistent ranking of the conslidated \textit{CDA} market having the largest NBBO spreads and the \textit{2M no LA} results having the smallest.
The relative NBBO spread movements from \textit{MarketSim} and \textit{BestGuess+MS} match WW in Environment 2, while in Environment 3 the with-LA results appear to match WW but the \textit{2M no LA} results do not have the jump in spread at $\delta=50$ that WW report.

\begin{figure}[H]
\centering
\begin{subfigure}[t]{0.3\textwidth}
    \centering
    \includegraphics[width=1\textwidth]{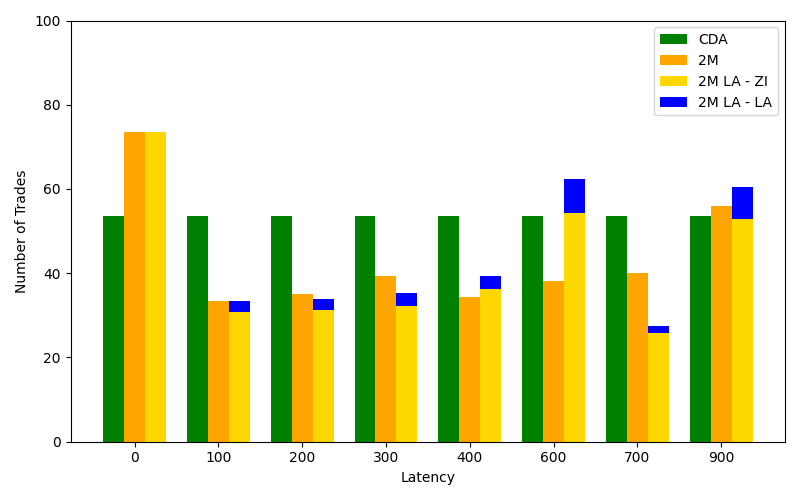}
    \caption{Env. 1 -- \textit{BestGuess}}
    \label{fig:env1_trans}
\end{subfigure}%
\begin{subfigure}[t]{.3\textwidth}
    \centering
    \includegraphics[width=1\textwidth]{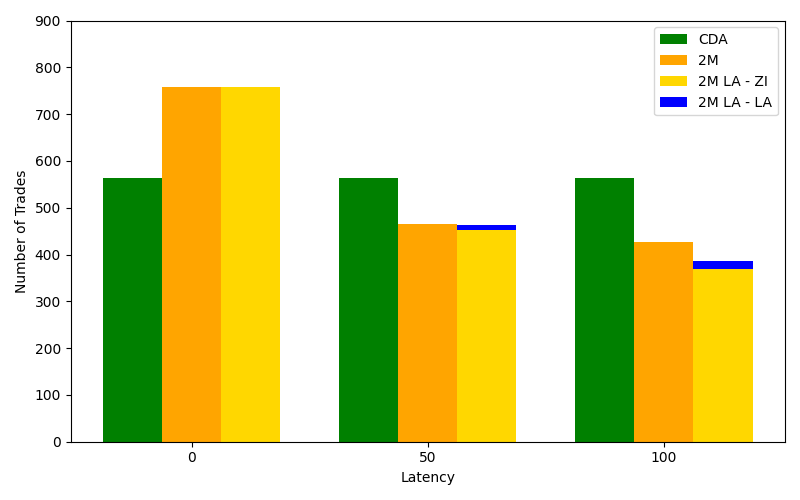}
    \caption{Env. 2 -- \textit{BestGuess}}
    \label{fig:env2_trans}
\end{subfigure}%
\begin{subfigure}[t]{0.3\textwidth}
    \centering
    \includegraphics[width=1\textwidth]{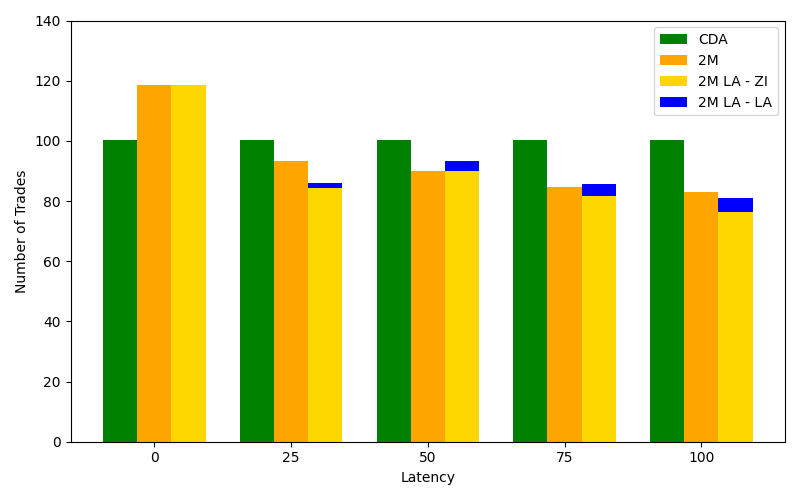}
    \caption{Env. 3 -- \textit{BestGuess}}
    \label{fig:env3_trans}
\end{subfigure}%
\hspace{1em}
\begin{subfigure}[t]{0.3\textwidth}
    \centering
    \includegraphics[width=1\textwidth]{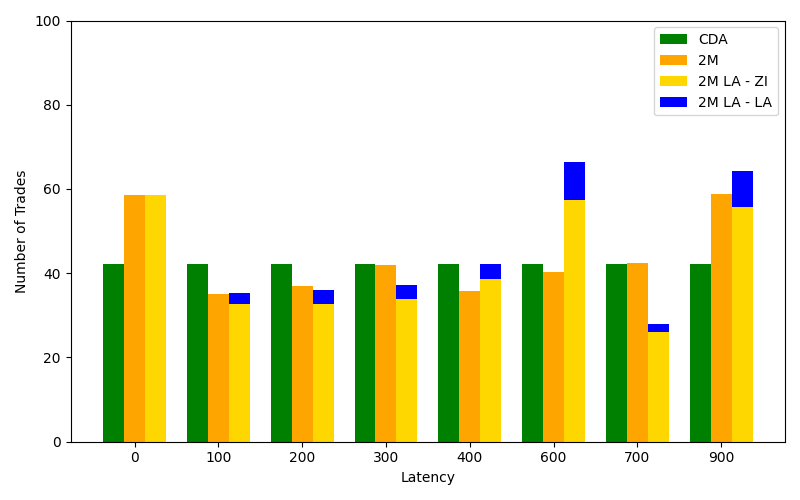}
    \caption{Env. 1 -- \textit{MarketSim}}
    \label{fig:env1_trans_ms}
\end{subfigure}%
\begin{subfigure}[t]{.3\textwidth}
    \centering
    \includegraphics[width=1\textwidth]{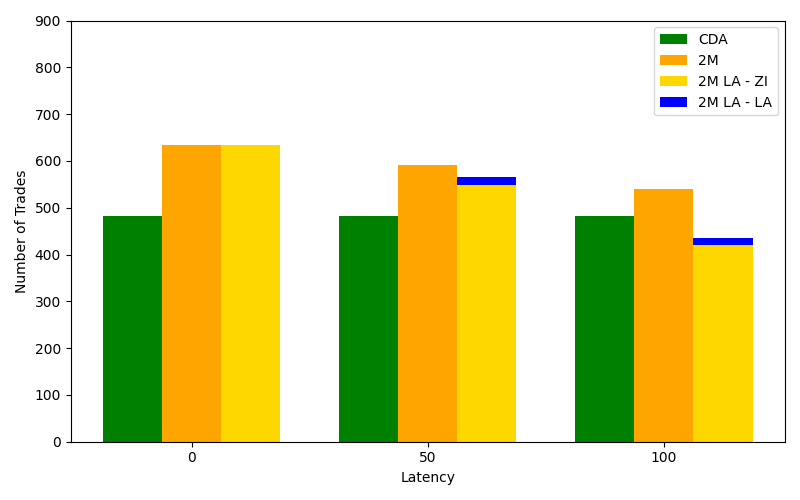}
    \caption{Env. 2 -- \textit{MarketSim}}
    \label{fig:env2_trans_ms}
\end{subfigure}%
\begin{subfigure}[t]{0.3\textwidth}
    \centering
    \includegraphics[width=1\textwidth]{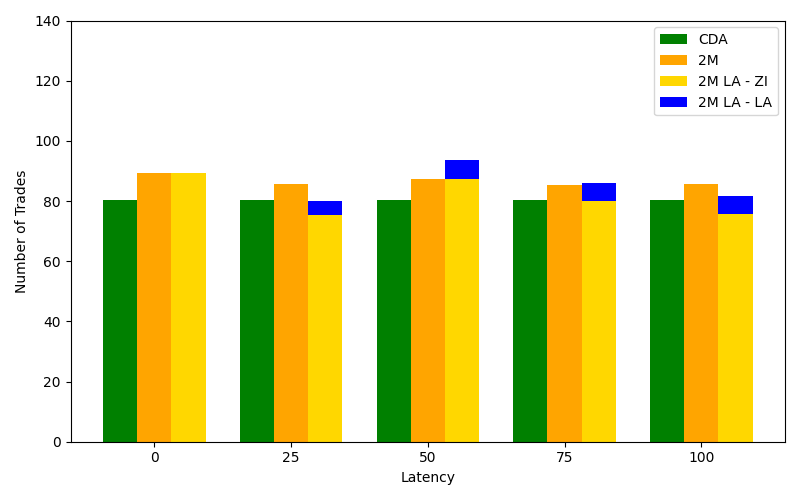}
    \caption{Env. 3 -- \textit{MarketSim}}
    \label{fig:env3_trans_ms}
\end{subfigure}%
\hspace{1em}
\begin{subfigure}[t]{0.3\textwidth}
    \centering
    \includegraphics[width=1\textwidth]{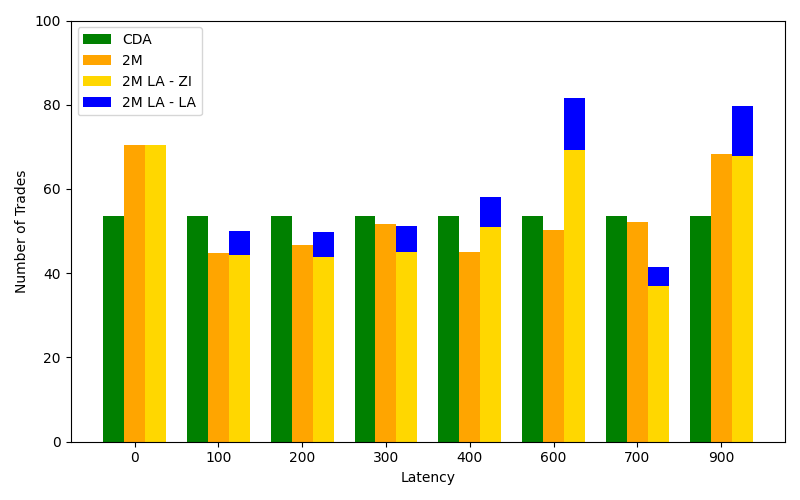}
    \caption{Env. 1 -- \textit{BestGuess+MS}}
    \label{fig:env1_trans_alt}
\end{subfigure}%
\begin{subfigure}[t]{.3\textwidth}
    \centering
    \includegraphics[width=1\textwidth]{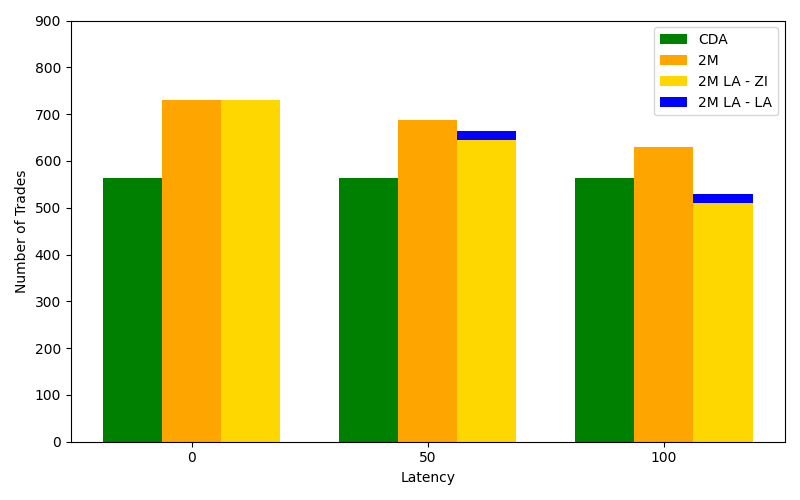}
    \caption{Env. 2 -- \textit{BestGuess+MS}}
    \label{fig:env2_trans_alt}
\end{subfigure}%
\begin{subfigure}[t]{0.3\textwidth}
    \centering
    \includegraphics[width=1\textwidth]{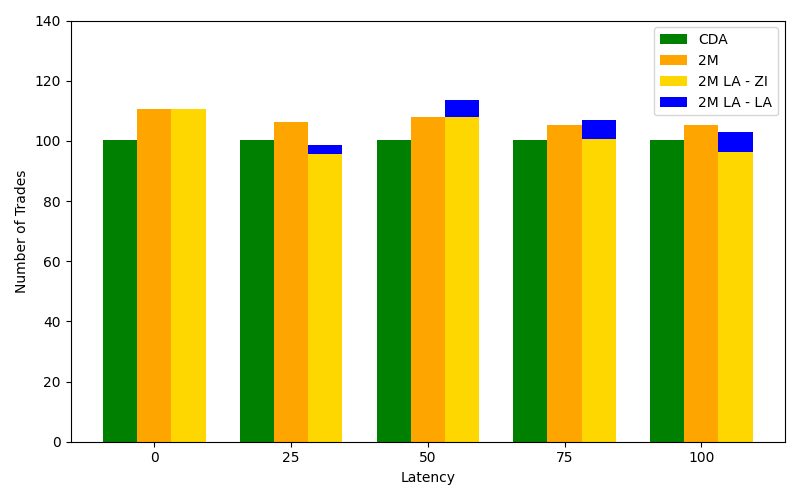}
    \caption{Env. 3 -- \textit{BestGuess+MS}}
    \label{fig:env3_trans_alt}
\end{subfigure}%
\caption{Mean number of transactions from our implementations of the model and \textit{MarketSim}.
These results should be compared to the corresponding original figures reported by WW \cite{wah_latency_2016} (Fig. 11, p.90). 
Please note, a single trade results in two transactions for this measurement (one per trader involved in the trade). Results from our simulations are averaged over 5,000 mixtures, 100 runs per mixture, for each experiment.}
\label{fig:ww_trans}
\end{figure}

\paragraph{Transactions}
Last, we look at the number of transactions generated by our simulation versus WW.
As highlighted earlier, it is important to note WW appear to report each trade as two transactions, one for each trader involved in the trade. Thus, a trade between two ZI agents is represented in Fig.~\ref{fig:ww_trans} as two `ZI only' transactions, while a trade between a ZI agent and the LA agent is shown as one `ZI only' transaction and one `LA only' transaction.

Visual appraisal of the relative changes in transaction numbers across experiments and between sets of results is more difficult in the bar chart format, but some differences are apparent between the \textit{BestGuess} results and WW.
The zero-latency results appear similar, but, as latency increases, the number of transactions in the \textit{BestGuess} results are lower than those reported by WW.
Overall, the fragmented markets without latency have the highest overall levels of trading in our results, while non-zero latency leads to fewer trades in most experiments. The effect of latency arbitrage is mixed, similar to WW's result.
The \textit{MarketSim} and \textit{BestGuess+MS} transaction results track much more closely with the WW results (and with each other).
There also appear to be fewer LA transactions in the \textit{BestGuess} results relative to the WW results, fitting with the lower LA surplus values we saw in our quantitative alignment exercise (Section~\ref{quantitative_alignment}). 

\subsection{Summary of results}

\begin{table}[]
    \centering
    \begin{tabular}{p{3.5cm}|p{5cm}|p{5cm}}
    \hline
        Feature & Effect reported by WW & \textit{BestGuess} replication \\
    \hline
        Latency arbitrage & \say{Generally degrades efficiency, but increased bid shading in low mean reversion environments alleviates inefficiencies caused by LA.} & LA generally degrades efficiency relative to fragmented markets without LA. \\
        Latency arbitrage & \say{LA exacerbates spreads, and execution times vary by environment based on bid shading in equilibrium} & LA exacerbates spreads but decreases execution times. \\
        Market fragmentation & \say{Can benefit continuous markets by admitting fewer inefficient trades (due to vagaries of the arrival sequence of orders), but LA defeats this benefit.} & Fragmentation generally improves surplus, with or without LA. \\
        Market fragmentation & \say{Execution times vary by environment, with consolidation improving liquidity in the fragmented model without LA when traders do not shade more.} & Fragmentation reduces execution times and NBBO spreads, but it increases the average BBO spread in most settings. \\
    \hline
    \end{tabular}
    \caption{Qualitative summary of model behavior from WW versus our \textit{BestGuess} results. WW summary text is reproduced from Table 5 in \cite{wah_latency_2016} (p.81).}
    \label{tab:summary_of_effects}
\end{table}

In the original paper, WW provide a summary of qualitative takeaways from their experimental results. We reproduce their takeaways in Table~\ref{tab:summary_of_effects} and compare against our \textit{BestGuess} replication.
WW report that latency arbitrage reduces total surplus and has a mixed effect on liquidity, while eliminating fragmentation can improve surplus and liquidity metrics. The conditional phrasing for these latter effects is important, as WW overall find mixed results on these effects depending on setting. They attribute this at least in part to changes in bid shading strategy in equilibrium by the ZI traders. 

Our \textit{BestGuess} implementation of the model meanwhile shows clear benefits of fragmentation in the experimental scenarios. 
Fragmentation in this implementation generally increases surplus, decreases execution times, and decreases NBBO spreads. The average exchange-level BBO spread is increased, however.
LA reduces surplus and exacerbates spreads relative to the fragmented market without LA, but LA also further reduces execution times.
The fragmented market with LA outperforms the single-CDA setting for most metrics in most settings.
These effects are also more consistent than those reported by WW, with less variation for specific experimental settings.

Comparing the results from our \textit{BestGuess} and \textit{BestGuess+MS} implementations of the model provide a clearer contrast, as we know the only difference between these two models is the implementation of the ZI traders' greedy strategy.
The greedy strategy taken from \textit{MarketSim} has the ZI traders base their decisions only on their primary exchange BBO and submit overly aggressive order prices to try to hit the observed price.
This results in reduced surplus and increased execution times relative to our \textit{BestGuess} implementation, where traders also utilize the NBBO and submit their orders at the observed market price.
The less strategic ZI actions taken in the \textit{BestGuess+MS} implementation also results in greater arbitrage opportunities, as evidenced by higher LA surplus values and more LA transactions.

Overall, the implementation of the greedy strategy for the ZI agents has a marked impact on a number of these takeaways, particularly for the effects of fragmentation relative to a consolidated market.
When the ZI agents submit less aggressive greedy orders and factor the NBBO into their strategy, the presence of fragmentation leads to reduced execution times in each experimental setting and generally increases trader surplus as well.
These results suggest that traders acting more strategically helps account for information uncertainty and that market fragmentation has clearer benefits when this is the case.
This has parallels with the findings from Duffin and Cartlidge's replication \cite{duffin_agent-based_2018} of Wah and Wellman's 2013 model \cite{wah_latency_2013}. They found that the background traders using the `minimally intelligent' ZIP trading algorithm led to results showing fragmented markets benefit from latency arbitrage.
The sensitivity of the qualitative takeaways to the specifics of the greedy ZI strategy in the 2016 model \cite{wah_latency_2016} at the very least highlights that this strategy should have been given more than a two-sentence description in the original paper.
More broadly, these results show that further exploration by the modeling community is needed to separate the effects of fragmentation and latency arbitrage from the specifics of trader agent implementation.

\section{Conclusion}
\label{conclusion}

Wah and Wellman \cite{wah_latency_2016} made a valuable contribution to agent-based market modeling by incrementally introducing elements of realistic market structure in order to assess their effects.
They note,
\begin{displayquote}
    As with any simulation model, our results are valid only to the extent our assumptions capture the essence of real-world markets \cite{wah_latency_2016} (p.91).
\end{displayquote}
Replication tests the extent the documented assumptions and results capture the essence of the \textit{model}.

We find that the specifics of the background trader greedy strategy proves especially important for the overall behavior of the model.
We cannot claim that others would necessarily have interpreted the details given by Wah and Wellman \cite{wah_latency_2016} the same as we did in our \textit{BestGuess} implementation. We argue that our interpretation better fits the spirit of their description. Specifically, as opposed to the \textit{MarketSim} implementation, the following properties hold in our \textit{BestGuess} implementation of the strategy:
\begin{enumerate}
    \item The ZI agents use all information to which they have access.
    \item The strategy with $\eta=1$ is equivalent to not using the strategy.
\end{enumerate}
Regardless, the sensitivity of the original study's qualitative takeaways to this implementation detail limits the range of applicability for those takeaways to real-world markets.
A more comprehensive exploration of the parameter space with our implementation could yield further insights into the range of behavior resulting from our assumptions.
Further work should also be done to validate the behavior of the model against stylized facts of empirical market behavior \cite{cont_empirical_2001,ratliff-crain_revisiting_2025}, test the framework with different ecologies of traders, and extend the framework to incorporate yet more realistic elements of fragmentation and modern market structure.

Our experience from this replication effort has takeaways for agent-based modeling and replication more broadly.
The first is that the replicator should test the reliability of their replication criteria for the model in question.
Some studies have checked this by testing the result distribution for normality in order to assess the appropriateness of a given statistical estimator \cite{radax_prospects_2010,seagren_replication_2015}.
An alternative approach, taken here, is to test the replication model against itself by taking many independent samples of the specified size and checking alignment between samples. This comes with computational tradeoffs, requiring many more runs than the original study.
The bootstrap confidence interval methodology we use here to test quantitative alignment proved a useful replication guide.
Our implementation of the model with the best quantitative alignment to WW's ZI and LA surplus metrics under this methodology also had the best qualitative alignment to the original model across the various metrics.

Next, increasing the complexity of the model corresponded with increased difficulty replicating the results. 
Even in our best replication implementation, we reject quantitative alignment for all model settings with non-zero latency, and we achieve relationally equivalent results for most but not all metrics and settings.
The existence of a public codebase from the original study's authors helped uncover important implementation details of the model, but the codebase failed to replicate the original model.
The complexity of the model and codebase made it challenging to determine what is driving the divergence.
As the model becomes more complex, the number of design choices increases and the sensitivity of the model to certain choices may also increase.
More complex models thereby need even more thorough documentation (and more acutely need to be independently replicated).

Edmonds and Hales provocatively said, \say{An unreplicated simulation is an untrustworthy simulation} \cite{edmonds_replication_2003}.
This would render most agent-based market models untrustworthy, in line with the SEC's expressed sentiment \cite{us_securities_and_exchange_commission_regulation_2024}.
By encouraging a culture of replication \cite{thiele_replicating_2015}, market modelers can build more trustworthy models that can then be extended to better capture the complexity of the financial system.

\section*{Acknowledgment}

The authors gratefully acknowledge helpful discussions with Anshul Anand, James Angel, Chris Bassler, Lashon Booker, Jean-Philippe Bouchaud, Robert Brooks, Eric Budish, Richard Byrne, Peter Carrigan, Christopher Danforth, Matthew Dinger, Peter Sheridan Dodds, Andre Frank, Bill Gibson, Frank Hatheway, Emily Hiner, Chuck Howell, Eric Hunsader, Robert Jackson, Neil Johnson, Will Kirkman, Michael Kometer, Blake LeBaron, Phil Mackintosh,
Chris Mascioli,
Matthew McMahon, Beth Meinert, Matthew Mihalcin, Rishi Narang, Mark Phillips, Hester Peirce, Joseph Saluzzi, Brendan Tivnan, Kevin Toner, Jason Veneman, Michael Wellman, and Thomas Wilk. The authors' affiliation with The MITRE Corporation is provided for identification purposes only, and is not intended to convey or imply MITRE's concurrence with, or support for, the positions, opinions, or viewpoints expressed by the authors. The authors declare no conflicts of interest.

\bibliographystyle{plain}
\bibliography{refs}

\appendix

\section{Using \textit{MarketSim} to replicate WWW}
\label{WWW}

In order to validate our understanding and implementation of the \textit{MarketSim} codebase, we attempt to use it to replicate the results from Wah, Wright, and Wellman (WWW) \cite{wah_welfare_2017}. 
WWW cite the \textit{MarketSim} GitHub repo\footnote{The original repo cited by WWW is here: \url{https://github.com/egtaonline/market-sim/tree/marketsim1}. We fork this repo to configure the replication experiments for WW and WWW here: \url{https://github.com/eratlif1/market-sim-WW-replication/tree/marketsim1-upstream}.} as the source code for the simulator used in their study.
Our expectation is therefore that \textit{MarketSim}, as published, should be able to replicate the WWW results. Success in doing so would help establish confidence in our implementation of the codebase for the purpose of replicating the earlier study by Wah and Wellman (WW) \cite{wah_latency_2016}.

As noted in Section~\ref{marketsim}, WWW is detailed with similar assumptions and logic as the \textit{CDA} configurations of the WW model. In the \textit{CDA} configuration, the main differences between WW and WWW are the environment parameters used in their respective experiments and the introduction of market maker agents in WWW.
WWW features experiments with and without market maker agents in five different configuration environments.
We limit our replication exercise to the environments \textbf{A} and \textbf{B} with 66 background traders (ZI) and without a market maker:
$$\textbf{A}\hspace{1em} \lambda=0.0005, \sigma_s^2=1\times 10^6, \sigma_{PV}^2=5\times 10^6$$
$$\textbf{B}\hspace{1em} \lambda=0.005, \sigma_s^2=1\times 10^6, \sigma_{PV}^2=5\times 10^6$$
These settings are respectively used for each $\textbf{A}\{T\}$ and $\textbf{B}\{T\}$, where $T\in\{1,4,12,24\}$ and the simulation length is $T\times1000$.

WWW uses a similar experimental design to WW. They use the `empirical game-theoretic analysis' (EGTA) approach to determine the surplus-maximizing equilibria ZI strategy mixtures. They then sample $M$ profiles from the equilibrium mixture, running $N$ simulations per mixture. In WWW, they specify the following:
\begin{displayquote}
    For each equilibrium, we estimated background-trader surplus by sampling 2,500 profiles according to the equilibrium mixture, running 25--100 simulations per sampled profile (at least 62,500 simulations in total) and then recording the aggregate surplus \cite{wah_welfare_2017} (p.625).
\end{displayquote}
They do not specify the specific number of runs used for each experiment, complicating the replication process.
The mixtures of background trader profiles for each $\textbf{A}\{T\}$ and $\textbf{B}\{T\}$ are based on Tables 1, 2, and 9 in \cite{wah_welfare_2017}, which, respectively, give the specific ZI strategies, the surplus-maximizing equilibria, and profile mixtures for the equilibria.

We sample 2,500 profiles and run 25 simulations for each experiment.
We then use the quantitative alignment methodology introduced in Section~\ref{quantitative_alignment_methodology} to test our results for quantitative alignment against the mean surplus values reported by WWW.
The WWW means fall within the 95\% confidence intervals (CIs) of our bootstrap sample means for six out of the eight environments tested.
The reported mean for Environment \textbf{B}24 fell within the 99\% CI but not the 95\% CI, and the reported mean for Environment \textbf{A}24 falls outside the range of our bootstrap samples.
By our replication criteria, we do not reject quantitative alignment between our \textit{MarketSim} results and WWW for six out of eight environments at the 95\% confidence level and seven out of eight environments at the 99\% confidence level.

It is not clear what is driving the difference observed in Environment \textbf{A}24.
As noted above, WWW state that each experiment was run for 25--100 simulations per sampled profile mixture, but they do not specify which experiment was run for what number of simulations per mixture.
It is possible WWW ran \textbf{A}24 for more than 25 simulations, that the original result is from the tail of possible sample means that can be produced by \textit{MarketSim} for the \textbf{A}24 experiment, or that there is some other small difference between how we are running the experiment and what was done in the original study.

\begin{table}[H]
    \centering
    \begin{tabular}{l|r|rrrr}
    \hline
    env & WWW & \multicolumn{4}{c}{Our Bootstrap ZI Surplus} \\
     & mean & mean & SE & 95\% CI diff. & 99\% CI diff. \\
    \hline
 A1 &               4439 &         4444.38 &           8.05 &            \textbf{(-10.97, 20.87)} &              \textbf{(-14.45, 25.33)} \\
 A4 &              16578 &        16572.59 &          15.61 &            \textbf{(-34.96, 26.31)} &              (\textbf{-45.73, 35.12)} \\
A12 &              33741 &        33699.51 &          23.06 &            \textbf{(-88.31, 0.22)} &             \textbf{(-103.14, 15.86)} \\
A24 &              42413 &        42516.26 &          28.99 &             (45.27, 160.30) &               (30.83, 173.83) \\
 B1 &              29150 &        29132.54 &          22.81 &            \textbf{(-61.62, 25.37)} &              \textbf{(-73.20, 40.95)} \\
 B4 &              40392 &        40352.06 &          30.62 &           \textbf{(-102.48, 20.33)} &             \textbf{(-120.91, 41.94)} \\
B12 &              40102 &        40075.14 &          30.65 &            \textbf{(-87.57, 29.82)} &             \textbf{(-103.35, 52.00)} \\
B24 &              40170 &        40234.19 &          29.84 &              (4.18, 124.34) &               \textbf{(-8.61, 139.32)} \\
\hline
\end{tabular}
    \caption{Test of quantitative alignment of \textit{MarketSim} against the results reported by Wah, Wright, and Wellman (WWW) \cite{wah_welfare_2017}. 
    Environment parameters are as specified by \cite{wah_welfare_2017}, summarized in Appendix~\ref{WWW}.
    Each `WWW' value is the mean ZI surplus value reported by Wah, Wright, and Wellman from 2,500 mixtures and 25--100 runs per mixture, for the given experimental settings.
    We ran \textit{MarketSim} for 25,000 mixtures, 25 runs per mixture. From these results, we calculate 1,000 bootstrap samples of 2,500 mixtures, 25 runs per mixture, for each experiment. We report the mean and standard error of the bootstrap sample means, along with the 95\% confidence intervals of the difference between the bootstrap sample means and the WWW mean. Bootstrap confidence intervals that contain the WWW mean are highlighted in \textbf{bold}. }
    \label{tab:www_quant_equivalence_ms}
\end{table}

\begin{figure}[H]
\centering
\begin{subfigure}[t]{0.4\textwidth}
    \centering
    \includegraphics[width=1\textwidth]{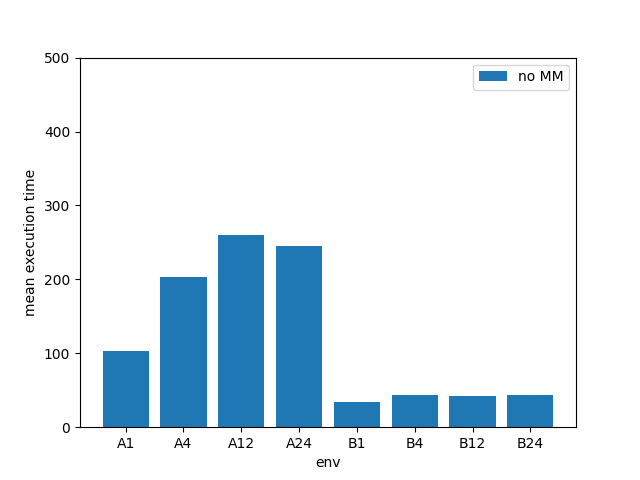}
    \caption{\textit{MarketSim} execution time results.}
    \label{fig:www_extime_ours}
\end{subfigure}%
\hspace{1em}
\begin{subfigure}[t]{0.4\textwidth}
    \centering
    \includegraphics[width=1\textwidth]{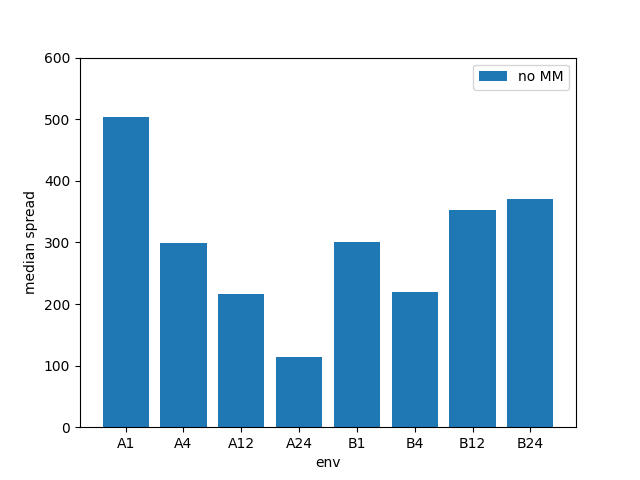}
    \caption{\textit{MarketSim} spread results}
    \label{fig:www_spread_ours}
\end{subfigure}%
\caption{Execution time and spread results from our \textit{MarketSim} runs of WWW experiments for Environments \textbf{A} and \textbf{B}, $N=66$, and no market maker. These results should be compared to the corresponding original figures from WWW \cite{wah_welfare_2017} (Fig.5, p.633).
Our results show the mean value produced by \textit{MarketSim} for 25,000 mixtures, 25 runs per mixture.}
\label{fig:www_results}
\end{figure}

We test relational equivalence by plotting the execution time and median spread metrics in Fig.~\ref{fig:www_results} for comparison against the corresponding figures in WWW (see Fig. 5 in \cite{wah_welfare_2017}, p.633).
Note that we did not run the experiments with a market maker from WWW, so our plots only correspond to the `no MM' results (shown in blue in each plot).
We see close alignment of the results, with similar quantitative values and relative differences between environments. 
The most notable difference is that $\textbf{A}12$ has a slightly higher execution time than $\textbf{A}24$ in our run of \textit{MarketSim}, while the opposite is true in WWW.
The mean execution times are similar between these two environments in both sets of results.
We are not able to otherwise visually distinguish the two sets of results.
For now, we take the similarity of results (quantitative alignment and relational equivalence in most settings) as positive sign we have implemented and are running \textit{MarketSim} correctly.

The fact \textit{MarketSim} can produce results that closely align to the \textit{WWW} results but not the WW results (see Sections \ref{marketsim} and \ref{qualitative}) raises questions about the logic used in the published results.
Wah's dissertation \cite{wah_computational_2016} describes common modeling assumptions for WW and WWW and does not appear to differentiate these models aside from their parameters and market configurations. 
The CDA market mechanism and ZI trader strategy in particular are given common descriptions for the two models.
We would therefore expect \textit{MarketSim} as published for the WWW study to reproduce the results from the \textit{CDA} configurations of WW.
As we are not able to reproduce the WW results using \textit{MarketSim} here, this suggests logic differences beyond what was specified in the original papers, perhaps in the ZI logic or in the logic used to calculate the output metrics at the end of the simulation.

\section{\textit{BestGuess+MS} vs \textit{BestGuess+MS+bug} model implementations}
\label{alt_vs_buggy}

\begin{figure}[H]
\centering
\begin{subfigure}[t]{0.3\textwidth}
    \centering
    \includegraphics[width=1\textwidth]{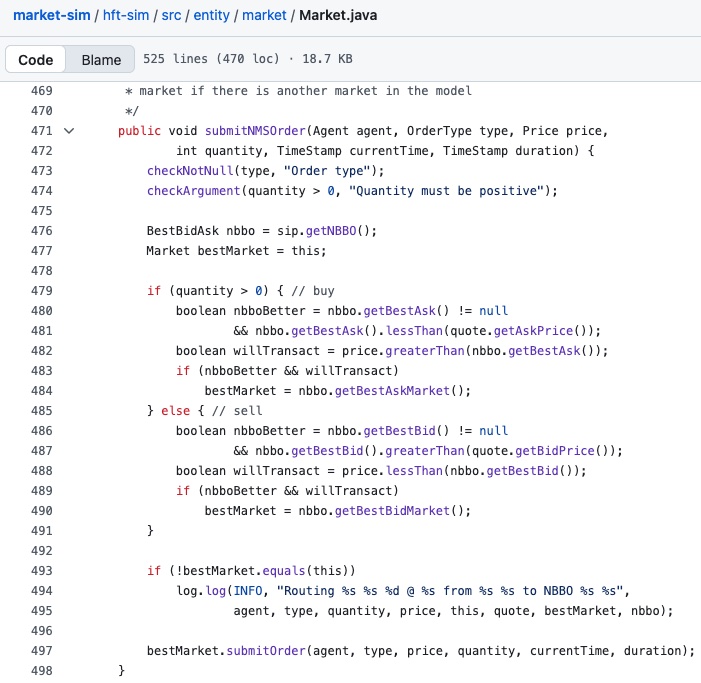}
\end{subfigure}%
\caption{ZI order routing logic from \textit{MarketSim} (\url{https://github.com/egtaonline/market-sim/blob/marketsim1/hft-sim/src/entity/market/Market.java\#L471}). Note that the code first checks that quantity is greater than zero, then uses whether quantity is greater than zero to determine whether the order is to buy or sell the asset. Thus, all orders will enter the `buy' block, thus being routed based on the best ask price in the market even if the incoming order is a sell order.}
\label{fig:marketsim_bug}
\end{figure}

In this section, we examine the results from including the order routing bug in addition to the \textit{MarketSim} greedy strategy logic.
We will refer to this implementation as \textit{BestGuess+MS+bug}, whereas our implementation that includes only the greedy strategy logic from \textit{MarketSim} is referred to as \textit{BestGuess+MS}.
First, the quantitative alignment results for the ZI surplus from \textit{BestGuess+MS+bug} versus the canonical results are shown in Table~\ref{tab:ww_quant_equivalence_bestguess_ms_bug}.
The quantitative alignment results for the LA surplus values are shown in Table~\ref{tab:ww_la_quant_equivalence_bestguess_ms_bug}.
If we compare these to the results from the main paper, we see that this model achieves closer quantitative alignment than \textit{BestGuess} and \textit{MarketSim}. The \textit{BestGuess+MS+bug} results are similar to \textit{BestGuess+MS}, but the \textit{BestGuess+MS} implementation aligns more closely for most experiments.
The exception is for \textit{2M LA} with $\delta=75$ in Environment 3, where we do not reject quantitative alignment for the \textit{BestGuess+MS+bug} LA surplus results. However, we view this alignment as potentially spurious given the \textit{2M LA} results are too high for $\delta=50$ and too low for $\delta=100$.
These results overall motivate our choice to focus on the \textit{BestGuess+MS} implementation in the main paper, under the belief that the order routing bug should not be included in the model unless it can help explain the original results in a way other interpretations can not.

\begin{table}[H]
    \centering
    \begin{tabular}{rlr|r|rrrr}
    \hline
    env & model & latency & WW & \multicolumn{4}{c}{Our Bootstrap ZI Surplus} \\
    &  &  & mean & mean & SE & 95\% CI diff. & 99\% CI diff. \\
    \hline
  1 &      CDA &        0 &              10383 &        10426.27 &          25.31 &       \textbf{(-5.53, 93.25)} &     \textbf{(-16.73, 109.80)} \\
  1 & 2M no LA &        0 &              11807 &        11932.08 &          19.51 &      (85.79, 163.41) &      (75.84, 171.66) \\
  1 & 2M no LA &      100 &              10373 &         9730.27 &          29.79 &   (-703.33, -587.89) &   (-720.61, -574.97) \\
  1 & 2M no LA &      200 &              10621 &        10012.32 &          29.14 &   (-665.62, -553.62) &   (-684.49, -538.06) \\
  1 & 2M no LA &      300 &              11244 &        10608.77 &          24.92 &   (-687.00, -588.50) &   (-709.19, -572.06) \\
  1 & 2M no LA &      400 &              10438 &         9970.25 &          32.10 &   (-534.39, -409.98) &   (-549.40, -389.80) \\
  1 & 2M no LA &      600 &              11128 &        10387.98 &          24.86 &   (-786.02, -689.18) &   (-801.83, -673.52) \\
  1 & 2M no LA &      700 &              11302 &        10637.86 &          24.04 &   (-709.43, -616.75) &   (-721.01, -596.78) \\
  1 & 2M no LA &      900 &              12358 &        11938.44 &          19.81 &   (-457.74, -380.17) &   (-466.94, -369.29) \\
  1 &    2M LA &      100 &               5919 &         7204.15 &          34.56 &   (1220.68, 1355.07) &   (1201.28, 1372.52) \\
  1 &    2M LA &      200 &               6358 &         7296.69 &          33.71 &    (871.47, 1002.68) &    (853.95, 1030.35) \\
  1 &    2M LA &      300 &               6398 &         7395.82 &          32.47 &    (935.52, 1063.26) &    (915.52, 1075.02) \\
  1 &    2M LA &      400 &               6130 &         7698.96 &          29.11 &   (1515.53, 1629.38) &   (1501.61, 1644.35) \\
  1 &    2M LA &      600 &               7459 &         8639.58 &          26.16 &   (1129.21, 1231.83) &   (1111.91, 1245.24) \\
  1 &    2M LA &      700 &               5256 &         6278.89 &          44.05 &    (933.03, 1104.25) &    (903.35, 1127.32) \\
  1 &    2M LA &      900 &               6819 &         8631.55 &          27.30 &   (1760.36, 1863.58) &   (1740.56, 1878.50) \\
  2 &      CDA &        0 &             136140 &       136131.67 &          63.79 &    \textbf{(-135.82, 114.54)} &    \textbf{(-181.92, 156.36)} \\
  2 & 2M no LA &        0 &             134339 &       134381.44 &          65.82 &     \textbf{(-90.28, 168.64)} &    \textbf{(-136.82, 207.63)} \\
  2 & 2M no LA &       50 &             135789 &       134638.56 &          65.76 & (-1274.73, -1021.46) &  (-1324.56, -979.64) \\
  2 & 2M no LA &      100 &             136542 &       135172.16 &          68.72 & (-1498.56, -1239.63) & (-1534.73, -1188.09) \\
  2 &    2M LA &       50 &             133177 &       130579.29 &          82.55 & (-2748.18, -2442.16) & (-2815.83, -2379.43) \\
  2 &    2M LA &      100 &             124012 &       122376.68 &          92.74 & (-1820.76, -1455.35) & (-1863.14, -1410.87) \\
  3 &      CDA &        0 &              27482 &        27471.48 &          39.97 &      \textbf{(-92.47, 64.16)} &     \textbf{(-119.32, 91.08)} \\
  3 & 2M no LA &        0 &              29424 &        29250.72 &          30.49 &   (-233.90, -117.28) &   (-250.45, -101.19) \\
  3 & 2M no LA &       25 &              29347 &        29021.88 &          30.98 &   (-386.76, -261.71) &   (-406.60, -243.15) \\
  3 & 2M no LA &       50 &              29479 &        29148.87 &          28.31 &   (-386.89, -277.74) &   (-400.12, -259.08) \\
  3 & 2M no LA &       75 &              29271 &        28944.63 &          29.64 &   (-382.92, -268.07) &   (-402.71, -252.81) \\
  3 & 2M no LA &      100 &              29277 &        28932.74 &          31.96 &   (-407.85, -284.66) &   (-426.89, -262.98) \\
  3 &    2M LA &       25 &              26612 &        25752.47 &          40.81 &   (-938.51, -784.25) &   (-963.98, -766.58) \\
  3 &    2M LA &       50 &              27953 &        27321.18 &          36.07 &   (-700.07, -562.18) &   (-719.21, -540.43) \\
  3 &    2M LA &       75 &              26388 &        26180.24 &          40.57 &   (-287.09, -132.65) &   (-317.42, -110.81) \\
  3 &    2M LA &      100 &              25070 &        25311.84 &          43.62 &     (155.05, 330.57) &     (134.14, 351.05) \\
    \hline
\end{tabular}
    \caption{Test of quantitative alignment for our \textit{BestGuess+MS+bug} implementation results versus WW. 
    Each WW ZI value is the mean ZI surplus reported by WW \cite{wah_latency_2016} (pp.~83,~85) for 500 mixtures (100 simulation runs each) sampled according to the welfare-maximizing strategy profile.
    Our results are calculated from 5,000 mixtures (100 runs each) based on the same strategy profile. From these, we construct 1,000 bootstrap samples of 500 mixtures each to produce confidence intervals for the mean difference. 
    We report the mean and standard error of the 1,000 bootstrap sample means, along with the 95\% and 99\% confidence intervals for the difference between our bootstrap sample means and the reported WW mean.
    Confidence intervals that contain zero are shown in \textbf{bold}, indicating where quantitative alignment is not rejected.}
    \label{tab:ww_quant_equivalence_bestguess_ms_bug}
\end{table}

\begin{table}[H]
    \centering
    \begin{tabular}{rlr|r|rrrr}
    \hline
    env & model & latency & WW LA & \multicolumn{4}{c}{Our Bootstrap LA Surplus} \\
    &  &  & mean & mean & SE & 95\% CI diff. & 99\% CI diff. \\
    \hline
  1 & 2M LA &      100 &               3487 &         1950.28 &          17.75 & (-1571.17, -1501.88) & (-1581.64, -1492.61) \\
  1 & 2M LA &      200 &               3164 &         1997.30 &          18.13 & (-1202.55, -1131.24) & (-1210.30, -1118.13) \\
  1 & 2M LA &      300 &               3224 &         2043.32 &          16.93 & (-1213.13, -1148.59) & (-1221.61, -1141.54) \\
  1 & 2M LA &      400 &               4018 &         2250.88 &          18.27 & (-1803.91, -1731.71) & (-1813.73, -1721.15) \\
  1 & 2M LA &      600 &               4349 &         3050.70 &          16.24 & (-1330.56, -1266.93) & (-1339.43, -1260.81) \\
  1 & 2M LA &      700 &               2958 &         1737.36 &          16.56 & (-1252.87, -1189.47) & (-1265.82, -1179.00) \\
  1 & 2M LA &      900 &               4825 &         2975.98 &          16.99 & (-1881.00, -1814.80) & (-1893.96, -1805.96) \\
  2 & 2M LA &       50 &               2417 &         4049.84 &          43.97 &   (1550.15, 1716.65) &   (1521.51, 1750.46) \\
  2 & 2M LA &      100 &               2888 &         3121.23 &          37.28 &     (163.52, 312.72) &     (137.37, 330.55) \\
  3 & 2M LA &       25 &                538 &         1156.56 &          13.64 &     (592.06, 645.30) &     (585.33, 654.13) \\
  3 & 2M LA &       50 &               1154 &         1542.84 &          16.44 &     (356.84, 422.29) &     (347.76, 431.54) \\
  3 & 2M LA &       75 &               1470 &         1469.18 &          17.12 &      \textbf{(-32.46, 34.18)} &      \textbf{(-44.31, 43.86)} \\
  3 & 2M LA &      100 &               1763 &         1387.12 &          16.28 &   (-406.43, -343.77) &   (-417.74, -332.81) \\
    \hline
\end{tabular}
    \caption{Test of quantitative alignment of LA surplus results from our \textit{BestGuess+MS+bug} replication versus WW.
    Each WW LA value is the mean LA surplus reported by WW \cite{wah_latency_2016} (pp.~83,~85) for 500 mixtures (100 simulation runs each) sampled according to the welfare-maximizing strategy profile.
    Our results are calculated from 5,000 mixtures (100 runs each) based on the same strategy profile. From these, we construct 1,000 bootstrap samples of 500 mixtures each to produce confidence intervals for the mean difference. 
    We report the mean and standard error of the bootstrap sample means, along with the 95\% and 99\% confidence intervals for the difference between our bootstrap sample means and the reported WW mean. 
    Confidence intervals that contain zero are shown in \textbf{bold}, indicating where quantitative alignment is not rejected.
    }
    \label{tab:ww_la_quant_equivalence_bestguess_ms_bug}
\end{table}

\begin{figure}[H]
\centering
\begin{subfigure}[t]{0.32\textwidth}
    \centering
    \includegraphics[width=1\textwidth]{env1_mean_surplus_alt.png}
    \caption{Env. 1 -- \textit{BestGuess+MS}}
    \label{fig:env1_surplus_alt_v_buggy}
\end{subfigure}%
\begin{subfigure}[t]{.32\textwidth}
    \centering
    \includegraphics[width=1\textwidth]{env2_mean_surplus_alt.png}
    \caption{Env. 2 -- \textit{BestGuess+MS}}
    \label{fig:env2_surplus_alt_v_buggy}
\end{subfigure}%
\begin{subfigure}[t]{0.32\textwidth}
    \centering
    \includegraphics[width=1\textwidth]{env3_mean_surplus_alt.png}
    \caption{Env. 3 -- \textit{BestGuess+MS}}
    \label{fig:env3_surplus_alt_v_buggy}
\end{subfigure}%
\hspace{1em}
\begin{subfigure}[t]{0.32\textwidth}
    \centering
    \includegraphics[width=1\textwidth]{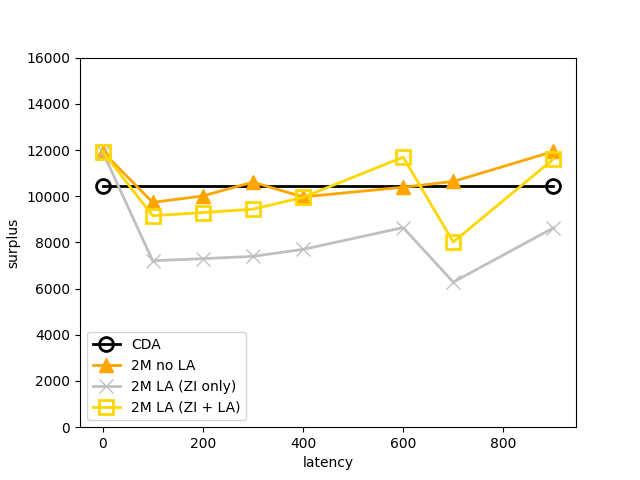}
    \caption{Env. 1 -- \textit{BestGuess+MS+bug}}
    \label{fig:env1_surplus_buggy}
\end{subfigure}%
\begin{subfigure}[t]{.32\textwidth}
    \centering
    \includegraphics[width=1\textwidth]{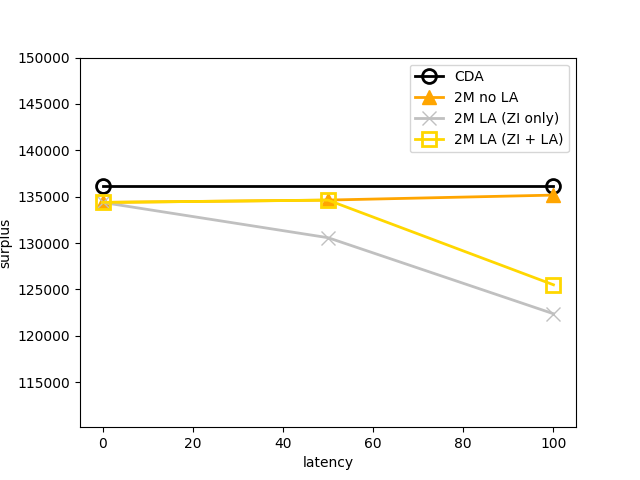}
    \caption{Env. 2 -- \textit{BestGuess+MS+bug}}
    \label{fig:env2_surplus_buggy}
\end{subfigure}%
\begin{subfigure}[t]{0.32\textwidth}
    \centering
    \includegraphics[width=1\textwidth]{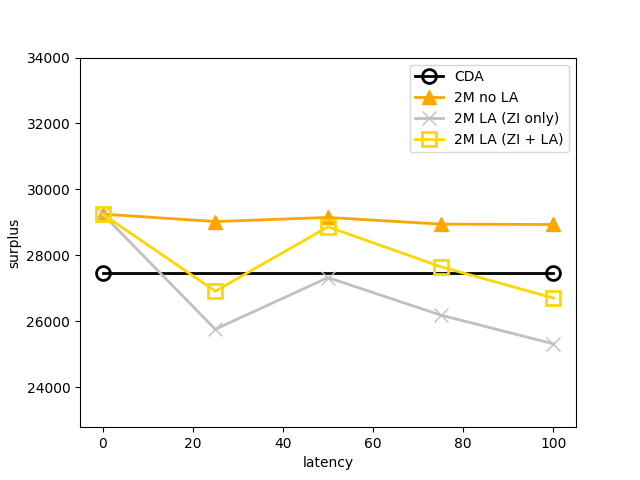}
    \caption{Env. 3 -- \textit{BestGuess+MS+bug}}
    \label{fig:env3_surplus_buggy}
\end{subfigure}%
\caption{Mean surplus results from our \textit{BestGuess+MS} and \textit{BestGuess+MS+bug} implementations of the model.
These results should be compared to the corresponding original figures from WW \cite{wah_latency_2016} (Fig. 4, p.82).
Our results show the mean surplus from 5,000 mixtures, 100 runs per mixture.}
\label{fig:ww_surplus_alt_v_buggy}
\end{figure}

\begin{figure}[H]
\centering
\begin{subfigure}[t]{0.32\textwidth}
    \centering
    \includegraphics[width=1\textwidth]{env1_mean_ex_time_alt.png}
    \caption{Env. 1 -- \textit{BestGuess+MS}}
    \label{fig:env1_ex_time_alt_v_buggy}
\end{subfigure}%
\begin{subfigure}[t]{.32\textwidth}
    \centering
    \includegraphics[width=1\textwidth]{env2_mean_ex_time_alt.png}
    \caption{Env. 2 -- \textit{BestGuess+MS}}
    \label{fig:env2_ex_time_alt_v_buggy}
\end{subfigure}%
\begin{subfigure}[t]{0.32\textwidth}
    \centering
    \includegraphics[width=1\textwidth]{env3_mean_ex_time_alt.png}
    \caption{Env. 3 -- \textit{BestGuess+MS}}
    \label{fig:env3_ex_time_alt_v_buggy}
\end{subfigure}%
\hspace{1em}
\begin{subfigure}[t]{0.32\textwidth}
    \centering
    \includegraphics[width=1\textwidth]{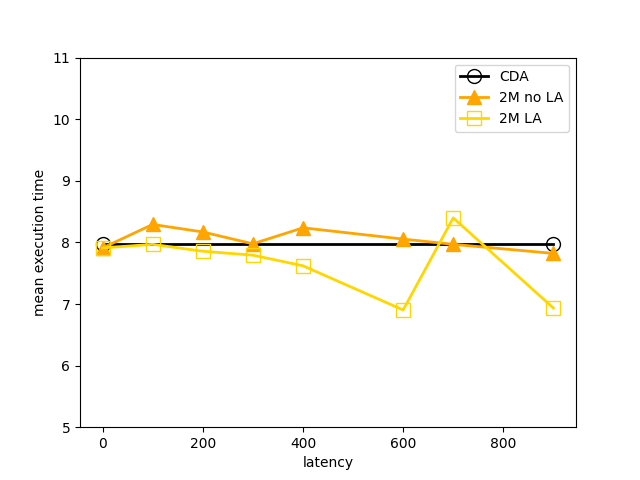}
    \caption{Env. 1 -- \textit{BestGuess+MS+bug}}
    \label{fig:env1_ex_time_buggy}
\end{subfigure}%
\begin{subfigure}[t]{.32\textwidth}
    \centering
    \includegraphics[width=1\textwidth]{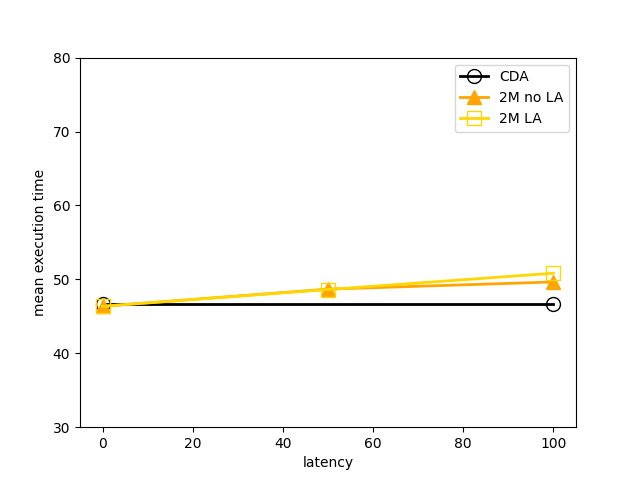}
    \caption{Env. 2 -- \textit{BestGuess+MS+bug}}
    \label{fig:env2_ex_time_buggy}
\end{subfigure}%
\begin{subfigure}[t]{0.32\textwidth}
    \centering
    \includegraphics[width=1\textwidth]{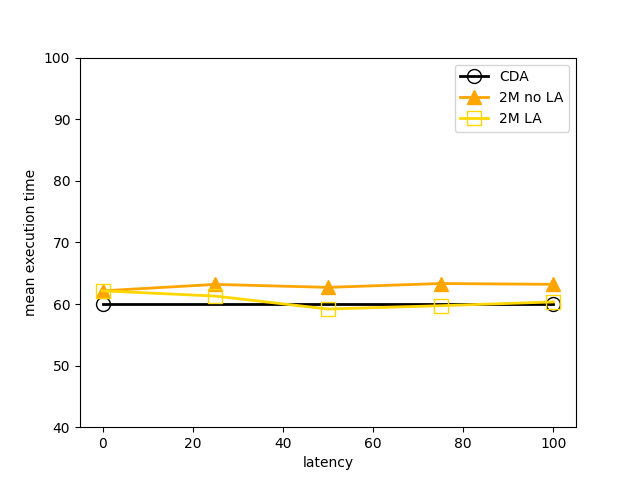}
    \caption{Env. 3 -- \textit{BestGuess+MS+bug}}
    \label{fig:env3_ex_time_buggy}
\end{subfigure}%
\caption{Mean execution time results from our \textit{BestGuess+MS} and \textit{BestGuess+MS+bug} implementations of the model.
These results should be compared to the corresponding original figures from WW \cite{wah_latency_2016} (Fig. 6, p.86). 
Our results show the mean execution time from 5,000 mixtures, 100 runs per mixture.}
\label{fig:ww_extime_alt_v_buggy}
\end{figure}

\begin{figure}[H]
\centering
\begin{subfigure}[t]{0.32\textwidth}
    \centering
    \includegraphics[width=1\textwidth]{env1_mean_median_bbo_spread_alt.png}
    \caption{Env. 1 -- \textit{BestGuess+MS}}
    \label{fig:env1_bbo_bg_ms_alt_alt_v_bug}
\end{subfigure}%
\begin{subfigure}[t]{.32\textwidth}
    \centering
    \includegraphics[width=1\textwidth]{env2_mean_median_bbo_spread_alt.png}
    \caption{Env. 2 -- \textit{BestGuess+MS}}
    \label{fig:env2_bbo_bg_ms_alt_alt_v_bug}
\end{subfigure}%
\begin{subfigure}[t]{.32\textwidth}
    \centering
    \includegraphics[width=1\textwidth]{env3_mean_median_bbo_spread_alt.png}
    \caption{Env. 3 -- \textit{BestGuess+MS}}
    \label{fig:env3_bbo_bg_ms_alt_alt_v_bug}
\end{subfigure}%
\hspace{1em}
\begin{subfigure}[t]{0.32\textwidth}
    \centering
    \includegraphics[width=1\textwidth]{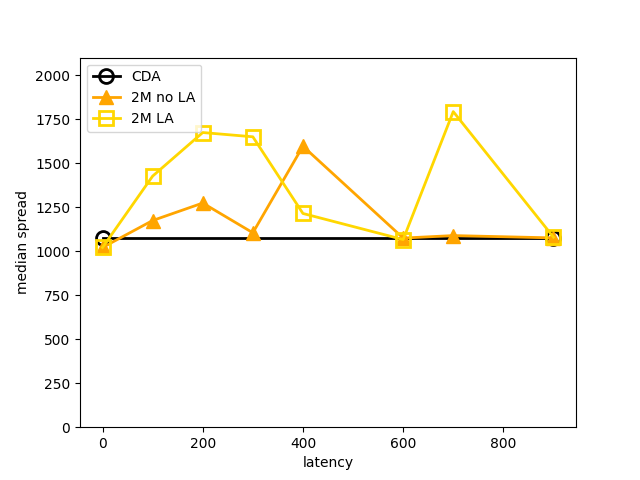}
    \caption{Env. 1 -- \textit{BestGuess+MS+bug}}
    \label{fig:env1_bbo_bg_ms_buggy_alt_v_bug}
\end{subfigure}%
\begin{subfigure}[t]{.32\textwidth}
    \centering
    \includegraphics[width=1\textwidth]{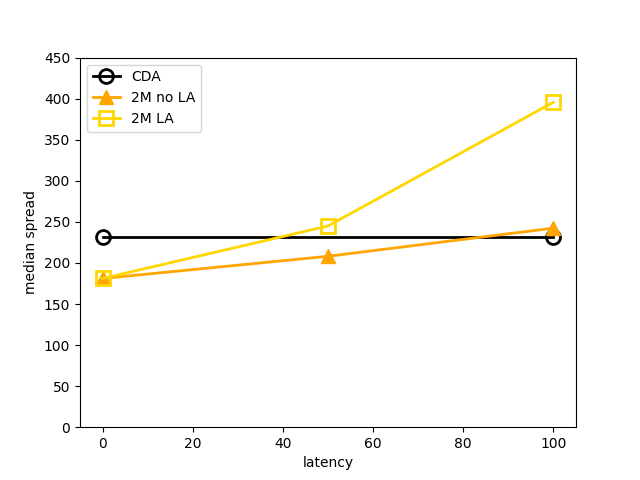}
    \caption{Env. 2 -- \textit{BestGuess+MS+bug}}
    \label{fig:env2_bbo_bg_ms_buggy_alt_v_bug}
\end{subfigure}%
\begin{subfigure}[t]{.32\textwidth}
    \centering
    \includegraphics[width=1\textwidth]{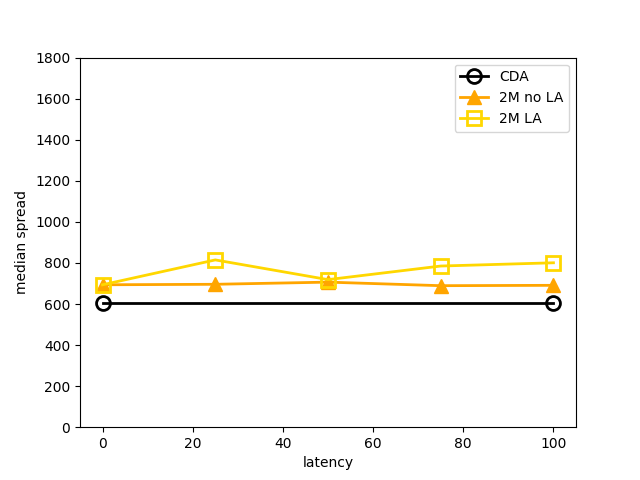}
    \caption{Env. 3 -- \textit{BestGuess+MS+bug}}
    \label{fig:env3_bbo_bg_ms_buggy_alt_v_bug}
\end{subfigure}%
\caption{Mean median spreads from the exchange BBOs for our \textit{BestGuess+MS} and \textit{BestGuess+MS+bug} implementations of the model.
These results should be compared to the corresponding original figures from WW \cite{wah_latency_2016} (Fig. 7, p.87). BBO spreads are averaged across the different exchanges in a given simulation run. Results from our simulation are averaged over 5,000 mixtures, 100 runs per mixture, for each experiment.}
\label{fig:ww_bbo_spreads_alt_v_buggy}
\end{figure}

We present the graphical results from the two implementations along with the WW results in Figs. \ref{fig:ww_surplus_alt_v_buggy}, \ref{fig:ww_extime_alt_v_buggy}, \ref{fig:ww_bbo_spreads_alt_v_buggy}, and \ref{fig:ww_trans_alt_v_buggy}. We overall see similar behavior between these implementations. Most results are relationally equivalent to each other, while the cases with differences (e.g. the mean execution times in Environment 3, Fig.~\ref{fig:ww_extime_alt_v_buggy}) favor \textit{BestGuess+MS} for more closely aligning to the original results.

\begin{figure}[H]
\centering
\begin{subfigure}[t]{0.32\textwidth}
    \centering
    \includegraphics[width=1\textwidth]{env1_mean_median_nbbo_spread_alt.png}
    \caption{Env. 1 -- \textit{BestGuess+MS}}
    \label{fig:env1_nbbo_alt_v_bug}
\end{subfigure}%
\begin{subfigure}[t]{.32\textwidth}
    \centering
    \includegraphics[width=1\textwidth]{env2_mean_median_nbbo_spread_alt.png}
    \caption{Env. 2 -- \textit{BestGuess+MS}}
    \label{fig:env2_nbbo_alt_v_bug}
\end{subfigure}%
\begin{subfigure}[t]{.32\textwidth}
    \centering
    \includegraphics[width=1\textwidth]{env3_mean_median_nbbo_spread_alt.png}
    \caption{Env. 3 -- \textit{BestGuess+MS}}
    \label{fig:env3_nbbo_alt_v_bug}
\end{subfigure}%
\hspace{1em}
\begin{subfigure}[t]{0.32\textwidth}
    \centering
    \includegraphics[width=1\textwidth]{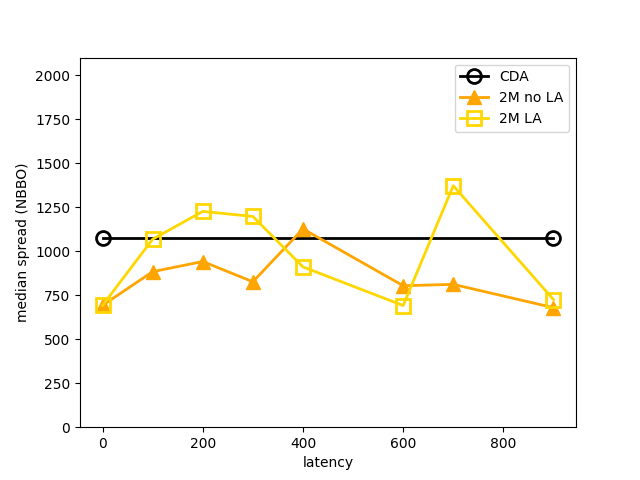}
    \caption{Env. 1 -- \textit{BestGuess+MS+bug}}
    \label{fig:env1_nbbo_alt_v_buggy}
\end{subfigure}%
\begin{subfigure}[t]{.32\textwidth}
    \centering
    \includegraphics[width=1\textwidth]{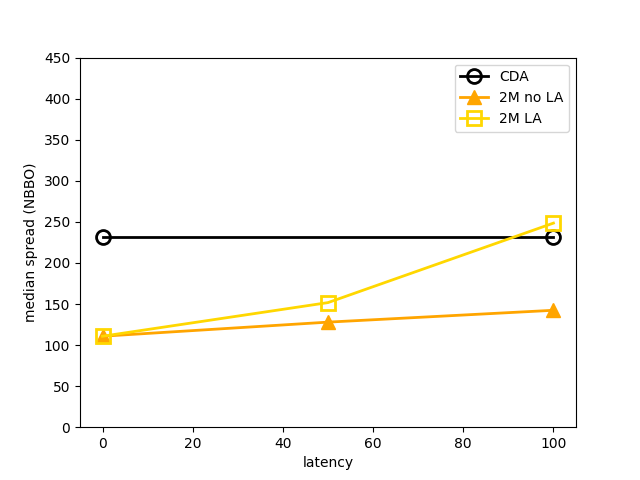}
    \caption{Env. 2 -- \textit{BestGuess+MS+bug}}
    \label{fig:env2_nbbo_alt_v_buggy}
\end{subfigure}%
\begin{subfigure}[t]{.32\textwidth}
    \centering
    \includegraphics[width=1\textwidth]{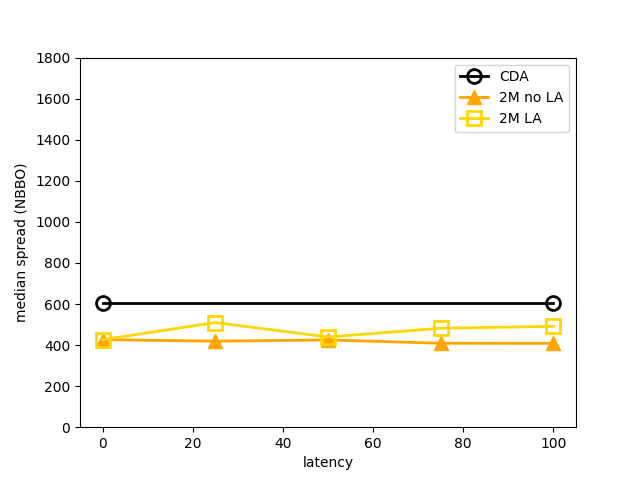}
    \caption{Env. 3 -- \textit{BestGuess+MS+bug}}
    \label{fig:env3_nbbo_alt_v_buggy}
\end{subfigure}%
\caption{Mean median SIP NBBO spread for our \textit{BestGuess+MS} and \textit{BestGuess+MS+bug} implementations of the model versus the figures reproduced from WW \cite{wah_latency_2016} (Fig. 7, p.87). Results from our simulation are averaged over 5,000 mixtures, 100 runs per mixture, for each experiment.}
\label{fig:ww_nbbo_spreads_alt_v_buggy}
\end{figure}

\begin{figure}[H]
\centering
\begin{subfigure}[t]{0.32\textwidth}
    \centering
    \includegraphics[width=1\textwidth]{env1_mean_transactions_alt.png}
    \caption{Env. 1 -- \textit{BestGuess+MS}}
    \label{fig:env1_trans_alt_v_buggy}
\end{subfigure}%
\begin{subfigure}[t]{.32\textwidth}
    \centering
    \includegraphics[width=1\textwidth]{env2_mean_transactions_alt.png}
    \caption{Env. 2 -- \textit{BestGuess+MS}}
    \label{fig:env2_trans_alt_v_buggy}
\end{subfigure}%
\begin{subfigure}[t]{0.32\textwidth}
    \centering
    \includegraphics[width=1\textwidth]{env3_mean_transactions_alt.png}
    \caption{Env. 3 -- \textit{BestGuess+MS}}
    \label{fig:env3_trans_alt_v_buggy}
\end{subfigure}%
\hspace{1em}
\begin{subfigure}[t]{0.32\textwidth}
    \centering
    \includegraphics[width=1\textwidth]{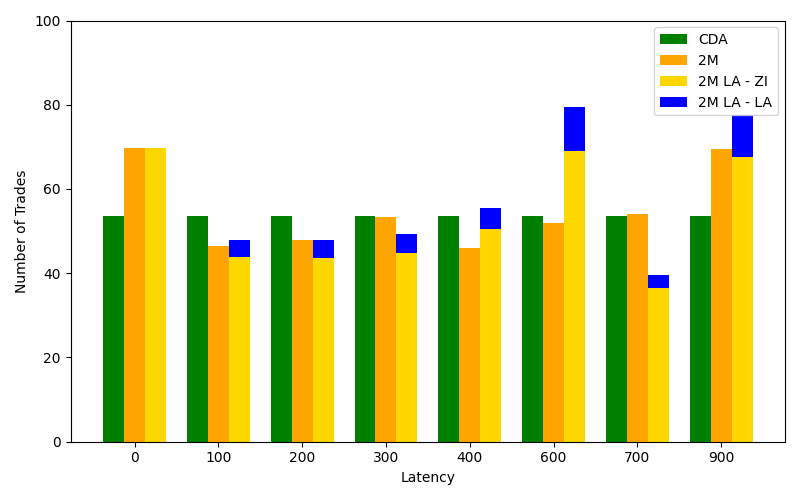}
    \caption{Env. 1 -- \textit{BestGuess+MS+bug}}
    \label{fig:env1_trans_buggy}
\end{subfigure}%
\begin{subfigure}[t]{.32\textwidth}
    \centering
    \includegraphics[width=1\textwidth]{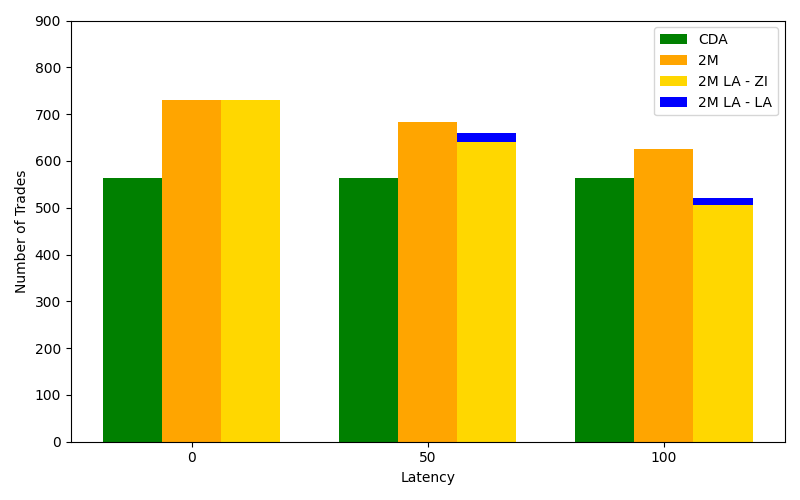}
    \caption{Env. 2 -- \textit{BestGuess+MS+bug}}
    \label{fig:env2_trans_buggy}
\end{subfigure}%
\begin{subfigure}[t]{0.32\textwidth}
    \centering
    \includegraphics[width=1\textwidth]{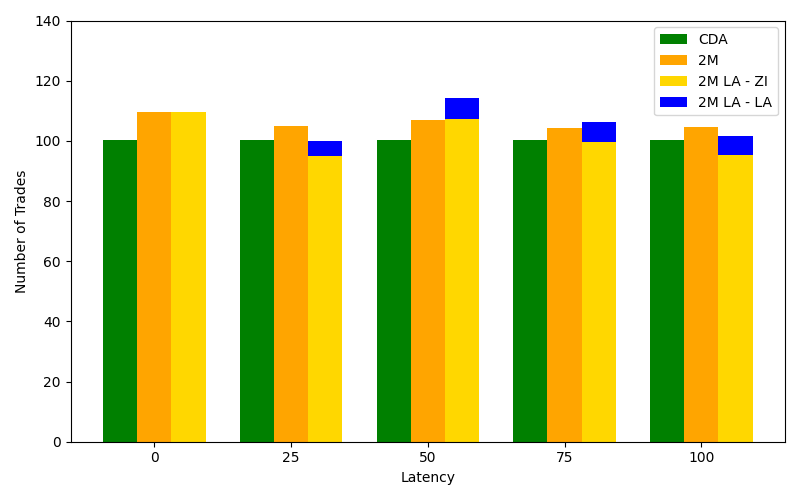}
    \caption{Env. 3 -- \textit{BestGuess+MS+bug}}
    \label{fig:env3_trans_buggy}
\end{subfigure}%
\caption{Mean number of transactions from our \textit{BestGuess+MS} and \textit{BestGuess+MS+bug} implementations of the model.
These results should be compared to the corresponding original figures from WW \cite{wah_latency_2016} (Fig. 11, p.86). Please note, a single trade results in two transactions for this measurement (one per trader involved in the trade). Results from our simulation are averaged over 5,000 mixtures, 100 runs per mixture, for each experiment.}
\label{fig:ww_trans_alt_v_buggy}
\end{figure}

\section{ODD Protocol}
\label{ww_ODD}
Below, we give the model description, which follows the ODD (Overview, Design concepts, Details) protocol \cite{grimm_standard_2006,grimm2010odd} for our replication of Wah and Wellman's 2016 model (WW) \cite{wah_latency_2016}. We follow as closely as possible the model as specified by the text of WW, and we include annotations where appropriate for specific pieces of logic. We also make note of any ambiguities from the text for which we needed to make a specific design decision (e.g. specifically how a number drawn from a continuous distribution is converted into an integer).

\subsection{Overview}

\subsubsection{Purpose}

This model examines the effects of fragmentation, latency, and latency arbitrage in a financial market.
The model features a single stock traded by investor agents in a market with one or more stock exchange.
A consolidated market with a single exchange is modeled first, after which an additional exchange, communication latency, and a latency arbitrage agent are incrementally added to examine their respective effects.
The hypothesis is that unequal access to cross-market information impacts the efficient allocation of value to investors and the liquidity of the market overall.

\subsubsection{Entities, state variables, and scales}
\begin{table}[H]
\centering
\renewcommand{\arraystretch}{1.05}
\begin{tabular}{p{0.26\textwidth}p{0.69\textwidth}}
\hline
\textbf{Entities} & \textbf{Scheduler}: schedules and kicks off events in the simulation \\
& \textbf{Exchange}: market operating a continuous double auction (CDA) \\
& \textbf{Security}: fungible asset traded on each exchange \\
& \textbf{Security Information Processor (SIP)}: data consolidator for the market; maintains and publishes the National Best Bid and Offer (NBBO) \\
& \textbf{Traders:} \\
& Zero Intelligence (ZI) background traders \\
& Latency Arbitrage (LA) trader \\
\hline
\textbf{State variables} & \textbf{Scheduler} \\
& Current time $t$ \\
& List of scheduled events (ZI arrivals and SIP updates) \\[0.3em]
& \textbf{Exchange} \\
& Limit order book (LOB): all resting orders with prices, quantities, and timestamps \\
& Best bid and offer (BBO): highest buy price and lowest sell price on the exchange \\
& BBO subscribers \\[0.3em]
& \textbf{SIP} \\
& Exchange BBO(s): Current BBO feeds from the exchange(s) \\
& National Best Bid and Offer (NBBO) \\
& NBBO subscribers \\[0.3em]
& \textbf{Security} \\
& Fundamental value $r_t$ \\[0.3em]
& \textbf{ZI trader} \\
& Valuation vector $\Theta_i$: marginal private benefits for buying and selling the asset \\
& Strategy parameters $\{R_{\min}, R_{\max}, \eta\}$ \\
& Primary exchange \\
& Primary exchange BBO \\
& Current holdings: quantity $q_i$ the trader is long or short \\
& Outstanding orders \\
& Profits from trading \\[0.3em]
& \textbf{LA trader (if present)} \\
& Exchange BBOs \\
& Profits from trading \\[0.3em]
\hline
\textbf{Scales} & $T$: Simulation length \\
& $N_{exch}\in[1,2]$: Number of exchanges \\
& $\delta\geq 0$: SIP latency \\
& $N_{ZI}$: Number of ZI traders \\
& $N_{LA}\in [0,1]$: Presence or absence of the LA trader \\
& $\lambda$: ZI arrival rate \\
& $q_{\max}$: maximum shares ZI trader can be long or short the asset \\
\hline
\end{tabular}
\caption{Entities, state variables by entity, and scales in the model ODD protocol.}
\end{table}

\subsubsection{Process overview and scheduling}
\label{process_overview_ODD}

Time is discrete, with steps $t\in{1,\ldots,T}$. A scheduler maintains a queue of exogenous events, ordered by time and then by insertion order (first-in, first-out). Only two types of events go through the scheduler queue: ZI trader arrivals and SIP NBBO update events (scheduled with latency $\lambda$ after an exchange BBO change).
All other actions in the model are triggered immediately (with zero modeled latency) as consequences of scheduled events. Non-scheduled actions include order submissions and cancellations, exchange matching and trade execution, exchange BBO updates, LA trader responses, and notifications of trades.


\subsection{Design Concepts}
\subsubsection{Basic principles}
Underlying the model is the concept that fragmentation in a market leads to communication latency. This latency affects different classes of traders unequally.
Retail and institutional investors are assumed to trade based on some understanding of the fundamental value of the security and additional private factors. These investors are less invested in speed technologies than the HFT traders and therefore have some exposure to stale information in the market relative to HFT traders. This is represented in the model through a simplified latency arbitrage scenario, with the SIP connecting the markets representing possibly stale information (when $\delta>0$) relative to the direct access the HFT arbitrage agent has to prices at each exchange. These assumptions (fragmentation, speed/information differentials, and high-level trading strategies) determine the structure and individual behaviors in the model, and the model provides insight into the impact these features have on the system overall through various market-level performance characteristics.

\subsubsection{Emergence}
The key results examined by WW are the surplus achieved by traders, mean execution time of orders that result in trades, the number of transactions, and the median spread between bid and ask prices in the market. These features are emergent from the interactions of traders adding orders in the simulation.

\subsubsection{Adaptation}
No details/parameters of the ZI or LA strategies change over the course of the simulation in order to improve profitability.

\subsubsection{Objectives}
Traders aim to maximize their surplus (private value plus profit), but they are not strategic beyond what they are hard-coded to do. 
The ZI traders aim to achieve immediate execution if possible through their routing strategy, which would serve to increase its surplus.
The LA agent submits orders immediately at prices ensured to execute and achieve immediate profit

\subsubsection{Learning}
The agents do not learn or adapt their traits over the course of the simulation.

\subsubsection{Prediction}
No prediction is done by the agents.

\subsubsection{Sensing}
Traders use data feeds in the market to sense the prices available. This determines whether to act in the LA agent's case, and it determines the order routing decision for the ZI agents.

\subsubsection{Interaction}
Traders interact through the matching of their orders in the LOBs at the exchanges. The resting orders at each exchange inform the routing decisions of the subsequent traders placing orders. Trader actions on the LOBs are communicated via BBO and NBBO updates, which are received as they are updated. These updates occur instantly for the BBOs and with latency $\delta$ for the NBBO after events in the model.

\subsubsection{Stochasticity}
The fundamental price of the security is a mean-reverting stochastic process. ZI trader's internal valuations and order prices are assumed to be stochastic perturbations on top of the security's fundamental value. Exact arrival times of ZI agents in the market are also stochastic, with interarrival times drawn from an exponential distribution.

\subsubsection{Collectives}
Equal sized groups of ZI traders are associated with each of the exchanges. This association determines which exchange's direct BBO feed the trader receives and determines the trader's default routing decision. These features in turn factor into the trader's specific actions. Aside from this difference between groups of ZI traders, there are no collectives in the model.

\subsubsection{Observation}
\label{observation}
The evolution of the order books, including trades executed in the market, are logged by the exchange BBOs and the SIP NBBO.
Every time an order is submitted to an exchange, this results in a trade or an order added to the LOB. As a result, the exchange's BBO and the SIP NBBO are updated.
Trades are tracked as a global variable that is updated by the exchange(s) any time a trade occurs.
The trade, BBO, and NBBO records are therefore updated in event-time.
These event-time observations are used at the end of the simulation to calculate the performance metrics, which are then logged.

\subsection{Initialization}
\label{init}

Before each model run, the traders, exchanges, and SIP are instantiated.
First, we create the exchanges and the SIP.
The SIP's NBBO and exchange BBOs are initialized to have null values as their best bid and best offer. 
The SIP subscribes to the direct BBO feeds from the exchanges, and the exchanges each subscribe to the NBBO feed from the SIP.

The security traded in the market is also initialized.
We pre-compute the security's fundamental value series $\{r_t\}$ for each integer time point in the simulation (i.e. $t\in[1,T]$) using the logic in Section \ref{security}.
The calculation of $\{r_t\}$ at initialization is for computational efficiency; it could be calculated at each time step, and we do not believe this would impact the model.

\paragraph{ZI initialization}
Next, we create the ZI traders. 
Each ZI trader is assigned one of the markets as its primary market. In the single-exchange configuration of the market, all ZI traders are assigned to the only exchange. In the two-market configurations, we alternate which exchange an agent is assigned to each time we create an agent. We therefore evenly split the ZI traders between the two exchanges, as there are no environments that feature an odd number of ZI agents.
Each ZI agent is assigned trading strategy parameters according to the specified mixture probabilities, specified in Table \ref{table:env_mixtures}.
 We sample ZI trader profiles according to the surplus-maximizing equilibrium mixture probabilities provided by WW for the given experiment (environment + latency + number of LA).
 The surplus-maximizing equilibrium mixture probabilities are determined by combining the information in Tables 6--11 in WW \cite{wah_latency_2016} (pp.~83--86).
 Let a `mixture' refer to a sample of $N_{ZI}$ profiles according to the mixture probabilities. A mixture is stochastically drawn based on the probabilities for a given run. Each profile from the mixture is assigned to one of the ZI agents.
ZI agents subscribe to the BBO feed from their assigned primary exchange and to the NBBO feed from the SIP.
ZI agents initialize their marginal private benefits vector $\Theta$ according to the logic detailed in Section \ref{ZI}.

\paragraph{Schedule initialization}
The Scheduler is initialized, with initial arrivals for the ZI agents scheduled. WW say, \say{The background traders arrive at the market according to a Poisson process with rate $\lambda$} \cite{wah_latency_2016} (p.75). They do not specify how the ZI arrival times are made to be discrete.
We draw the ZI arrival times independently from an exponential distribution ($\lceil Exponential(1/\lambda)\rceil$).
We choose to take the ceiling such that no event will be scheduled at time $t=0$. We account for this by letting arrivals occur on times 1 through $T$ (inclusive), giving $T$ total time steps.

\paragraph{LA initialization}
Finally, if the given model configuration includes an LA agent, it is also initialized. Mainly, the LA agent subscribes to both exchanges' BBO feeds.

\subsection{Input data}
The model does not use input data aside from the ZI strategy mixture probabilities (discussed above) and the parameters, which are specified in the next subsection.

\subsection{Parameters}

The three environments specified by WW are given in Table \ref{table:ww16_env}. The parameter $N_{ZI}$ is the number of ZI traders, $\lambda$ is the arrival rate of ZI traders, $\kappa$ is the strength of mean reversion of the security's fundamental value, $T$ is the simulation length, and $\Delta_\delta$ is the granularity of latency steps across experiments for a given environment. 
The parameters that are fixed across environments are shown in Table \ref{table:default_params}.

\begin{table}[H]
\centering
\begin{tabular}{|r|r|r|r|r|r|}
\hline
Environment & $N_{ZI}$ & $\lambda$ & $\kappa$ & $T$ & $\Delta_{\delta}$ \\
\hline
    1 & 24 & 0.05 & 0.05 & 15,000 & 100 \\
    2 & 238 & 0.005 & 0.02 & 10,000 & 10 \\
    3 & 58 & 0.005 & 0.02 & 5,000 & 10 \\
\hline
\end{tabular}
\caption{Parameter settings for the three market environments from WW \cite{wah_latency_2016} (p. 80). Please note, the $\lambda$ parameter for Environments 1 and 2 has an apparent typo ($\lambda=0.0005$) in WW's Table 3 \cite{wah_latency_2016}; the 0.005 value is specified in the text (\cite{wah_latency_2016}, p.80) and in Wah's dissertation \cite{wah_computational_2016}.}
\label{table:ww16_env}
\end{table}

\begin{table}[H]
\centering
\begin{tabular}{|p{0.15\textwidth}|p{0.15\textwidth}|p{0.6\textwidth}|}
\hline
Parameter &  Default &  Description \\
\hline
    $\overline{r}$ & 100,000 & Mean fundamental value of security. \\
    $\sigma_{shock}^2$ & 5,000,000 & Variance of the random shock to the security's fundamental value. \\
    $\sigma_{PV}^2$ & 5,000,000 & Variance of ZI private valuations; used to draw private marginal utility of the agent's net position. \\
    $\alpha$ & 0.001 & Threshold from which the LA agent determines whether an opportunity is worth pursuing. \\
    $q_{\max}$ & 10 & Maximum position (long or short) for ZI traders \\
\hline
\end{tabular}
\caption{Default parameters used across model configurations. These were specified by WW \cite{wah_latency_2016} (p. 80) except for $\alpha$. The $\alpha$ value is based on the 2013 paper \cite{wah_latency_2013} and Wah's dissertation \cite{wah_computational_2016}, and it is further confirmed by looking at the default value $\alpha$ in \textit{MarketSim}.}
\label{table:default_params}
\end{table}

\begin{table}[H]
\centering
\begin{tabular}{|l|r|r|r|}
\hline
ID & $R_{min}$ & $R_{max}$ & $\eta$ \\
\hline
$ZI_1$ & 0 & 125 & 1 \\
$ZI_2$ & 0 & 250 & 1 \\
$ZI_3$ & 0 & 500 & 1 \\
$ZI_4$ & 250 & 500 & 1 \\
$ZI_5$ & 0 & 1000 & 1 \\
$ZI_6$ & 500 & 1000 & 0.4 \\
$ZI_7$ & 500 & 1000 & 1 \\
$ZI_8$ & 0 & 1500 & 0.6 \\
$ZI_9$ & 1000 & 2000 & 0.4 \\
$ZI_{10}$ & 0 & 2500 & 0.4 \\
$ZI_{11}$ & 0 & 2500 & 1 \\
\hline
\end{tabular}
\caption{ZI strategy combinations, taken from WW \cite{wah_latency_2016}, p. 80.}
\label{Table:zi_strategies}
\end{table}

\begin{landscape}
\begin{table}[H]
\centering
\begin{tabular}
{|l|l|r|r|r|r|r|r|r|r|r|r|r|r|}
\hline
Env & Model   &  Latency  &  $ZI_1$   &  $ZI_2$   &  $ZI_3$   &  $ZI_4$   &  $ZI_5$   &  $ZI_6$   &  $ZI_7$   &  $ZI_8$   &  $ZI_9$   &  $ZI_{10}$    &  $ZI_{11}$ \\
\hline
1   & CDA     &  -        &  0        &   0       &   0       &   0       &   0       &   0       &   0       &   0       &   0.507       &   0.493       &   0    \\
1   & 2M      &  0        &  0    &   0       &   0       &   0       &   0       &   0           &   0       &   0       &   0           &   1.0         &   0     \\
1   & 2M (no LA)  &  100  &  0    &   0       &   0       &   0       &   0       &   0.602       &   0       &   0       &   0.239       &   0.159       &   0     \\
1   & 2M (LA) &  100      &  0    &   0       &   0       &   0       &   0       &   0.237       &   0       &   0       &   0.537       &   0.226       &   0     \\
1   & 2M (no LA) &  200   &  0    &   0       &   0       &   0       &   0       &   0.381       &   0       &   0       &   0.338       &   0.281       &   0     \\
1   & 2M (LA) &  200      &  0    &   0       &   0       &   0       &   0       &   0           &   0       &   0       &   0.679       &   0.321       &   0     \\
1   & 2M (no LA)  &  300  &  0    &   0       &   0       &   0       &   0       &   0.692       &   0       &   0       &   0.036       &   0.272       &   0     \\
1   & 2M (LA) &  300      &  0    &   0       &   0       &   0       &   0       &   0           &   0       &   0       &   0.655       &   0.345       &   0     \\
1   & 2M (no LA)  &  400  &  0    &   0       &   0       &   0       &   0       &   0           &   0       &   0       &   0.595       &   0.405       &   0     \\
1   & 2M (LA) &  400      &  0    &   0       &   0       &   0       &   0       &   0.47        &   0       &   0       &   0.258       &   0.272       &   0     \\
1   & 2M (no LA)  &  600  &  0    &   0       &   0       &   0       &   0       &   0.81        &   0       &   0       &   0           &   0.19        &   0     \\
1   & 2M (LA)     &  600  &  0    &   0       &   0       &   0       &   0       &   0           &   0.029   &   0       &   0           &   0.971       &   0     \\
1   & 2M (no LA)  &  700  &  0    &   0       &   0       &   0       &   0       &   0.739       &   0       &   0       &   0           &   0.261       &   0     \\
1   & 2M (LA)     &  700  &  0    &   0       &   0       &   0       &   0       &   0.006       &   0       &   0       &   0.826       &   0.168       &   0     \\
1   & 2M (no LA)  &  900  &  0    &   0       &   0       &   0       &   0       &   0           &   0       &   0       &   0           &   1.0         &   0     \\
1   & 2M (LA)     &  900  &  0    &   0       &   0       &   0       &   0       &   0.131       &   0       &   0       &   0           &   0.869       &   0     \\
\hline
2   & CDA     &  -        &  0        &   0       &   0       &   0       &   0       &   0       &   0       &   0       &   0       &   0.659       &   0.341    \\
2   & 2M      &  0        &  0.146    &   0       &   0       &   0       &   0       &   0       &   0       &   0       &   0       &   0.854       &   0     \\
2   & 2M (no LA) &  50    &  0        &   0.162   &   0       &   0       &   0       &   0       &   0       &   0       &   0       &   0.832       &   0     \\
2   & 2M (LA) &  50       &  0        &   0       &   0.188   &   0       &   0       &   0       &   0       &   0       &   0       &   0.812       &   0     \\
2   & 2M (no LA) &  100   &  0.051    &   0       &   0       &   0       &   0       &   0       &   0       &   0       &   0       &   0.76        &   0.189  \\
2   & 2M (LA) &  100      &  0        &   0       &   0       &   0       &   0       &   0       &   0       &   0       &   0.233   &   0.767       &   0     \\
\hline
3   & CDA     &  -        &  0    &   0       &   0       &   0       &   0       &   0           &   0       &   0       &   0.248       &   0.752       &   0    \\
3   & 2M      &  0        &  0    &   0       &   0.017   &   0       &   0       &   0           &   0       &   0       &   0.004       &   0.979       &   0     \\
3   & 2M (no LA)  &  25   &  0    &   0       &   0       &   0       &   0       &   0           &   0       &   0       &   0           &   0.854       &   0.146 \\
3   & 2M (LA) &  25       &  0    &   0       &   0       &   0       &   0       &   0           &   0       &   0       &   0.21        &   0.79        &   0     \\
3   & 2M (no LA)  &  50   &  0    &   0       &   0       &   0       &   0       &   0           &   0       &   0       &   0           &   0.948       &   0.052 \\
3   & 2M (LA) &  50       &  0    &   0       &   0       &   0       &   0       &   0           &   0       &   0.065   &   0.043       &   0.892       &   0     \\
3   & 2M (no LA) &  75    &  0    &   0       &   0       &   0       &   0       &   0           &   0       &   0       &   0           &   0.823       &   0.177 \\
3   & 2M (LA) &  75       &  0    &   0       &   0       &   0       &   0       &   0           &   0       &   0       &   0.142       &   0.858       &   0     \\
3   & 2M (no LA) &  100   &  0    &   0       &   0       &   0       &   0       &   0           &   0       &   0       &   0           &   0.839       &   0.161 \\
3   & 2M (LA) &  100      &  0    &   0       &   0.015   &   0       &   0       &   0           &   0       &   0       &   0.231       &   0.754       &   0     \\
\hline
\end{tabular}
\caption{ZI surplus-maximizing equilibrium mixtures for each experiment (environment+configuration+latency). ZI strategy settings ($ZI_i$) are as specified in Table~\ref{Table:zi_strategies}.}
\label{table:env_mixtures}
\end{table}
\end{landscape}


\subsection{Simulation Driver}
\label{sim_driver}
The simulation driver initializes the simulation by creating the scheduler, exchanges, traders, security, and SIP. It kicks off the initialization functions for each of these according to the logic in Section \ref{init}. This results in a population of exchanges, traders, a security with a time-dependent fundamental value, and data feeds in the market. The simulation driver then kicks off the schedule to run through events. The scheduler adds events along the way as messages are sent by agents and as ZI agents reschedule themselves for arrival. When the scheduler finishes all events, the simulation driver kicks off the functions to save the data from the simulation for analysis.

\subsection{Scheduler}
\label{scheduler}
The scheduler maintains the evolving schedule of events, kicking them off based on their time.
WW say, \say{[E]vents are maintained in a queue ordered by time of occurrence} \cite{wah_latency_2016}, and Wah's dissertation says, \say{Multiple events may be scheduled for the same time step, in which case they are executed deterministically in the order in which they are enqueued} \cite{wah_computational_2016} (p.16--17).
We therefore assume events are prioritized first by time, then by order of scheduling (i.e. first-in-first-out for each time step).

An event that has sub-events will kick off those related actions directly instead of going through the scheduler. When the simulation is started, the scheduler kicks off each event based on its time and the order in which it was enqueued for the current time step. Once there are no more events at the current time step, the scheduler skips ahead to the next time step which has a scheduled event.
No events are scheduled for $t>T$.

\subsection{Exchanges}
\label{exchange}

Each exchange in the model operates a continuous double auction (CDA)\footnote{
Wah and Wellman \cite{wah_latency_2013,wah_latency_2016} also examined an alternative clearing mechanism, periodic call auctions, as a possible remedy to latency arbitrage. 
All exchanges in the U.S. stock market currently use CDAs, however, and we leave the periodic call auction models out of scope for now.
} for the asset.
A CDA is a two-sided market where agents submit orders to buy and sell the asset. 
The market is `continuous' in the sense that orders and trades occur as soon as they are received and processed by the exchange's matching engine\footnote{There is a time resolution below which the time of two different messages cannot be differentiated, but for modern exchanges this is on the magnitude of microseconds or even nanoseconds \cite{angel_when_2014,van_oort_adaptive_2023}.}.
A limit order\footnote{
A market order is an alternative type of order which executes immediately at the best available price. All agents in the model exclusively submit limit orders, however, and so market orders are out of scope.
} to buy (sell) the stock specifies a maximum (minimum) amount the agent is willing to trade at and a quantity to buy (sell). 
In the model, all orders are for a single unit of the stock. 
Each exchange maintains a limit order book (LOB), constructed with the following elements:
\begin{enumerate}
    \item Orders: map from order IDs to their respective orders
    \item Bids: map from price to the queue of buy order IDs at that price
    \item Asks: map from price to the queue of sell order IDs at that price
\end{enumerate}
The exchange publishes a BBO to its subscribers, a feed containing the best bid and offer prices in the LOB at that moment. 
LOBs are initialized with no bids or asks at the start. The BBO has null prices for its best available bid and ask whenever their respective side of the LOB is empty (including at initialization).

When an exchange receives an add message for order $O_i$ with limit price $p$, it does the following. First, it checks whether the added price can be matched against existing orders in the book. More explicitly, if order $O_i$ is a buy order, the exchange checks whether there is a resting sell order $O_j$ with price $q\leq p$. Similarly, if $O_i$ is a sell order, the exchange checks whether there is a resting buy order $O_j$ with price $q\geq p$. In either of these cases, the orders are matched at price $q$, resulting in a transaction. If multiple resting orders exist on the LOB that could match with the incoming order, the oldest order is given priority (i.e. price-time priority). In this simulation, all orders are for a single quantity of the asset, and so each trade is always for a single unit. Receipts are sent to the traders associated with $i$ and $j$, and $O_j$ is removed from the LOB.
If $O_i$ instead does \textit{not} immediately match, it is added to the LOB. 
A trader can request to withdraw a resting order from the LOB. Upon receiving a withdraw request, the exchange removes the corresponding order from the LOB.

After any change to the exchange's LOB due to a withdraw, add, or trade, the exchange publishes its updated best bid and ask prices to its subscribers. This is published regardless of whether the best bid or ask prices were changed by the LOB update.
The exchange loops over its subscribers in the order they subscribed to the exchange's BBO feed, directly calling the subscriber's \texttt{update\_bbo()} function.

The market can be configured with one or two exchanges.
Each ZI trader agent has a `primary' exchange. In the two-market configurations, the ZI agents are split evenly between the two exchanges and subscribe only to the BBO feed of their primary exchange.
The SIP subscribes to both exchange feeds, as does the latency arbitrage (LA) agent if it is present.

\subsection{The SIP}
\label{sip}
The role of the SIP in the simulation is to consolidate information from across the fragmented market and publish the national best bid and offer (NBBO). This process is kicked off whenever the SIP receives a BBO update from an exchange.
We assume that when latency $\delta$ is zero the SIP should update its NBBO (and subscribers) without going through the scheduler. In our implementation, this means when the SIP receives an \texttt{update\_bbo()} call from an exchange and $\delta=0$, the SIP directly calls its own \texttt{update\_nbbo()} function. This prevents agents acting on the same time step from having stale information due to the SIP being queued behind their already-scheduled action.
When $\delta>0$, we assume an \texttt{update\_nbbo()} call is scheduled at time step $t+\delta$ using the Scheduler, with no priority preference. In other words, if a ZI agent is already scheduled for that time step, the NBBO update will occur after that agent makes its trading action (and any resultant actions).

When \texttt{update\_nbbo()} is called, the NBBO is calculated by finding the best BID (ASK) at one exchange and then seeing whether the next exchange has a higher (lower) BID (ASK). If neither exchange has a BID (ASK) order, the NBBO BID (ASK) is empty. WW do not specify how ties for the best BID or ASK should be handled. We arbitrarily award ties to the first exchange in the list.
NBBO updates specify the time $t$ of the update, the best BID price and exchange, and the best ASK price and exchange. Once calculated, NBBO updates are published to the subscribers of the SIP in the order in which they subscribed. This is done by looping through the subscribers and directly calling each subscriber's \textit{update\_nbbo} function.

\subsection{Security}
\label{security}

The security being traded in the market is mostly abstract, except it possesses an evolving `common fundamental value' known by the traders. As noted in Section~\ref{init}, we pre-compute the fundamental value series $\{r_t\}$ for efficiency reasons.
This series is calculated following the below logic:
\begin{lstlisting}
    values = {}
    r = r_bar
    for i in [1,...,T]:
        shock = N(0, sigma_shock^2)
        r = max(0, kappa * r_bar + (1-kappa) * r + shock)
        fundamental_value[i] = r
\end{lstlisting}
The mean is $\overline{r}$ (\texttt{r\_bar}), the degree of mean-reversion is $\kappa$ (\texttt{kappa}), and $\mu_t\sim N\left(0,\sigma_{\text{shock}}^2\right)$ is a Gaussian random shock drawn independently for each time step.

WW do not specify what $r_0$ should be, so we assume $r_0=\overline{r}$. WW also do not specify how or whether the fundamental value should be made to be an integer (since prices are discrete in the model). We choose not to discretize the fundamental value when it is being calculated.
A ZI trader arriving at time $t$ looks up $r_t$ as part of its trading strategy (Section~\ref{ZI}). 
We assume $r_t$ should be rounded to the nearest integer whenever it is looked up by a trader.

The ZI traders use the estimated terminal value of the security in their trading strategy. We model this logic as a function of the security, which the ZI agents call.
The estimate is based on the fundamental value $r_t$ at the time $t$ the trader looks it up. The estimated terminal value $\hat{r}_t$ is then the expected fundamental value at time $T$, accounting for the mean-reversion process.
WW do not specify how $\hat{r}_t$ is made to be an integer, nor whether each component to the price should be converted to int or just the final price. We assume that only the estimated terminal value should be rounded after the calculation of $\hat{r}_t$. In other words,
$$\hat{r}_t = \texttt{round}\left(\left(1 - (1 - \kappa)^{T-t}\right)\overline{r} + (1 - \kappa)^{T-t}r_t\right).$$

\subsection{Zero Intelligence (ZI)}
\label{ZI}

\subsubsection{Initialization}

As noted in Section \ref{init}, each ZI trader subscribes to the SIP NBBO and to the BBO feed from its primary exchange.
Its holdings, profits, and number of transactions are all initialized to zero.
Next, for the given ZI trader $ZI_i$, we construct $\Theta_i$, the vector of differences in private benefits of changes to the trader's net position:
$$\Theta_i=\{\theta_i^q\},$$
where $-q_{\max}<q\leq q_{\max}$ and the $\theta_i$'s are sorted such that $\theta^q\geq \theta^{q+1}$ $\forall{q}$.
No discretization logic is specified by WW, nor do they state anywhere that the internal component for the ZI agents should be integers. Hence, we do not round or otherwise discretize the $\theta$ values when they are initialized.

\subsubsection{Arrival and trading strategy}

The ZI traders are scheduled to initially arrive at the market on a fixed schedule drawn from a Poisson distribution, as detailed in Section \ref{init}.
Each ZI trader arrives to place an order at least once. 
We assume the first thing the ZI trader $ZI_i$ does upon arrival at time $t$ is schedule its next arrival for time $t+\lceil b_i\rceil$, where $b_i\sim Exponential\left(\frac{1}{\lambda}\right)$.
WW do not specify how ZI arrival times should be made discrete. We choose to take the ceiling so the trader does not schedule itself to arrive again on the same time step.

After scheduling its next arrival, $ZI_i$ withdraws any outstanding orders from previous turns that have not yet resulted in a trade.
It then determines its trading action for this turn.
It randomly chooses (with equal probability) whether to buy or sell on this turn.
Traders are not allowed to have a net position of $q_{\max}$ shares long or short of the asset. Thus, if the decision to trade or buy would result in a net position $|q_|>q_{\max}$, the trader does not submit an order and exits.

Otherwise, $ZI_i$ looks up the estimated terminal fundamental value $\hat{r}_t$ of the security (see Section~\ref{security}).
It then calculates its internal valuation $v_i(t)$ of the asset based on $\hat{r}_t$ and its marginal private benefits vector $\Theta_i$:
\[ v_i(t)=\hat{r}_t + \begin{cases} 
      \theta_i^{q_t+1} & \text{if buying one unit} \\
      \theta_i^{q_t} & \text{if selling one unit} \\
   \end{cases}.
\]
We choose to round $v_i(t)$ to the nearest integer, as WW do not specify how prices are made to be discrete.
$ZI_i$'s order price $p$ is `shaded' some amount away from its private valuation in order to achieve surplus value.
\[ p\sim \begin{cases} 
    \max\left(0, U[v_i(t)-R_{max},v_i(t)-R_{min}]\right) & \text{if buying} \\
    \max\left(0, U[v_i(t)+R_{min},v_i(t)+R_{max}]\right) & \text{if selling} \\
   \end{cases},
\]
where $U$ here denotes an integer uniformly chosen at random within the specified range.
The requested surplus for $ZI_i$ is then $s_i(t)=|v_i(t)-p_i(t)|$.

Before submitting its order, $ZI_i$ determines whether there is a greedy opportunity based on $p$, the trader's $\eta$ parameter, and the current market information. As detailed in the main paper, we assess two interpretations of this greedy strategy; one implementation based on our `best guess' based on the text, and one implementation based on \textit{MarketSim}. We detail these two implementations below.

\paragraph{\textit{BestGuess} greedy strategy}
WW say the ZI agents have access to quotes on their primary market and the NBBO. Let \texttt{NBBO} denote the current $NBBO_t$ received from the SIP as of time $t$, and let \texttt{BBO} denote $ZI_i$'s primary exchange $BBO_t$ as of time $t$. Based on these sources, $ZI_i$ determines the best bid price and the best ask:
\begin{lstlisting}
    max_bid = NBBO.bid.price
    min_ask = NBBO.ask.price.

    if BBO.bid != None and (max_bid == None or BBO.bid >= max_bid):
        max_bid = bbo.bid
    if BBO.ask != None and (min_ask == None or BBO.ask <= min_ask):
        min_ask = BBO.ask
\end{lstlisting}

$ZI_i$ determines whether there is a greedy opportunity based on $p$, the trader's $\eta$ parameter, and \texttt{max\_bid} and \texttt{bid\_ex} if selling (or \texttt{min\_ask} and \texttt{ask\_ex} if buying). If so, $ZI_i$ will update its order price $p$ to the observed available price it is trying to immediately execute at.
\begin{lstlisting}
requested_surplus = |valuation - p|
if side == `buy':
    if min_ask != None and
            (requested_surplus * eta <= valuation - min_ask):
        p = min_ask
else: // `sell'
    if max_bid != None and
            (requested_surplus * eta <= max_bid - valuation):
        p = max_bid
\end{lstlisting}

\paragraph{\textit{MarketSim} greedy strategy}
In the \textit{MarketSim} implementation of the greedy strategy, $ZI_i$ does not look at the NBBO when determining whether there is a greedy opportunity, and it updates $p=v_i(t)$ when there is a greedy opportunity:

\begin{lstlisting}
requested_surplus = |self.valuation - price|
if side == `buy':
    if BBO.ask != None and 
            requested_surplus * eta <= valuation - BBO.ask:
        price = valuation
else: // `sell'
    if BBO.bid != None and
            requested_surplus * eta <= BBO.bid - valuation:
        price = valuation
\end{lstlisting}

\paragraph{Order submission}
After the price is finalized (for either greedy strategy implementation), the order is then submitted to the exchange determined by the below logic.
Assume \texttt{order.price} is set to $p$ and \texttt{order.exchange} is set to $ZI_i$'s primary exchange. Also assume \texttt{BBO} is $ZI_i$'s primary exchange's current BBO and \texttt{NBBO} is the current NBBO received from the SIP.
\begin{lstlisting}
submit_nms_order(order):
    NBBO_price_better = False
    will_transact = False
    if order.type == `BUY':
        if (NBBO.ask.price != Null and (BBO.ask == Null or 
                NBBO.ask.price < BBO.ask)):
            NBBO_price_better = True
        if order.price >= NBBO.ask.price:
            will_transact = True
        alt_ex = NBBO.ask.exchange
    else:
        if (NBBO.bid.price != Null and (BBO.bid == Null or 
                NBBO.bid.price > BBO.bid)):
            NBBO_price_better = True
        if order.price <= NBBO.bid.price:
            will_transact = True
        alt_ex = NBBO.bid.exchange

    if NBBO_price_better and will_transact:
        order.exchange = alt_ex
        
    order.exchange.add_order(order)
\end{lstlisting}

\subsubsection{Calculation of surplus}
At the end of the simulation, we calculate the total surplus $S_i$ for an agent $ZI_i$ with a terminal net position $q_i^T$ and valuation vector $\Theta_i=\{\theta_j\}$.
This is calculated as the sum of the following: the terminal value of its net position, the net cash flow of the trader from its trades over the course of the simulation, and the private benefits it receives from its net position.
Let \texttt{q\_T} denote the terminal net position of $ZI_i$, let \texttt{r\_T} denote the terminal fundamental value of the security, and let \texttt{profits} be the net cash flow from trades. Then, at the end of the simulation, we calculate the following for the trader's surplus:
\begin{lstlisting}
surplus = profits
surplus += q_T * r_T
private_value = 0
if q_T > 0:
    for q in [1, q_T]:
        private_value += thetas[q]
elif q_T < 0:
    for q in [q_T+1, 0]:
        private_value -= thetas[q]

surplus += private_value
\end{lstlisting}

\subsubsection{Latency Arbitrage (LA)}
\label{LA}

The latency arbitrageur (LA) in the two-market model operates as follows. At initialization, it subscribes to the BBO updates from both exchanges. 
During the simulation, whenever the LA agent's \texttt{update\_bbo()} function is called by an exchange, it updates its internal BBO for that exchange and then evaluates whether there is an arbitrage opportunity:
\begin{lstlisting}
if executing_strategy:
    return

max_bid = None
min_ask = None
for exchange, BBO in BBOs:
    if max_bid == None or
            (BBO.bid != None and BBO.bid > max_bid):
        max_bid = BBO.bid.
        bid_ex = exchange
    if min_ask == None or
            (BBO.ask != None and BBO.ask < min_ask):
        min_ask = BBO.ask
        ask_ex = exchange
if max_bid != None and min_ask != None and
        max_bid > (1 + self.alpha) * min_ask:
    executing_strategy = True
    midpoint = (max_bid + min_ask) / 2
    submit_order(ask_ex, `buy', price=floor(midpoint), q=1)
    submit_order(bid_ex, `sell', price=ceil(midpoint), q=1)
    executing_strategy = False
\end{lstlisting}
Note that we implement a lock to ensure the LA strategy can complete in full before it submits new orders. Otherwise it might receive a quote update from the exchange and think there's a new arbitrage opportunity, kicking off more orders prior to the other leg of the original strategy finishing.
When its orders result in trades (which they are guaranteed to do, since it is not subject to any latency) the LA agent adds the cash flow to its total profits. For a given arbitrage opportunity it acts on, it achieves a profit of $BID^* - ASK^*$ from the resulting trades.
The surplus for the LA trader is just its profits (i.e. net cash flow).

\subsection{Metric calculation}
\label{metrics}

After the scheduler finishes its scheduled events, we calculate the following metrics on the events from the simulation: the total ZI surplus, LA profits, mean execution time, median spread, and the number of transactions.

\paragraph{Spread}
WW describe the spread as the amount by which the best ask ($ASK$) exceeds the best bid ($BID$).
They report the median spreads for the BBO and the NBBO.
They do not address how to handle a locked or crossed market when calculating spread. 
We assume we should remove observations where the BID or ASK are empty or where the market was crossed (i.e. require ASK $\geq$ BID)\footnote{\textit{MarketSim} calculates spread as infinity when $BID$ or $ASK$ are null or when $ASK < BID$: \url{https://github.com/egtaonline/market-sim/blob/marketsim1/hft-sim/src/entity/infoproc/BestBidAsk.java\#L37}}.
WW also do not clarify whether the median is calculated on the spread from each quote update or if the spread should be calculated once per time step. We assume it should be calculated for each update (i.e. tick time, rather than clock time).
The NBBO median spread is then taken as the median value from the spreads calculated for each NBBO update where the BID and ASK are both non-empty and ASK $\geq$ BID. 
One other important note on the NBBO spread is it is not clear whether NBBO updates scheduled for $t>T$ should be included in the calculation (which can happen when latency $\delta>0$).
We assume no events are executed for $t>T$, and thus these NBBO updates are ignored in our implementation.

When there is a single exchange in the market, the BBO spread is the median of all spreads on BBO updates where the BID and ASK are both non-empty\footnote{Note that $ASK > BID$ at the exchange-level by definition of the CDA.}. For configurations with two exchanges, the median is calculated for each exchange BBO and then the average is taken between the two.

\paragraph{Execution time}
WW say, \say{We measure execution time by the interval between order submission and transaction for orders that eventually trade} \cite{wah_latency_2016} (p.79). We assume the execution time should be calculated for both legs of the trade, therefore meaning at least one leg will have an execution time of zero for each trade.
Thus, the mean execution time is the average time to execution (i.e. execution time minus submission time) for all orders that resulted in trades.

\paragraph{Transactions}
The logic to calculate the number of transactions is inferred from Fig. 11 in WW \cite{wah_latency_2016} (p.90) and the behavior of the model. Namely, transactions are the total \textit{orders} that resulted in trades throughout the simulation, separated by the type of trader whose order it was. Thus, a trade between two ZI traders results in two ZI transactions, while a trade between a ZI trader and the LA agent is logged as one ZI transaction and one LA transaction.

\paragraph{Surplus}
Finally, surplus is calculated as follows. The total ZI surplus is calculated by summing the terminal \texttt{surplus} values for each $ZI_i$ in the simulation (see Section \ref{ZI} for how the \texttt{surplus} is calculated for each ZI trader). The LA trader only has the profit it achieved from the net cash flow of its trades. We take the LA agent's terminal profits as its total surplus.

\end{sloppypar}

\end{document}